\pdfoutput=1
\documentclass[12pt,a4paper]{article}

\usepackage[T1]{fontenc}\usepackage{lmodern}
\usepackage[utf8]{inputenc}
\usepackage[english]{babel}
\usepackage{comment}

\usepackage[a4paper,left=20mm,right=20mm,top=22mm,bottom=22mm]{geometry}

\usepackage[table,dvipsnames]{xcolor} %

\usepackage{hyperref}
\hypersetup{
    colorlinks = true,
    linkcolor  = black,  %
    citecolor  = black,  %
    urlcolor   = black   %
}

\usepackage{amsmath,amssymb,amsfonts}
\IfFileExists{dsfont.sty}{\usepackage{dsfont}}{\let\mathds\mathbf}
\usepackage{algorithm}
\usepackage{algorithmic}

\usepackage{graphicx}
\usepackage{float}

\usepackage{array,multirow}

\newcolumntype{R}[1]{>{\raggedleft\arraybackslash}b{#1}}  %
\newcolumntype{L}[1]{>{\raggedright\arraybackslash}b{#1}} %
\newcolumntype{C}[1]{>{\centering\arraybackslash}b{#1}}   %

\usepackage{rotating}
\usepackage{booktabs}
\usepackage{longtable}
\usepackage{diagbox}

\usepackage{caption}
\usepackage{subcaption}

\usepackage{natbib}

\usepackage{url}
\usepackage{adjustbox}
\usepackage{enumerate}

\newtheorem{theorem}{Theorem}

\newtheorem{remark}{Remark}
\newtheorem{lm}{Lemma}
\newtheorem{assumption}{Assumption}

\numberwithin{equation}{section}

\usepackage{tikz}
\usetikzlibrary{arrows.meta, positioning, calc, backgrounds}

\begin{document}
\hypersetup{hidelinks}
\renewcommand{\theHequation}{\theHsection.\arabic{equation}}

\title{Efficient and Robust Variable Selection in Large-Scale Data via $L^k$-Quantile Regression with Distributed Computation and Inference}
\author{Abdellah Atanane\textsuperscript{1,2,*}\quad Abdallah Mkhadri\textsuperscript{2}\quad Karim Oualkacha\textsuperscript{1}\\[0.7em]
\small\textsuperscript{1}Department of Mathematics, Université du Québec à Montréal,\\
\small 201 avenue Président-Kennedy, Montreal, Quebec, H2X 3Y7, Canada\\[0.3em]
\small\textsuperscript{2}Department of Mathematics, Cadi Ayyad University,\\
\small Abdelkrim Khattabi Avenue, Marrakesh 40000, Marrakesh-Safi, Morocco\\[0.5em]
\small\textsuperscript{*}Corresponding author: \texttt{atanane.abdellah@courrier.uqam.ca}}
\date{}
\maketitle

\begin{abstract}
Large-scale data analysis poses substantial challenges, including storage constraints, the presence of skewed distributions, heteroskedastic variance structures, and intricate dependencies between outcomes and high-dimensional covariates. Distributed storage offers an effective solution by alleviating the burden on any single system. The $L^{k}$-quantile regression framework, which extends the concepts of both quantile and expectile regression, has gained popularity due to its robustness and effectiveness, especially for $1<k\leq 2$. In the context of large-scale distributed data, this work investigates penalized $L^k$-quantile regression incorporating the smoothly clipped absolute deviation (SCAD) and adaptive LASSO (aLASSO) penalties, both of which are effective for variable selection and achieving sparse solutions in high-dimensional settings. To address the distributed nature of the data, a communication-efficient surrogate likelihood---termed the CSL\(^k\) (Communication-efficient Surrogate for \( L^k \)-quantile regression)---is proposed as a proxy for the global loss function. A distributed optimization algorithm, based on the parallel alternating direction method of multipliers (ADMM), is developed to minimize the proposed penalized CSL\(^k\) objective function. Oracle properties of the resulting estimators are established {for fixed dimension and a fixed number of machines,} under suitable choices of tuning parameters. The finite-sample performance of the method is evaluated through comprehensive simulation studies, and its practical utility is demonstrated through an application to a real dataset.
\end{abstract}


\section{Introduction}

As modern datasets continue to grow rapidly in both sample size $N$ and parameter dimension $p$, it has become infeasible to process all the data on a single machine, necessitating its distribution across multiple machines. Distributed statistical inference has therefore attracted considerable attention, encompassing various topics such as Bayesian approaches \citep{wang2013}, $M$-estimation \citep{lee2015}, nonparametric regression \citep{szabo2019}, among others. Traditional centralized methods for parameter estimation and hypothesis testing are no longer practical due to excessive communication costs and privacy concerns. A primary challenge in the distributed setting is inter-machine communication. Thus, any effective distributed statistical learning framework must achieve statistical optimality while minimizing communication overhead \citep{Wang2024}.

Numerous algorithms have been proposed for analyzing massive distributed datasets. Among them, the One-Shot (OS) method stands out for its simplicity and communication efficiency. The OS approach first computes local estimators on each machine and then aggregates them (e.g., by averaging) to obtain a global estimator. Notable extensions include averaging over subspaces \citep{fan2019} and de-biased estimators \citep{battey2018distributed}. Further developments are found in \cite{lee2015}, \cite{zhang2015divide}, and \cite{li2020distributed}. A key advantage of the OS method is that it requires only a single round of communication. However, studies such as \cite{zhang2013} and \cite{rosenblatt2016} have shown that its performance deteriorates when per-machine sample sizes are limited or when model complexity increases. To address these limitations, multi-round distributed methods have been proposed. These methods alternate between local computation and global aggregation over several iterations, leading to improved statistical efficiency. Notable contributions include the communication-efficient surrogate likelihood (CSL) approach by \cite{jordan2019} and frameworks for high-dimensional regression and Bayesian inference by \cite{fan2023communication}.

In parallel, an alternative line of research has focused on subsampling and divide-and-conquer strategies as a means to reduce both computational and communication burdens.
For quantile regression with massive data, \cite{chao2024distributed} proposed a distributed optimal subsampling framework that achieves statistical efficiency while operating on carefully selected local subsets.
In the generalized linear model setting, \cite{yin2025outcome} developed an outcome-dependent subsampling divide-and-conquer procedure, showing that informative subsampling can substantially improve estimation accuracy under limited computational resources. Related divide-and-conquer ideas have also been extended to the expectile framework, where \cite{zeng2025divide} studied generalized approximately expectile regression for massive datasets and established asymptotic properties of the resulting estimators.

Beyond computational constraints, distributed data environments often exhibit additional structural challenges,
including data heterogeneity, non-random data allocation across machines, and privacy restrictions.
Addressing these issues, \citet{liu2025privacy} investigated privacy-preserving estimation
for non-randomly distributed data and demonstrated that ignoring such heterogeneity
may lead to biased inference.
These considerations further motivate the development of robust and communication-efficient
distributed inference methods that remain valid under complex data-generating mechanisms.

Another fundamental challenge in large-scale data analysis is heterogeneity, often arising from heteroscedasticity or non-location-scale error structures. In such settings, the conditional mean is insufficient to capture the full relationship between predictors and response. Quantile regression (QR) by \cite{koenker1978regression} and expectile regression (ER) by \cite{newey1987asymmetric} have emerged as important alternatives, each focusing on different parts of the conditional distribution. QR and ER are based on minimizing asymmetric \(L^1\) and \(L^2\) loss functions, respectively.

A central challenge in contemporary regression  is to strike a balance between robustness and statistical efficiency. One approach is to combine multiple convex losses, leading to hybrid methods that leverage the strengths of each component. Recent examples include composite quasi-likelihood techniques \citep{bradic2011penalized}, hybrid-risk models for financial systems \citep{fan2018single}, and the hybrid quantile--expectile framework of \citet{atanane2025efficient}. As an alternative to hybridization, the $L^k$-quantile regression model \citep[see][]{zha2021, jiang2021k, li2022improved, lin2022k, atanane2026distributed} achieves a similar robustness--efficiency trade-off by smoothing the quantile loss through a power parameter $k>1$. This article builds on the latter perspective. For a given \( \tau \in [0,1] \), the population \(\tau\)-th \(L^k\)-quantile of a random variable \(Y\), assuming the existence of the $(k-1)$-th moment, is defined as

\begin{equation}\label{tau th Lkquantile}
Q_{\tau,k}(Y) = \arg \min_{\theta \in \mathbb{R}} \mathbb{E}\left\{ \ell_{\tau,k}(Y - \theta) -\ell_{\tau,k}(Y)\right\},  
\end{equation}
where the generalized asymmetric loss function \( \ell _{\tau,k}(\cdot)\) is given by
\begin{equation}\label{Lk quantile loss}
\ell_{\tau,k}(s) =\Psi_{\tau}(s)\cdot |s|^k, \quad 1 \leq k \leq 2,
\end{equation}
and \( \Psi_\tau(s)=|\tau - \mathds{1}\{s<0\}| \), with \( \mathds{1}\{A\} \) denoting the indicator function. Clearly, QR and ER correspond to the special cases \(k = 1\) and \(k = 2\), respectively \citep{jiang2021k}.

Following its initial proposal, extensive research has been conducted on QR, ER, and \(L^k\mathrm{QR}\). For instance, \cite{breckling1988, schnabel2009, sobotka2012} have explored the theoretical and practical aspects of QR and ER. \cite{sobotka2013} established asymptotic properties of the ER estimator and constructed corresponding confidence intervals. \cite{ziegel2016} demonstrated that expectiles are elicitable and coherent. \cite{zhao2018variable} showed the robustness of sample expectiles under heavy-tailed distributions, and their smooth behavior across levels of \( \tau \). A number of recent studies have explored communication-efficient strategies for distributed estimation across expectile, quantile, and related regression frameworks. In the expectile setting, \citet{pan2020} examined conditional expectile estimation with missing covariates, while \citet{pan2021} and  \citet{pan2021efficient} developed distributed expectile regression procedures based on SCAD and adaptive LASSO penalties for data randomly distributed across machines. For quantile regression, \citet{Xi2020} proposed a sparse high-dimensional distributed estimator by transforming the check loss into a least squares problem, and \citet{Wang2024} introduced a double-smoothing approach to address the non-smooth nature of quantile loss, enabling efficient estimation and support recovery in high dimensions. Related work on composite quantile regression includes the adaptive distributed smooth CQR estimator of \citet{WANG2025108110}, the robust communication-efficient CQR method with variable selection proposed by \citet{WANG2021107262}, and the communication-efficient quantile matrix regression framework of \citet{YANG2023107812}. Beyond the quantile and expectile settings, \citet{LI2025108154} extended communication-efficient methodologies to high-dimensional longitudinal data, highlighting the increasing versatility of distributed inference techniques in large-scale applications.

More recently, increasing attention has been given to \( L^k\mathrm{QR} \). \cite{jiang2021k} applied it for heteroscedasticity detection, while \cite{arnroth2024bayesian} proposed a Bayesian version, highlighting its robustness and efficiency. Recent works such as \cite{gao2024communication} and \cite{li2024optimalp} have extended \( L^k\mathrm{QR} \) to large datasets, employing score projection techniques or subsampling strategies in low-dimensional settings. However, these methods rely on projection scores that become unstable near zero, leading to suboptimal efficiency. As residuals approach zero, the projection matrix becomes ill-conditioned, resulting in only marginal gains over global estimators \citep{li2024optimalp}. These limitations also affect our method, as discussed later in this paper. Despite these advances, the study of \( L^k\mathrm{QR} \) in distributed environments, particularly in terms of communication efficiency and variable selection, remains largely unexplored and warrants further investigation. 

 We address this gap by proposing a communication-efficient \( L^k\mathrm{QR} \) approach for {large-scale distributed data.} The global \( L^k \) loss is approximated by a surrogate loss on a designated machine, with local nodes transmitting only gradient information, thus reducing communication costs. We establish the asymptotic normality of the resulting estimator, showing that it matches the full-data estimator in efficiency under the conditions in Remark~6. Our proposed \( \mathrm{CSL}^k \) loss function alleviates the computational burden on the central node, and we construct a distributed estimator via \(M\)-estimation and convex analysis. To tackle optimization challenges, we incorporate the ADMM algorithm across worker machines. Furthermore, we propose a distributed iterative procedure. At each step, machines interact with a central processor to exchange and update information. Through multiple rounds of communication, each machine refines its estimates. The method exploits the structure of the data and the computing power available at each node. Compared to OS and CSL approaches that rely heavily on a single machine, {the accuracy depends on the configuration and local sample size; the numerical comparisons do not establish uniform superiority or convergence for every finite-budget implementation.} 

 While our method builds upon the ideas of \( L^k\mathrm{QR} \) \citep{gao2024communication} and incorporates CSL-inspired techniques \citep{jordan2019}, it should not be seen as a mere extension. The surrogate loss function we introduce differs substantially, as it leverages the full capacity of all machines while relaxing the smoothness assumptions (e.g., twice differentiability) typically required by CSL-based frameworks. As noted in \citet{gao2024communication} and \citet{li2024optimalp}, such assumptions often fail in nonsmooth or hybrid loss settings, introducing significant theoretical challenges---particularly in establishing asymptotic results. To overcome these issues, we employ tools from \(M\)-estimation theory and convex analysis to derive the asymptotic properties of the proposed estimator under the new \( \mathrm{CSL}^k \) loss.

 Compared to subsampling- or divide-and-conquer-based approaches
\citep{chao2024distributed,yin2025outcome,zeng2025divide},our framework directly targets communication efficiency through surrogate loss construction,
while retaining full-sample asymptotic efficiency {under the fixed-dimensional conditions stated below.}

Our work distinguishes itself by extending distributed composite-type estimation to the penalized \( L^k \)-quantile framework, enabling both robust estimation and sparsity recovery under communication constraints. To facilitate practical use, an \textsf{R} package, \texttt{dlkqr}, implementing the proposed algorithm {is publicly available at} \url{https://github.com/abdellah-atanane/dlkqr}.

The remainder of this paper is organized as follows. In \mbox{Section~\ref{Sec. 2}}, we describe how to fit the linear \(L^k\mathrm{QR}\) model in a distributed framework and introduce the surrogate \( \mathrm{CSL}^k \) loss. In \mbox{Section~\ref{Sec. 4}}, we present a distributed algorithm to compute the estimator. {\mbox{Section~\ref{sec:51-implementation}} details the implementation, regularization path and computation.} \mbox{Section~\ref{Sec. 3}} establishes its asymptotic properties. \mbox{Section~\ref{Sec. 5}} contains simulation studies. An application to data from the Stanford HIV Drug Resistance Database is presented in \mbox{Section~\ref{Sec. 6}}. Concluding remarks appear in \mbox{Section~\ref{Sec. 7}}, and all technical proofs are presented in  the Appendix.

\section{Methodology} \label{Sec. 2}

\noindent Let \( Y \in \mathbb{R} \) denote a response variable, and let \( \boldsymbol{X} = (X_1, \dots, X_p)^\top \in \mathbb{R}^p \) denote a \( p \)-dimensional vector of covariates. Suppose we have a set of \( N \) independent and identically distributed (i.i.d.) random samples \( \{ \boldsymbol{X}_i, Y_i \}_{i=1}^N \) with the same law as \(  (\boldsymbol{X},Y) \). Consider the linear model:

\begin{equation*}
  Y_i = \boldsymbol{X}_i^\top \boldsymbol{\beta} + \varepsilon_i,  
\end{equation*}
where \( \boldsymbol{\beta} = (\beta_1, \dots, \beta_p)^\top \) is the regression parameter vector, the true value of which is \( \boldsymbol{\beta}^\star = (\beta^\star_{1}, \dots, \beta^\star_{p})^\top \) and \( \varepsilon_i \) is an unknown random error term that satisfies the $\tau$-th $L^k$-quantile defined in \eqref{tau th Lkquantile} is zero for $i=1,\ldots,N$, i.e., $Q_{\tau,k}(\varepsilon_i|\boldsymbol{X}_i) := \arg \min_{\theta \in \mathbb{R}} \mathbb{E}\left\{ \ell_{\tau,k}(\varepsilon_i - \theta) -\ell_{\tau,k}(\varepsilon_i)|\boldsymbol{X}_i\right\}=0.
$
Building on both quantile and expectile regression, \cite{jiang2021k} proposed the $L^k\mathrm{QR}$ estimator
\begin{equation}\label{Lk quantile estimator}
    \boldsymbol{\tilde{\beta}}_N(\tau,k)
    = \arg\min_{\boldsymbol{\beta}\in \mathbb{R}^{p}}
    \frac{1}{N}\sum_{i=1}^{N}\!\left[\ell_{\tau,k}\!\big(Y_i-\boldsymbol{X}_i^\top \boldsymbol{\beta}\big)-\ell_{\tau,k}(Y_i)\right],
\end{equation}
where $\ell_{\tau,k}(\cdot)$ is defined in \eqref{Lk quantile loss}. For notational simplicity, we henceforth drop $\tau$ and $k$ from the parameters and the error term. 
Following this definition, the estimator's key properties (existence/uniqueness, consistency, asymptotic normality, and related tests) are developed in \cite{jiang2021k} and \cite{lin2022k}.

\noindent
However, when the entire dataset is distributed across multiple machines, solving the optimization problem in equation \eqref{Lk quantile estimator} becomes impractical. We consider a distributed context in which the samples \( \{ \boldsymbol{X}_i, Y_i \}_{i=1}^N \) are divided among \( V \) machines. We assume that each machine has an identical sample size of \( n \), so that \( N = nV \). The data on the \( \nu \)-th machine, where \( \nu = 1, \dots, V, \) is represented as \( \{ Y_{\nu i}, \boldsymbol{X}_{\nu i} = (X_{\nu i1}, \dots, X_{\nu ip})^\top \}_{i=1}^n \).
Define the local and the global objective functions, respectively, as follows:
\begin{equation}\label{local loss function}
\mathcal{R}_\nu (\boldsymbol{\beta}) = \frac{1}{n} \sum_{i=1}^n[  \ell_{\tau,k}(Y_{\nu i} - \boldsymbol{X}_{\nu i}^\top \boldsymbol{\beta})-\ell_{\tau,k}(Y_{\nu i})],\quad \nu=1,2,\ldots,V,
\end{equation}
and 
\begin{equation}\label{global loss function}
\mathcal{R}_N (\boldsymbol{\beta}) = \frac{1}{n V} \sum_{i=1}^n\sum_{\nu=1}^V [\ell_{\tau,k}(Y_{ \nu i} - \boldsymbol{X}_{\nu i}^\top \boldsymbol{\beta})-\ell_{\tau,k}(Y_{\nu i})].
\end{equation}
Note that \( \mathcal{R}_N (\boldsymbol{\beta}) =\frac{1}{V} \sum_{\nu=1}^V  \mathcal{R}_\nu (\boldsymbol{\beta})   \). The best statistical error can be obtained using the global optimization approach based on the global loss function in equation  \((\ref{global loss function})\). However, since each machine only has access to its local data,  minimizing directly $\mathcal R_N(\boldsymbol{\beta})$ is costly. Therefore, we rely on the CSL methodology described in \cite{jordan2019} to construct a proxy for the objective function in equation  \eqref{global loss function}, which only requires communication of the gradient of the local data from the other machines. In a distributed setting, starting with an initial estimator $\boldsymbol{\beta}^0$, only the gradient vectors \( \nabla \mathcal{R}_\nu(\boldsymbol{\beta}^0) \), \( \nu = 1, \dots, V \), can be easily communicated, where \( \nabla \) denotes the gradient with respect to \( \boldsymbol{\beta} \). Inspired by \cite{jordan2019}, we propose a communication-efficient surrogate likelihood based on the $L^k\mathrm{QR}$ loss function ($\mathrm{CSL}^k$) as 

\begin{equation}\label{csl loss without penalty}
  \mathcal{R}_{N}(\boldsymbol{\beta}; \boldsymbol{\beta}^0 ) := \mathcal{R}_\nu(\boldsymbol{\beta}) +  [\nabla \mathcal R_{N,\nu}(\boldsymbol{\beta}^0)]^\top \boldsymbol{\beta},   
\end{equation}
where $\nabla \mathcal{R}_{N,\nu}(\cdot):= \nabla \mathcal{R}_{N}(\cdot)- \nabla \mathcal{R}_{\nu}(\cdot),$ $\nabla \mathcal{R}_\nu (\boldsymbol{\beta}) = -\frac{1}{n} \sum_{i=1}^n  \boldsymbol{X}_{\nu i}  \ell'_{\tau,k}(Y_{\nu i} - \boldsymbol{X}_{\nu i}^\top \boldsymbol{\beta}),\quad \nu=1,2,\ldots,V;$
 $\nabla \mathcal{R}_N (\boldsymbol{\beta}) = -\frac{1}{nV} \sum_{i=1}^n \sum_{\nu=1}^V \boldsymbol{X}_{\nu i} \ell'_{\tau,k}(Y_{ \nu i} - \boldsymbol{X}_{ \nu i}^\top \boldsymbol{\beta}),$ with 

\begin{equation}\label{ell prime}
\ell'_{\tau,k}(s) = k\,\Psi_\tau(s)\,|s|^{k-1}\mathrm{sgn}(s),\quad s\in \mathbb{R},\, 1<k\leq 2,
\end{equation}
where $\mathrm{sgn} (\cdot)$ is the sign function.
Then, we define the $\mathrm{CSL}^k$ estimator as

\begin{equation}\label{estimator argmin}
    \boldsymbol{\hat \beta}=\arg\min\limits_{\boldsymbol{\beta}\in\mathbb{R}^p} \mathcal{R}_N(\boldsymbol{\beta};\boldsymbol{\beta}^0).
\end{equation}
Note that $\nabla \mathcal{R}_N(\boldsymbol{\beta}) = \frac{1}{V} \sum_{\nu =1}^V \nabla \mathcal{R}_\nu(\boldsymbol{\beta})$ and that the problem \eqref{estimator argmin} can  be solved on a given machine $\nu$, while the other machines only communicate their local gradients.
{\par \label{loc:gradient}
\noindent The local gradient $\nabla\mathcal R_\nu(\boldsymbol{\beta}^0)$ uses only site $\nu$; the global gradient $\nabla\mathcal R_N(\boldsymbol{\beta}^0)$ is their average for balanced partitions. Their difference $\boldsymbol{g}_\nu^0=\nabla\mathcal R_N(\boldsymbol{\beta}^0)-\nabla\mathcal R_\nu(\boldsymbol{\beta}^0)$ is the correction vector. It is evaluated at a common outer iterate and held fixed during the corresponding inner solve.}

\begin{remark}
The proposed approach is inspired by the CSL method \citep{jordan2019}. However, challenges arise in deriving its theoretical properties. {For $1<k<2$, the loss has a continuous derivative but unbounded curvature near zero, so smooth-loss arguments requiring bounded curvature need separate verification. At $k=2$, its derivative is Lipschitz, although the one-sided curvatures differ when $\tau\ne1/2$. This is a property of the present loss, not an impossibility result attributed to \citet{jordan2019}.} In the Appendix, we address this challenge by using the technical tools associated with $M$-estimation and convex processes,  thereby extending the applicability of the CSL method to $L^k\mathrm{QR}$.

\end{remark}
\noindent
Set $m:=\# \{l\in[p] : \beta^\star_l \neq 0\},$  the cardinality of the active variables and assume that $m<<p$ (sparsity assumption). For variable selection when both $p$ and $N$ are large with $p<<N$, we add a penalty term to the $\mathrm{CSL}^k$ function in equation  \eqref{csl loss without penalty}. This avoids overfitting the model and improves its accuracy. We then  propose the following penalized $\mathrm{CSL}^k$ function.
{\par \label{loc:scope}
\noindent The oracle theory assumes fixed $p$ and $V$, with nonsingular limiting design matrices. A growing-dimensional extension requires uniform empirical-process bounds, restricted curvature, sparsity and beta-min conditions, and joint restrictions on $p,n,V$ and pilot error. The fixed-dimensional arguments do not cover this case. Fixed nonlinear basis expansions can be treated as covariates under the same assumptions. For nonparametric quantile regression, see \citet{wang2024communication} on random features. Nonlinear parameterizations require separate analysis.}

\begin{equation}\label{GEL penalized}
\mathcal{R}_{N}(\boldsymbol{\beta}; \boldsymbol{\beta}^0 ) = \mathcal{R}_\nu(\boldsymbol{\beta}) +  [\nabla \mathcal R_{N,\nu}(\boldsymbol{\beta}^0)]^\top \boldsymbol{\beta}  +\gamma_\lambda (|\boldsymbol{\beta}|) ,
\end{equation}

\noindent
where $\gamma_\lambda(|\boldsymbol{\beta}|):=\sum_{l=1}^{p} \gamma_{\lambda}(|\beta_l|)$, and  $\gamma_{\lambda}(\cdot)$ is a penalty function, and $\lambda$ is the regularization parameter. The smaller the value of $\lambda$, the weaker the punishment  and the more variables are selected. In contrast, the larger the value of $\lambda$, the fewer explanatory variables are selected. In this paper, we consider two types of penalties: the SCAD penalty and the adaptive LASSO penalty. First, we introduce the SCAD penalty function as follows:
\begin{equation}\label{scad penalty}
    {\gamma_{\lambda}(|s|) = \begin{cases}
    \lambda|s|, & |s|<\lambda,\\
    \dfrac{2a\lambda|s|-|s|^2-\lambda^2}{2(a-1)}, & \lambda\le |s|\le a\lambda,\\
    \dfrac{(a+1)\lambda^2}{2}, & |s|>a\lambda.
    \end{cases}}
\end{equation}
where $a > 2$ is a positive constant. A typical choice is $a = 3.7$, as proposed by \cite{fan2001}. As discussed in  \cite{zou2008}, the SCAD penalty function \eqref{scad penalty} can be locally linearly approximated (LLA) as

\begin{equation}\label{taylor for penalty}
    \gamma_{\lambda}(|\beta_l|) \approx \gamma_{\lambda}(|\beta^0_l|) + \gamma'_{\lambda}(|\beta^0_l|)(|\beta_l| - |\beta^0_l|),
\end{equation}
for $\beta_l\approx \beta^0_l$, where $\gamma'_{\lambda}(s) = \lambda \mathds{1}\left\{ \left|s\right| \le \lambda \right\} + \max(0, a\lambda - |s|)/(a - 1)\mathds{1}\left\{ |s| > \lambda\right\}$.
\noindent
Substituting the penalty \eqref{taylor for penalty} into \eqref{GEL penalized}, we obtain the SCAD penalized $\mathrm{CSL}^k$ function as

\begin{equation}\label{scad loss}
    \mathcal{ R}_N^{\text{(S)}}(\boldsymbol{\beta};\boldsymbol{ \beta}^0) = \mathcal R_\nu(\boldsymbol{\beta}) + [\nabla \mathcal{R}_{N,\nu}(\boldsymbol{\beta}^0) ]^\top \boldsymbol{\beta}  + \sum_{l=1}^{p} \gamma'_{\lambda}(\left|\beta^0_l\right|)|\beta_l| .
\end{equation}
By \eqref{scad loss}, we propose the first estimator of $\boldsymbol{\beta}$ as follows
\begin{equation}\label{SCAD pb}
    \boldsymbol{\hat \beta}^{\text{(S)}}= \arg \min_{\boldsymbol{\beta} \in \mathbb{R}^p} \mathcal{ R}_N^{ \text{(S)} }(\boldsymbol{\beta};\boldsymbol{ \beta}^0).
\end{equation}
Next, we will consider the penalized CSL$^k$  function with the adaptive LASSO penalty \citep{zou2006}, which can be rewritten from \eqref{scad penalty} as follows:
\begin{equation} \label{AL loss}
   \mathcal{{R}}_N^{\text{(A)}}(\boldsymbol{\beta};\boldsymbol{ \beta}^0) = \mathcal{R}_\nu(\boldsymbol{\beta}) + [ \nabla \mathcal{R}_{N,\nu}(\boldsymbol{ \beta}^0) ]^\top \boldsymbol{\beta}   + \lambda \sum_{l=1}^{p}  w_l \left|\beta_l\right|,
\end{equation}
where $ w_l,\, l = 1, 2, \ldots, p$ is a predetermined weight. As outlined in \cite{Huaijun2021} and \cite{Liu2023-fy}, we can choose $  w_l = \left(\left|\boldsymbol{\hat \beta}_l^{\text{(L)}}\right| + 1/n\right)^{-\iota},\; l = 1, 2, \ldots, p$ for some appropriately chosen $\iota > 0$, where $\boldsymbol{\hat \beta}^{\text{(L)}} = (\boldsymbol{\hat \beta}_l^{\text{(L)}}, l = 1, 2, \ldots, p)^\top$ denotes the  LASSO penalized $L^k$QR estimator. By (\ref{AL loss}), we propose the second estimator of $\boldsymbol{\beta}$ as follows
\begin{equation}\label{AL problem}
      \boldsymbol{\hat \beta}^{\text{(A)}} = \arg \min_{\boldsymbol{\beta} \in \mathbb{R}^p} \mathcal{ R}_N^{\text{(A)}}(\boldsymbol{\beta};\boldsymbol{ \beta}^0).
\end{equation}
Note that, in general, for the current $t$-th iteration ($t>0$) $\boldsymbol{\beta}^t$ (instead of $\boldsymbol{\beta}^0$), one can unify the problems \eqref{SCAD pb} and \eqref{AL problem}  in the following optimization problem over $\boldsymbol{\beta}$

\begin{equation}\label{admm pb}
   \mathcal R_N(\boldsymbol{\beta};\boldsymbol \beta^t):=\mathcal R^{(S\, or\, A)}_N(\boldsymbol{\beta};\boldsymbol \beta^t) = \mathcal R_\nu(\boldsymbol{\beta}) + [\nabla \mathcal R_{N,\nu}(\boldsymbol{\beta}^t) ]^\top \boldsymbol{\beta}  + \lambda \left\| \boldsymbol{w} \circ \boldsymbol{\beta} \right\|_1,
\end{equation}
where $\boldsymbol{w} = (w_1, \ldots, w_p)^\top$ is the vector of non-negative weights, and $\left\| \boldsymbol{w} \circ \boldsymbol{\beta} \right\|_1 = \sum_{l=1}^{p} w_l \left|\beta_l\right|,$ where $\circ$ denotes the Hadamard product. For SCAD penalized $L^k$-quantile linear regression, $w_l = \lambda^{-1} \gamma'_{\lambda}(\left|\boldsymbol{\beta}^{t}_l\right|)$. For adaptive LASSO penalized $L^k$-quantile  regression, $w_l = ({|\boldsymbol{\hat \beta}_l^{(L)}|} + 1/n)^{-\iota}, l=1,2,\ldots,p;\, \iota>0$.

\section{A Proximal Algorithm}
 \label{Sec. 4}

In practice, optimizing the penalized $\mathrm{CSL}^k$ function given in equation \eqref{admm pb} is challenging and complex due to the large amount of data and high dimensionality. Recently, the Alternating Direction Method of Multipliers (ADMM) algorithm has gained popularity due to its effectiveness in solving large-scale optimization problems. Below, we propose an augmented proximal ADMM algorithm as an efficient tool for solving large-scale $L^k\mathrm{QR}$ problems with sparsity constraints.\\
For ease of notation, write  $\boldsymbol{Z} = \boldsymbol{Y} - \boldsymbol{X} \boldsymbol{\beta}$, $\boldsymbol{Y} = (Y_{\nu 1}, Y_{\nu 2}, \ldots, Y_{\nu n})^\top $ is an $(n\times 1)$-dimensional vector and $\boldsymbol{X} = (\boldsymbol{X}_{\nu 1}, \boldsymbol{X}_{\nu 2}, \ldots, \boldsymbol{X}_{\nu n})^\top$ is a matrix of dimension $n \times p$. The given  problem in equation \eqref{admm pb} can be transformed by convexity into the following form 

\begin{equation}\label{problem under constarint}
\begin{aligned}
\begin{cases}
 \min\limits_{(\boldsymbol{Z},\boldsymbol{\beta} ) \in \mathbb{R}^n \times \mathbb{R}^p } M_{\tau,k}(\boldsymbol{Z}) + [\nabla \mathcal R_{N,\nu}(\boldsymbol{\beta}^t)]^\top \boldsymbol{\beta} +\lambda \left\| \boldsymbol{w} \circ \boldsymbol{\beta} \right\|_1, \\
 \text{s.t.} \quad \boldsymbol{X} \boldsymbol{\beta} + \boldsymbol{Z} = \boldsymbol{Y},
\end{cases}
\end{aligned}
\end{equation}
where $M_{\tau,k}(\boldsymbol{Z}) := \frac{1}{n} \sum_{i=1}^n \ell_{\tau,k}(Z_{i})$.
The augmented Lagrangian for solving $\eqref{problem under constarint}$ using ADMM is defined as

\begingroup\small
\begin{equation*}
L_\mu(\boldsymbol{\beta}, \boldsymbol{Z}, \boldsymbol{h}) = M_{\tau,k}(\boldsymbol{Z}) + [\nabla \mathcal R_{N,\nu}(\boldsymbol{\beta}^t)]^\top \boldsymbol{\beta} +\lambda \left\| \boldsymbol{w} \circ \boldsymbol{\beta} \right\|_1 - (\boldsymbol{X} \boldsymbol{ \beta} + \boldsymbol{Z} - \boldsymbol{Y})^\top \boldsymbol{h} + \frac{\mu}{2} \left\| \boldsymbol{X} \boldsymbol{\beta} + \boldsymbol{Z} - \boldsymbol{Y} \right\|_2^2,
\end{equation*}
\endgroup
where $\boldsymbol{h} \in \mathbb{R}^n$ is the Lagrangian multiplier, $\mu > 0$ is a tuning augmentation parameter, and $\| \cdot \|_{2}$ is the classical $L^2$ norm.\\
The ADMM algorithm alternates between the following iterative steps with given $\boldsymbol{Z}^0$ and $\boldsymbol{h}^0$.

\begin{equation*}
\begin{aligned}
\begin{cases}
\boldsymbol{\beta}^{t+1} &= \arg\min\limits_{\boldsymbol{\beta} \in \mathbb{R}^p} L_\mu(\boldsymbol{\beta}, \boldsymbol{Z}^t, \boldsymbol{h}^t), \\
\boldsymbol{Z}^{t+1} &= \arg\min\limits_{\boldsymbol{Z} \in \mathbb{R}^n} L_\mu(\boldsymbol{\beta}^{t+1}, \boldsymbol{Z}, \boldsymbol{h}^t), \\
\boldsymbol{h}^{t+1} &= \boldsymbol{h}^t - \mu (\boldsymbol{X} \boldsymbol{ \beta}^{t+1} + \boldsymbol{Z}^{t+1} - \boldsymbol{Y}), \label{h update}
\end{cases}
\end{aligned}
\end{equation*}
where $(\boldsymbol{\beta}^{t+1}, \boldsymbol{Z}^{t+1}, \boldsymbol{h}^{t+1})$ denotes the $(t + 1)$-th iteration of the algorithm on the $\nu$-th machine for $\nu \in [V]$. 
Similar iterative schemes were used by \cite{boyd2011}. By some simple computation, the iterations are
\begingroup\small
\begin{equation*}
{
\begin{aligned}
\text{$\boldsymbol{\beta}$ \bf Step:} & \quad \boldsymbol{\beta}^{t+1} = \arg\min_{\boldsymbol{\beta} \in \mathbb{R}^p}\Big\{ [\nabla \mathcal R_{N,\nu}(\boldsymbol{\beta}^t)]^\top \boldsymbol{\beta}  + \lambda \| \boldsymbol{w} \circ \boldsymbol{\beta} \|_1   - (\boldsymbol{X} \boldsymbol{\beta})^\top \boldsymbol{h}^t + \frac{\mu}{2} \| \boldsymbol{X} \boldsymbol{\beta} + \boldsymbol{Z}^t - \boldsymbol{Y} \|_2^2\Big\}, \\
\text{$\boldsymbol{Z}$ \bf Step:} & \quad \boldsymbol{Z}^{t+1} = \arg\min_{\boldsymbol{Z} \in \mathbb{R}^n}\Big\{ M_{\tau,k}(\boldsymbol{Z}) -  \boldsymbol{Z}^\top \boldsymbol{h}^t + \frac{\mu}{2} \| \boldsymbol{X} \boldsymbol{ \beta}^{t+1} + \boldsymbol{Z} - \boldsymbol{Y} \|_2^2\Big\}, \\
\text{$\boldsymbol{h}$ \bf Step:} & \quad \boldsymbol{h}^{t+1} = \boldsymbol{h}^t - \mu (\boldsymbol{X} \boldsymbol{ \beta}^{t+1} + \boldsymbol{Z}^{t+1} - \boldsymbol{Y}).
\end{aligned}
}
\end{equation*}
\endgroup

\noindent  
The update of  
\[
\boldsymbol{Z}^{t+1} = (Z^{t+1}_1, \ldots, Z^{t+1}_n)^\top
\]
is performed by a Newton--Raphson algorithm, similar to that used in the $L^k$-quantile regression model \citep[e.g.,][]{jiang2021k,gao2024communication}, to efficiently solve the proximal subproblem associated with each coordinate. This iterative scheme leverages both the first and second derivatives of the objective to refine the estimates and ensure fast local convergence.

\noindent  
In particular, each coordinate \(Z^{t+1}_i\) is updated by solving the univariate minimization problem  
\begin{equation}\label{eq:zi-step}
Z^{t+1}_{i}
= \arg\min_{Z_i \in \mathbb{R}}
\;\ell_{\tau,k}(Z_i)
+ \frac{n\mu}{2}\!\left\{
Z_i - \Big( Y_{ i} - \boldsymbol{X}_{i}^\top \boldsymbol{\beta}^{t+1} + h_i^t/\mu \Big)
\right\}^{\!2},\quad i \in [n].
\end{equation}

\noindent  
Defining the proximal operator of \( \ell_{\tau,k} (\cdot)\) as  
\begin{equation}\label{eq:prox}
\mathrm{Prox}_{\ell_{\tau,k}}(s,\xi)
:= \arg\min_{\theta \in \mathbb{R}}\;
\ell_{\tau,k}(\theta)
+ \frac{\xi}{2}(\theta - s)^2 ,
\end{equation}
we can equivalently express the update in \eqref{eq:zi-step} as  
\begin{equation}\label{eq:z-update}
Z^{t+1}_{i}
= \mathrm{Prox}_{\ell_{\tau,k}}\!\left(
Y_{ i} - \boldsymbol{X}_{i}^\top \boldsymbol{\beta}^{t+1} + h_i^t/\mu,\;
n\mu
\right),\quad \nu \in [V].
\end{equation}

\noindent  
The optimization in \eqref{eq:zi-step} is carried out componentwise, as each \(Z_i\) is independent. For general \(1 < k \le 2\), the proximal operator lacks a closed-form solution; thus, the Newton--Raphson algorithm is applied to each univariate problem to compute \(Z^{t+1}_i\). The quadratic term ensures strong convexity, providing a unique minimizer and rapid convergence.

\noindent  
In contrast, for \(k = 1.5\) and \(k = 2\), the proximal operator in \eqref{eq:prox} admits closed-form expressions, which are directly used to obtain \(Z^{t+1}_i\) without iteration.
The explicit formulas are stated below in \mbox{Remark~\ref{remark zi 6}}, while their derivations are provided in Appendix~\ref{appendix C}.

\begin{remark}\label{remark zi 6}
\cite{efron1991regression} noted that the loss \eqref{Lk quantile loss} with \(k = 1.5\) (i.e., \(k = \tfrac{3}{2}\)) offers an appealing compromise between the robustness of quantile regression (\(k = 1\)) and the normal--theoretic efficiency of expectile regression (\(k = 2\)).
\noindent
For a fixed \(\xi > 0\) and \(k = 1.5\), after some algebra, the solution to \eqref{eq:zi-step} is given by
\[
\mathrm{Prox}_{\ell_{\tau,k}}(s,\xi)
=
\begin{cases}
\displaystyle 
\frac{\big(k\tau - \sqrt{(k\tau)^2 + 4\xi^2 s}\big)^2}{4\xi^2}, & s \ge 0,\\[10pt]
\displaystyle 
-\,\frac{\big(k(1-\tau) - \sqrt{(k(1-\tau))^2 - 4\xi^2 s}\big)^2}{4\xi^2}, & s < 0.
\end{cases}
\]
\noindent
Moreover, when \(k = 2\), the proximal operator admits a simpler closed form since \(\ell_{\tau,2}(s) = \lvert \tau - \mathds{1}\{s < 0\}\rvert\,s^2\) is quadratic on each side. In this case,
\[
\mathrm{Prox}_{\ell_{\tau,2}}(s,\xi)
=
\frac{\xi\,s}{\xi + 2\big(\tau\,\mathds{1}\{s \ge 0\} + (1-\tau)\,\mathds{1}\{s < 0\}\big)}.
\]
This expression directly yields \(Z_i^{t+1}\) in closed form without any iterative refinement.
\end{remark}

{
\noindent For one outer iterate, let $\boldsymbol{g}_\nu=\nabla\mathcal R_N(\boldsymbol{\beta}^0)-\nabla\mathcal R_\nu(\boldsymbol{\beta}^0)$ be fixed. For fixed nonnegative LLA weights, the proximal update is
\begingroup\small
\[
\boldsymbol{\beta}^{t+1}=\arg\min_{\boldsymbol{\beta}\in\mathbb R^p}\left\{
 \boldsymbol{g}_\nu^\top\boldsymbol{\beta}+\lambda\|\boldsymbol{w}\circ\boldsymbol{\beta}\|_1
 -(\boldsymbol{X}\boldsymbol{\beta})^\top \boldsymbol{h}^t+\frac\mu2\|\boldsymbol{X}\boldsymbol{\beta}+\boldsymbol{Z}^t-\boldsymbol{Y}\|_2^2
 +\frac12(\boldsymbol{\beta}-\boldsymbol{\beta}^t)^\top \mathcal{K}(\boldsymbol{\beta}-\boldsymbol{\beta}^t)\right\}.
\]
\endgroup
\label{loc:prox}Choose $\mathcal{K}=\mu(\zeta\mathbb{I}_p-\boldsymbol{X}^\top\boldsymbol{X})$ with $\zeta\ge\rho_{\max}(\boldsymbol{X}^\top\boldsymbol{X})$ and $\mu>0$, where $\rho_{\max}$ is the largest eigenvalue. This matrix is positive semidefinite (definite if the inequality is strict) and makes the update separable:
\[
\boldsymbol{\eta}^t=\boldsymbol{\beta}^t+\frac{\boldsymbol{X}^\top\{\mu\boldsymbol{Y}-\mu\boldsymbol{X}\boldsymbol{\beta}^t-\mu\boldsymbol{Z}^t+\boldsymbol{h}^t\}-\boldsymbol{g}_\nu}{\mu\zeta},\qquad
\beta_l^{t+1}=\mathcal S\!\left(\eta_l^t,\frac{\lambda w_l}{\mu\zeta}\right).
\]
Here $\mathcal S(u,a)=\operatorname{sgn}(u)(|u|-a)_+$. In particular, the coordinate is exactly zero whenever $|\eta_l^t|\le\lambda w_l/(\mu\zeta)$, including at $k=2$. SCAD obtains this threshold through its LLA weights; a zero SCAD weight gives no threshold on that coordinate. Numerical support reporting must also state a tolerance. The computational benchmark reports support using $|\widehat\beta_l|>10^{-6}$. 
}

\noindent
We summarize the iterative procedure executed on each local machine in \mbox{Algorithm~\ref{alg:admm}}. 
\mbox{Algorithm~\ref{alg:admm}} is a general ADMM-type routine, adapted from the classical framework of 
\cite{boyd2011} and specialized to the computation of the CSL estimator associated with the 
$L^{k}\mathrm{QR}$ loss.

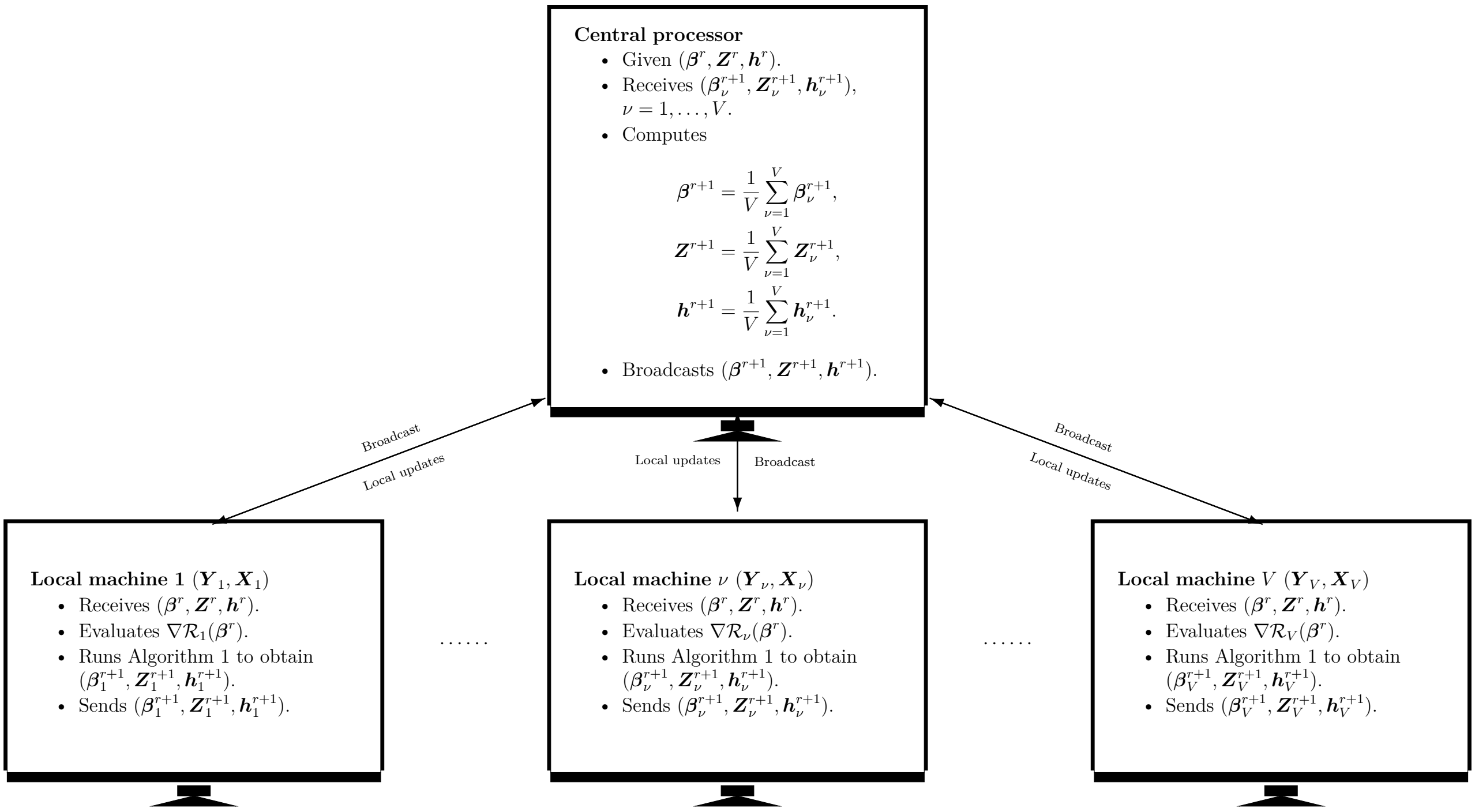
\begin{figure}[htbp]
\centering

\resizebox{0.99\textwidth}{!}{%
\begin{tikzpicture}[
    monitor/.style={
        inner sep=6pt,
        align=left,
        text width=7cm,
        minimum height=4.8cm,
        draw=none
    },
    arrow/.style={-{Latex[length=3mm,width=2mm]}, thick,
                  shorten >=12pt, shorten <=12pt}
]

\node[monitor] (central) {
\textbf{Central processor}\\[1mm]
\begin{minipage}{6.8cm}
\begin{itemize}
  \setlength\itemsep{1pt}
  \setlength\parsep{0pt}
  \setlength\parskip{0pt}
  \item Given $(\boldsymbol{\beta}^r,\boldsymbol{Z}^r,\boldsymbol{h}^r)$.
  \item Receives $(\boldsymbol{\beta}_{\nu}^{r+1},\boldsymbol{Z}_{\nu}^{r+1},
        \boldsymbol{h}_{\nu}^{r+1})$, $\nu=1,\dots,V$.
  \item Computes
    \[
      \begin{aligned}
      \boldsymbol{\beta}^{r+1} &= \frac{1}{V}\sum_{\nu=1}^{V}\boldsymbol{\beta}_{\nu}^{r+1},\\
      \boldsymbol{Z}^{r+1} &= \frac{1}{V}\sum_{\nu=1}^{V}\boldsymbol{Z}_{\nu}^{r+1},\\
      \boldsymbol{h}^{r+1} &= \frac{1}{V}\sum_{\nu=1}^{V}\boldsymbol{h}_{\nu}^{r+1}.
      \end{aligned}
    \]
  \item Broadcasts $(\boldsymbol{\beta}^{r+1},\boldsymbol{Z}^{r+1},\boldsymbol{h}^{r+1})$.
\end{itemize}
\end{minipage}
};

\node[monitor, below left=3.0cm and 4.2cm of central] (m1) {
\textbf{Local machine 1} $(\boldsymbol{Y}_1,\boldsymbol{X}_1)$\\[1mm]
\begin{minipage}{6.8cm}
\begin{itemize}
  \setlength\itemsep{1pt}
  \setlength\parsep{0pt}
  \setlength\parskip{0pt}
  \item Receives $(\boldsymbol{\beta}^r,\boldsymbol{Z}^r,\boldsymbol{h}^r)$.
  \item Evaluates $\nabla\mathcal{R}_1(\boldsymbol{\beta}^r)$.
  \item Runs \mbox{Algorithm~\ref{alg:admm}} to obtain
        $(\boldsymbol{\beta}_1^{r+1},\boldsymbol{Z}_1^{r+1},\boldsymbol{h}_1^{r+1})$.
  \item Sends $(\boldsymbol{\beta}_1^{r+1},\boldsymbol{Z}_1^{r+1},\boldsymbol{h}_1^{r+1})$.
\end{itemize}
\end{minipage}
};

\node[monitor, below=3.0cm of central] (mm) {
\textbf{Local machine $\nu$} $(\boldsymbol{Y}_{\nu},\boldsymbol{X}_{\nu})$\\[1mm]
\begin{minipage}{6.8cm}
\begin{itemize}
  \setlength\itemsep{1pt}
  \setlength\parsep{0pt}
  \setlength\parskip{0pt}
  \item Receives $(\boldsymbol{\beta}^r,\boldsymbol{Z}^r,\boldsymbol{h}^r)$.
  \item Evaluates $\nabla\mathcal{R}_{\nu}(\boldsymbol{\beta}^r)$.
  \item Runs \mbox{Algorithm~\ref{alg:admm}} to obtain
        $(\boldsymbol{\beta}_{\nu}^{r+1},\boldsymbol{Z}_{\nu}^{r+1},
          \boldsymbol{h}_{\nu}^{r+1})$.
  \item Sends $(\boldsymbol{\beta}_{\nu}^{r+1},\boldsymbol{Z}_{\nu}^{r+1},
          \boldsymbol{h}_{\nu}^{r+1})$.
\end{itemize}
\end{minipage}
};

\node[monitor, below right=3.0cm and 4.2cm of central] (mM) {
\textbf{Local machine $V$} $(\boldsymbol{Y}_{V},\boldsymbol{X}_{V})$\\[1mm]
\begin{minipage}{6.8cm}
\begin{itemize}
  \setlength\itemsep{1pt}
  \setlength\parsep{0pt}
  \setlength\parskip{0pt}
  \item Receives $(\boldsymbol{\beta}^r,\boldsymbol{Z}^r,\boldsymbol{h}^r)$.
  \item Evaluates $\nabla\mathcal{R}_{V}(\boldsymbol{\beta}^r)$.
  \item Runs \mbox{Algorithm~\ref{alg:admm}} to obtain
        $(\boldsymbol{\beta}_{V}^{r+1},\boldsymbol{Z}_{V}^{r+1},
          \boldsymbol{h}_{V}^{r+1})$.
  \item Sends $(\boldsymbol{\beta}_{V}^{r+1},\boldsymbol{Z}_{V}^{r+1},
          \boldsymbol{h}_{V}^{r+1})$.
\end{itemize}
\end{minipage}
};

\begin{scope}[on background layer]
\foreach \name in {central,m1,mm,mM} {
  \draw[fill=black]
    ($(\name.north west)+(-0.35,0.25)$) rectangle
    ($(\name.south east)+(0.35,-0.30)$);
  \draw[fill=white]
    ($(\name.north west)+(-0.28,0.18)$) rectangle
    ($(\name.south east)+(0.28,-0.35)$);
  \draw[fill=black]
    ($(\name.south west)+(-0.28,-0.35)$) rectangle
    ($(\name.south east)+(0.28,-0.55)$);
  \draw[fill=black]
    ($(\name.south)+(0,-0.55)$) ++(-0.35,-0.08) rectangle ++(0.70,-0.22);
  \draw[fill=black]
    ($(\name.south)+(0,-0.85)$) --
    ++(-0.9,-0.22) --
    ++(1.8,0) -- cycle;
}
\end{scope}

\node at ($(m1.east)!0.5!(mm.west)$) {$\cdots\cdots$};
\node at ($(mm.east)!0.5!(mM.west)$) {$\cdots\cdots$};

\draw[arrow] (central.south west) --
  node[above, sloped, pos=0.45, yshift=4pt] {\scriptsize Broadcast}
  (m1.north);
\draw[arrow] (central.south) --
  node[right, pos=0.5, xshift=6pt] {\scriptsize Broadcast}
  (mm.north);
\draw[arrow] (central.south east) --
  node[above, sloped, pos=0.45, yshift=4pt] {\scriptsize Broadcast}
  (mM.north);

\draw[arrow] (m1.north) --
  node[below, sloped, pos=0.55, yshift=-4pt] {\scriptsize Local updates}
  (central.south west);
\draw[arrow] (mm.north) --
  node[left, pos=0.5, xshift=-6pt] {\scriptsize Local updates}
  (central.south);
\draw[arrow] (mM.north) --
  node[below, sloped, pos=0.55, yshift=-4pt] {\scriptsize Local updates}
  (central.south east);

\end{tikzpicture}%
}

\caption{Communication scheme of the distributed penalized $L^{k}\mathrm{QR}$ algorithm 
(\mbox{Algorithm~\ref{alg:distributed}}). Each local machine performs the ADMM updates of \mbox{Algorithm~\ref{alg:admm}}, 
sends its local results to the central processor, and receives aggregated quantities 
for the next iteration.}
\label{fig:dist-LkQR}
\end{figure}

To illustrate the information flow between the machines and the central processor, 
we provide in \mbox{Figure~\ref{fig:dist-LkQR}} a schematic representation of the communication pattern 
underlying the distributed estimation procedure. Each local machine performs the updates 
specified in \mbox{Algorithm~\ref{alg:admm}}, transmits its local results to the central processor, 
and receives the aggregated quantities required for the next iteration.

Building on this structure, we introduce in \mbox{Algorithm~\ref{alg:distributed}} our proposed distributed 
algorithm for solving large-scale $L^{k}\mathrm{QR}$ problems efficiently and 
communication-effectively.

\begin{algorithm}[H]
\caption{LADMM algorithm for the $\nu$-th machine ($\nu\in [V]$) with power parameter $k\in(1,2]$}
\label{alg:admm}
\begin{algorithmic}[1]
\REQUIRE $\boldsymbol{\beta}^0, \boldsymbol{Z}^0, \boldsymbol{h}^0,\ \zeta=\rho_{\max}(\boldsymbol{X}^\top \boldsymbol{X}),\ \mu,\ \lambda,\ \tau,\ k$;
\FOR{$t = 0, 1, \ldots, L-1$}
\STATE \emph{$\boldsymbol{\beta}$--step:} \quad $\displaystyle {\beta_l^{t+1}}= \mathcal{S}\!\left({\eta_l^{t}},\ \frac{\lambda w_l}{\mu\,\zeta}\right),\qquad l=1,2,\ldots,p$;
\STATE Define the shifted residuals $\displaystyle s_i^{\,t}:=Y_{i}-\boldsymbol{X}_{i}^\top \boldsymbol{\beta}^{t+1}+h_i^{t}/\mu,\quad i=1,\ldots,n$;
\STATE \emph{$\boldsymbol{Z}$--step:} For each $i=1,\ldots,n$,
\begingroup\footnotesize
\[
Z_{i}^{t+1} \;=\;
\begin{cases}
\mathrm{Prox}_{\ell_{\tau,k}}(s_i^{\,t},\, n\mu), & \text{if } k=1.5 \quad\text{(closed form; see \mbox{Remark~\ref{remark zi 6}})},\\[6pt]
\displaystyle \frac{n\mu\,s_i^{\,t}}{\,n\mu +2 \tau\,\mathds{1}\{s_i^{\,t}\ge 0\} + 2(1-\tau)\,\mathds{1}\{s_i^{\,t}<0\}\,}, 
& \text{if } k=2 \quad\text{(closed form)},\\[10pt]
\arg\min\limits_{z\in\mathbb{R}}\ \big\{\ell_{\tau,k}(z) + \tfrac{n\mu}{2}\,(z-s_i^{\,t})^2\big\}, 
& \text{if } 1<k\le 2,\ k\notin\{1.5,2\}\ \text{(Newton--Raphson)}.
\end{cases}
\]
\endgroup
\STATE \emph{$\boldsymbol{h}$--step:} \quad $\displaystyle \boldsymbol{h}^{t+1}= \boldsymbol{h}^{t}-\mu\big(\boldsymbol{X}\boldsymbol{\beta}^{t+1}+\boldsymbol{Z}^{t+1}-\boldsymbol{Y}\big)$;
\ENDFOR
\ENSURE $(\boldsymbol{\beta}^{L}, \boldsymbol{Z}^{L}, \boldsymbol{h}^{L})$ for the machine $\nu$.
\end{algorithmic}
\end{algorithm}

\begin{algorithm}[H]
\caption{Distributed algorithm for solving large-scale penalized $L^k\mathrm{QR}$}
\label{alg:distributed}
\begin{algorithmic}[1]
\REQUIRE $\boldsymbol{\beta}^0, \boldsymbol{Z}^0, \boldsymbol{h}^0$, \( \{ Y_{\nu i}, \boldsymbol{X}_{\nu i} = (X_{\nu i1}, \dots, X_{\nu ip})^\top \}_{i=1}^n \), $\zeta=\rho_{\max}(\boldsymbol{X}^\top \boldsymbol{X})$, $\mu,\lambda$, $\tau$, the number of iterations $L$, and the number of rounds of communication $U$.
\FOR{$r = 0, 1, \ldots, U-1$}
\STATE Each machine evaluates $\nabla \mathcal{R}_\nu(\boldsymbol{\beta})$, $\nu \in \{1,2,\ldots,V\}$, and sends it to the central processor;
\STATE The central processor computes $\nabla \mathcal{R}_N(\boldsymbol{\beta}^{\,r}) = \frac{1}{V} \sum_{\nu=1}^V \nabla \mathcal{R}_\nu(\boldsymbol{\beta}^{\,r})$ and broadcasts it to machines;
\STATE Update $(\boldsymbol{\beta}^{\,r+1}_{\nu}, \boldsymbol{Z}^{\,r+1}_{\nu}, \boldsymbol{h}_{\nu}^{\,r+1})$, $\nu \in [V]$, via ~\mbox{Algorithm~\ref{alg:admm}} on each machine and send them to the central processor; \label{STEP 4}
\STATE The central processor computes $\boldsymbol{\beta}^{\,r+1}=\frac{1}{V} \sum_{\nu=1}^V \boldsymbol{\beta}^{\,r+1}_{\nu},\ \boldsymbol{Z}^{\,r+1} = \frac{1}{V} \sum_{\nu=1}^V \boldsymbol{Z}^{\,r+1}_{\nu},\ \boldsymbol{h}^{\,r+1} = \frac{1}{V} \sum_{\nu=1}^V \boldsymbol{h}^{\,r+1}_{\nu}$ and broadcasts them to machines; \label{STEP 5}
\ENDFOR
\ENSURE $\boldsymbol{\hat{\beta}}^{(S)}\ (\boldsymbol{\hat{\beta}}^{(A)}) = \boldsymbol{\beta}^U$.
\end{algorithmic}
\end{algorithm}

\begin{remark}\label{remark sub method}
Averaging at the end of each outer round (Step~$\ref{STEP 5}$ in $\mbox{Algorithm~\ref{alg:distributed}}$) has two key advantages. First, it is computationally cheap: the central node aggregates the local
model parameters into a single iterate, enforcing consensus with negligible overhead. Second, it often stabilizes and accelerates convergence by reducing variance across local updates,
especially when each machine performs a substantial local solve in parallel (Step~$\ref{STEP 4}$).
This stands in contrast to the canonical CSL variant of \citet{jordan2019}, where the central machine carries most of the optimization while others primarily send gradients {(the Sub estimator in Section~\ref{Sec. 5})}.\\
However, a simple average $\frac{1}{V}\sum_{\nu=1}^V \widehat{\boldsymbol{\beta}}_\nu$ implicitly assumes balanced splits ($n_\nu = N/V$). This equal-size (or same-order) assumption is common in
distributed estimators---including, for instance, distributed expectile and related linear models    \citep[e.g.,][]{pan2021,Liu2023-fy,pan2025distributed}. When local sizes differ, this update no longer targets the global objective $\mathcal{R}_N(\cdot)$ and can be inefficient or biased toward small shards. In practice, the aggregation in Step~$\ref{STEP 5}$ should be size-weighted as follows
\(
\boldsymbol{\beta}^{r+1}\;=\;\sum_{\nu=1}^V \frac{n_\nu}{N}\,\boldsymbol{\beta}^{r+1}_\nu,\) with \( N=\sum_{\nu=1}^V n_\nu\) or, when feasible, \emph{variance-aware} (coordinate-wise or block-wise inverse-variance weights) to account for heterogeneity across machines \citep{DobribanSheng2021,LuMaoLiHou2022}.
Moreover, auxiliary variables $\boldsymbol{Z}_\nu$ and $\boldsymbol{h}_\nu$ are data-indexed and generally non-conformable when $n_\nu$ differs; they should therefore remain local and not be
averaged across machines. Equal-size (or same-order) partitions are commonly assumed in the divide-and-conquer literature \citep[e.g.,][]{battey2018distributed}, but weighted aggregation preserves statistical efficiency when partitions are unbalanced.
\end{remark}

\section{Implementation}\label{sec:51-implementation}
\noindent
A central element of any penalized estimator is the construction of a principled and computationally feasible regularization path. In distributed environments, this task must be performed in a manner that is both statistically meaningful and communication-efficient. In this section, we describe how we construct the support of the tuning parameter~$\lambda$ for the penalized $L^k$QR estimator.

\medskip

\noindent
We start from the null model, i.e., $\boldsymbol{\beta}=\boldsymbol{0}$. Evaluating the (sub)gradient of the empirical loss at this point gives
\[
\frac{1}{N}\sum_{i=1}^N \ell'_{\tau,k}\!\bigl(Y_i\bigr)\,X_{il}
=\frac{1}{V}\sum_{\nu=1}^V 
\underbrace{\Bigg[\frac{1}{n}\sum_{i=1}^{n}\ell'_{\tau,k}\!\bigl(Y_{\nu i}\bigr)\,X_{\nu i,l}\Bigg]}_{=:~u^{(\nu)}_l},
\qquad l = 1,\dots,p.
\]
For each machine $\nu$, we define the local vectors $\boldsymbol{u}^{(\nu)} = \bigl(u^{(\nu)}_1,\dots,u^{(\nu)}_p\bigr)^\top\in\mathbb{R}^p$, and their global average $\boldsymbol{u}
= \frac{1}{V}\sum_{\nu=1}^V \boldsymbol{u}^{(\nu)}$. In an $\ell_1$-penalized problem, the KKT condition for optimality at $\boldsymbol{\beta}=\boldsymbol{0}$ requires
\[
\Big|\tfrac{1}{N}\sum_{i=1}^N \ell'_{\tau,k}(Y_i)\,X_{il}\Big|
\le \lambda,
\qquad l=1,\dots,p.
\]
Thus, the smallest value of $\lambda$ for which the null model is optimal is
\[
\lambda_{\max}
=\max_{1\le l\le p} |u_l|
= \max_{1\le l\le p}
\Big|\tfrac{1}{V}\sum_{\nu=1}^V u^{(\nu)}_l\Big|.
\]
This quantity plays a dual role: it guaranties a proper starting value for the regularization path, and it identifies the scale at which predictors first begin to enter the model. Crucially, $\lambda_{\max}$ is obtained with a single communication round involving one averaged vector in~$\mathbb{R}^p$, which makes it ideally suited to distributed settings.

Each outer iteration of the distributed ADMM algorithm requires one reduction of gradient-like vectors in $\mathbb{R}^p$ and occasional scalar or vector reductions for computing $\lambda_{\max}$ or diagnostic quantities. Under equal-size partitions, all reductions simplify to averaging, keeping communication overhead minimal and predictable.

To explore a range of sparsity levels, we set $\lambda_{\min} = \varpi\,\lambda_{\max}$, with $\varpi = 10^{-2}$ when $N<p$, and $\varpi = 10^{-4}$ when $N\ge p$. We then generate a decreasing sequence,
$\lambda_1=\lambda_{\max} > \lambda_2 > \cdots > \lambda_K=\lambda_{\min}$,
using $K$ points equally spaced on the logarithmic scale. This produces a geometrically decaying regularization path, enabling stable warm-starting and efficient computation of the entire penalized solution path.

{
\par \label{rev:tuning}
\noindent We recommend a prespecified grid $\{1.2,1.5,2\}$ and training-only cross-validation jointly with $\lambda$. Use identical folds across candidates and keep the final test set untouched. Since different $k$ define different estimands and differently scaled losses, raw candidate-specific losses must not be compared directly across $k$. Instead, select a common, prespecified validation criterion, such as mean absolute prediction error, when prediction is the goal. If a particular $L^k$ functional is scientifically required, fix $k$ and tune only $\lambda$. Tail summaries are diagnostics rather than an automatic rule for choosing $k$.
Each site returns only a validation-loss sum and validation count for each candidate/fold. With $G$ powers, $J$ penalty values and $F$ folds, naive tuning needs $GJF$ fits and the corresponding communication budgets; warm starts can reduce computation but do not eliminate candidate evaluations. Fold construction and predictor preprocessing must use training data only.

\begin{remark}\label{remark on alg}
Convergence analyses for penalized quantile regression include smoothed parallel ADMM \citep{wu2023partition} and fixed-weight proximal ADMM for the unsmoothed loss \citep[Theorem~3]{Huaijun2021}. For $1<k<2$, the derivative of the $L^k$ loss is not globally Lipschitz, unlike at $k=2$. Thus results requiring Lipschitz gradients cannot be invoked directly \citep{guo2017convergence,WangYinZeng2019}. For fixed corrections and nonnegative weights, the inner problem is convex; classical ADMM theory \citep{boyd2011} requires a saddle point, with suitable proximal matrices and subproblem accuracy for the proximal variant. Newton steps near zero may need safeguarding; bisection on the monotone scalar first-order equation is an alternative.
For a fixed SCAD objective, exact LLA steps yield descent by majorization--minimization \citep{zou2008}, not necessarily global optimality. Our routine updates weights within inner iterations and changes corrections and aggregates local solutions; a global convergence guarantee for the entire routine remains beyond this analysis. Ill-conditioning, insufficient inner accuracy and changing surrogates can cause oscillation or stalled iterates. Primal and dual residuals, KKT residuals and objective values help assess convergence; small coefficient changes alone do not certify it. Iteration caps and failure reporting are therefore useful in practice.
\end{remark}

\label{loc:surrogate}\par 
\noindent When $n<p$, the corrected local SCAD objective may be unbounded below if the gradient correction has a nonzero component in the null space of the local design. Existence of a minimizer is thus relevant to the convergence discussion.

\par \label{rev:init}
\noindent For fixed $p,V$, an unpenalized local $L^k$ fit is a natural pilot when the local design is full rank and the usual consistency conditions hold. A root-$n$ pilot then has the same rate order as root-$N$ consistency because $N=nV$. The intended improvement is support recovery and the pooled oracle covariance, under the corresponding score-moment conditions, rather than a faster consistency rate. The covariance comparison is detailed in Remark~\ref{remark:initial}.

}

\section{Asymptotic Theory} \label{Sec. 3}

\noindent We provide  notations to illustrate the asymptotic properties of 
$\boldsymbol{\hat{\beta}}$. The true value of $\boldsymbol{\beta}$ is denoted by 
$\boldsymbol{\beta}^\star$. Let $i$ be an element of the set $[N]$. 
\begin{equation} \label{errors}
\varepsilon_{ i} = Y_{i} - \boldsymbol{X}_{i}^\top \boldsymbol{\beta}^\star, 
\qquad 
\varepsilon^0_{ i} = Y_{ i} - \boldsymbol{X}_{ i}^\top \boldsymbol{\beta}^0,
\end{equation}
\noindent
where $\boldsymbol{\beta}^0$ denotes an initial estimator of $\boldsymbol{\beta}^\star$ used to construct the proposed procedures. 
We then establish the asymptotic properties of the proposed estimators 
$\boldsymbol{\hat \beta}^{\text{(S)}}$ and $\boldsymbol{\hat \beta}^{\text{(A)}}$, respectively.
\noindent
We assume that our sample set $\{\boldsymbol{X}_i, {Y}_i\}_{i=1}^{N}$ comes from the following data generation process
\begin{equation}\label{process without lost of generality}
\boldsymbol{Y}_i = \boldsymbol{X}_i^\top \boldsymbol{\beta}^\star + \varepsilon_i = \boldsymbol{X}_{i}^{(1)\top} \boldsymbol{\beta}_{1}^\star + \boldsymbol{X}_{i}^{(2)\top} \boldsymbol{\beta}_{2}^\star + \varepsilon_i, \quad i = 1, 2, \ldots, N,
\end{equation}
where 
$\boldsymbol{X}_i = \left( \boldsymbol{X}_{i}^{(1) \top}, \boldsymbol{X}_{i}^{(2) \top} \right)^\top$,
$\boldsymbol{\beta} = \left( \boldsymbol{\beta}_{1}^\top, \boldsymbol{\beta}_{2}^\top \right)^\top$, $\boldsymbol{X}^{(1)}_{i} \in \mathbb{R}^m$, $\boldsymbol{X}^{(2)}_{i} \in \mathbb{R}^{p-m}$. 
The true regression coefficients are $\boldsymbol{\beta}_1 = \boldsymbol{\beta}_{1}^{\star}$, where each component is non-zero, and $\boldsymbol{\beta}_2 = \boldsymbol{\beta}^{\star}_{2} = \boldsymbol 0$, so   $\boldsymbol{\beta}^\star = \left( \boldsymbol{\beta}_{1}^{\star 
 \top}, \boldsymbol{\beta}_{2}^{\star \top} \right)^\top$. We make the following assumptions throughout the paper. {The theoretical statements concern fixed dimension $p$ and a fixed number of machines $V$.}

\begin{assumption}\label{assumption 1}
Let \(1<k\le 2\) and \(\tau\in(0,1)\). Given the design \(\{\boldsymbol X_i\}\), the sequences
\(\{\varepsilon_i\}_{i=1}^N\) and \(\{\varepsilon_i^0\}_{i=1}^N\) are i.i.d.\ across \(i\) and share the same
conditional density \(f_{\varepsilon}(\cdot \mid \boldsymbol X_i)\), which is continuous and strictly positive
on a neighborhood of \(0\), a.s. Their \(\tau\)-th \(L^k\)-quantiles are zero a.s.,
\[
Q_{\tau,k}(\varepsilon_i \mid \boldsymbol X_i)
=
Q_{\tau,k}(\varepsilon_i^0 \mid \boldsymbol X_i)
=
0 \quad \text{a.s.},
\]
and the following conditional moment bounds hold a.s.:
\[
\mathbb{E}\!\left(|\varepsilon_i|^{2k}\mid \boldsymbol X_i\right) < \infty,
\qquad
\mathbb{E}\!\left(|\varepsilon_i^0|^{2k}\mid \boldsymbol X_i\right) < \infty.
\]
\end{assumption}

\begin{assumption} \label{assumption 2}
\noindent
\begin{enumerate}[(i)]

\item \label{assumption 2-i}
There exists a positive definite matrix $\boldsymbol{\Sigma}$ such that
\[
\boldsymbol{\Sigma}_{n} := \frac{1}{n}\sum_{i=1}^n \boldsymbol{X}_{\nu i}\boldsymbol{X}_{\nu i}^\top 
\;\longrightarrow \boldsymbol{\Sigma} \ \text{as } n \to +\infty \ \text{uniformly for each } \nu\in[V].
\]

    Denote the top-left \(m \times m\) sub-matrix of \(\boldsymbol{\Sigma}\) by \(\boldsymbol{\Sigma}_{11}\).
    \item \label{assumption 2-ii} 
Assume that the  deterministic design satisfies
\[
\frac{1}{nV}\sum_{i=1}^{n}\sum_{\nu=1}^{V}\|\boldsymbol X_{\nu i}\|^{4}=O(1),
\]
as \(n\to\infty\). 

\end{enumerate}

\end{assumption}

\begin{assumption}\label{assumption 3}
    $\lambda_n=\lambda(n)\to 0$ and $\sqrt{n}\lambda_n\to+\infty$ as $n\to +\infty.$
\end{assumption}

\begin{assumption}\label{assumption 4}
    $\sqrt{n}\lambda_n\to{0}$ and $n^{(\iota+1)/2}\lambda_n\to +\infty$ for some chosen $\iota>0.$ {These are the adaptive-LASSO conditions, distinct from the SCAD conditions in Assumption~\ref{assumption 3}.}
\end{assumption}

\begin{remark}
$\mbox{Assumption~\ref{assumption 1}}$ is often used for extreme value estimates. \cite{pollard1991} proved that  $\mbox{Assumption~\ref{assumption 2}}$ implies $\max\limits_{1\leq i\leq n}\| \boldsymbol{X}_{\nu i} \|/\sqrt{n}\to 0$. The proposed estimator $\boldsymbol{\hat{\beta}}$ has asymptotic properties, which are demonstrated in the Appendix and summarized in $\mbox{Theorems~\ref{theorem scad} and \ref{theorem al}}.$ 
\end{remark}

We establish the $\sqrt{N}$-consistency of the proposed SCAD-penalized estimator $\widehat{\boldsymbol{\beta}}^{(S)}$ (\mbox{Theorem~\ref{theorem scad}}--\ref{consistency scad}), under the condition that the tuning parameter satisfies $\lambda=\lambda(n)\to 0$ as $n\to\infty$. 
We further show that $\widehat{\boldsymbol{\beta}}^{(S)}$ enjoys the oracle property (\mbox{Theorem~\ref{theorem scad}}--\ref{sparsity scad}), provided that ~\mbox{Assumption~\ref{assumption 3}} holds. 
Finally, we establish the asymptotic normality of $\widehat{\boldsymbol{\beta}}^{(S)}$ (\mbox{Theorem~\ref{theorem scad}}-\ref{normality scad}).

\begin{theorem}[Consistency and oracle properties of the SCAD estimator]
\label{theorem scad}
Let ~\mbox{Assumptions~\ref{assumption 1} and \ref{assumption 2}} hold. Then:
\begin{enumerate}[(i)]

\item \label{consistency scad}
\textbf{Consistency.}  
The SCAD estimator satisfies
\[
\sqrt{N}\,\|\widehat{\boldsymbol{\beta}}^{(S)} - \boldsymbol{\beta}^\star\|_2 = O_p(1).
\]
Furthermore, if $~\mbox{Assumption~\ref{assumption 3}}$ holds, then the 
$\sqrt{N}$-consistent local minimizer
\[
\widehat{\boldsymbol{\beta}}^{(S)} 
= \begin{pmatrix} 
\widehat{\boldsymbol{\beta}}^{(S)}_1 \\[2pt] 
\widehat{\boldsymbol{\beta}}^{(S)}_2 
\end{pmatrix}
\]
satisfies the following oracle properties.

\item \label{sparsity scad}
\textbf{Sparsity {(with probability tending to one)}.}  
\[
\widehat{\boldsymbol{\beta}}^{(S)}_2 = \boldsymbol{0}_{p-m}.
\]

\item \label{normality scad}
\textbf{Asymptotic normality.}  
\[
\sqrt{N}\,(\widetilde{\boldsymbol{\Sigma}}_{11})^{-1/2}
\left( 
\widehat{\boldsymbol{\beta}}^{(S)}_1 - \boldsymbol{\beta}^\star_1
\right)
\xrightarrow{\mathcal D} \mathcal{N}(\boldsymbol{0}_m, \mathbb{I}_m).
\]

\end{enumerate}

\noindent
Here, $\overset{\mathcal D}\to$ denotes convergence in distribution, and
\[
\widetilde{\boldsymbol{\Sigma}}_{11}
=\frac{\upsilon(\tau)}{\varkappa^2(\tau)}\,\boldsymbol{\Sigma}_{11}^{-1},
\qquad
\upsilon(\tau)= V\varrho(\tau)+(V-1)\varsigma(\tau)+2(1-V)\vartheta(\tau),
\]
\[
\varkappa(\tau)=\mathbb{E}\!\left[\ell''_{\tau,k}(\varepsilon_i)\right], 
\qquad
\varrho(\tau)=\mathbb{E}\!\left[(\ell'_{\tau,k}(\varepsilon_i))^2\right],
\]
\[
\varsigma(\tau)=\mathbb{V}\!ar\!\left[\ell'_{\tau,k}(\varepsilon^0_i)\right],
\qquad
\vartheta(\tau)=\mathbb{C}\!ov\!\left[\ell'_{\tau,k}(\varepsilon^0_i),\,\ell'_{\tau,k}(\varepsilon_i)\right],
\]
where $\ell'_{\tau,k}(\cdot)$ is defined in~\eqref{ell prime}, and
\begin{equation}\label{ell 2 prime}
\ell''_{\tau,k}(s) 
= k(k-1)\,\Psi_\tau(s)\,|s|^{k-2},
\qquad s\in\mathbb{R},\; 1<k\le 2.
\end{equation}

\end{theorem}

\begin{remark}[On the role of the initial estimator in distributed SCAD estimation]
\label{remark:initial}
A key feature of the asymptotic variance in ~\mbox{Theorem~\ref{theorem scad}} is the appearance of the quantities $\varrho(\tau)$, $\varsigma(\tau)$, and $\vartheta(\tau)$, which depend on both the residuals 
$\varepsilon_i = Y_i - \boldsymbol X_i^\top\boldsymbol \beta^\star$ and the residuals computed from the initial estimator 
$\varepsilon_i^0 = Y_i - \boldsymbol X_i^\top\boldsymbol \beta^0$.  
If the initial estimator satisfies 
\[
\boldsymbol{\beta}^0 
= \boldsymbol{\beta}^\star + O_p(N^{-1/2}),
\]
{then $\varepsilon_i^0-\varepsilon_i=-\boldsymbol{X}_i^\top(\boldsymbol{\beta}^0-\boldsymbol{\beta}^\star)$. Under Assumption~\ref{assumption 1}, if the corresponding score second moments and cross-moments converge to the same limit, we obtain}

\[
\varsigma(\tau)
\,\approx\, 
\varrho(\tau),
\qquad
\vartheta(\tau)
\,\approx\, 
\varrho(\tau).
\]
Consequently, the variance factor in ~\mbox{Theorem~\ref{theorem scad}},
\[
\upsilon(\tau)
=
V\varrho(\tau)
+(V-1)\varsigma(\tau)
+2(1-V)\vartheta(\tau),
\]
simplifies to 
\[
\upsilon(\tau) \approx \varrho(\tau),
\]
and the limiting covariance matrix becomes
\[
\widetilde{\boldsymbol{\Sigma}}_{11}
\;\approx\;
\frac{\varrho(\tau)}{\varkappa^2(\tau)}\,\boldsymbol{\Sigma}_{11}^{-1},
\]
which matches the oracle asymptotic variance one would obtain if the active set 
$\{l\in[p] : \beta_l^\star \neq 0\}$ were known.  

This phenomenon mirrors similar results in nonconvex penalized expectile and quantile regression \citep[e.g.][]{fan2001,zou2008,zhao2018variable}, where a sufficiently accurate initial estimator causes the SCAD estimator to achieve full 
oracle efficiency.  In our distributed setting, this demonstrates that, despite relying on a surrogate loss and nonconvex penalties, the proposed SCAD penalized $L^{k}\mathrm{QR}$ estimator retains the same asymptotic efficiency as its centralized counterpart {when its pilot and score moments satisfy the conditions stated above.}
\end{remark}

\begin{remark}\label{remark residual}
Note that  the second (appearing in \mbox{Theorem~\ref{theorem scad}}) and third derivatives of the loss function $\ell_{\tau,k}(\cdot)$ are not defined when $\varepsilon_i = 0$. This issue is not 
restrictive in our context, since under a continuous distribution, the probability of a residual being exactly zero is zero  \citep[see][]{jiang2021k,li2024optimalp}
\end{remark}

\begin{remark} \label{remark 3}
\mbox{Theorem~\ref{theorem scad}}-$\ref{consistency scad})$ shows that the SCAD estimator $\widehat{\boldsymbol{\beta}}^{\text{(S)}}$ is $\sqrt{N}$-consistent.  
Furthermore, the asymptotic normality derived in ~\mbox{Theorem~\ref{theorem scad}}-$\ref{normality scad})$ agrees with the theoretical results obtained for nonconvex penalized expectile  models in \citet{zhao2018variable}.  
These results are also consistent with the general oracle framework of \citet{fan2001} and the nonconcave penalization theory of \citet{zou2008}, reinforcing that our distributed SCAD penalized $L^{k}\mathrm{QR}$ estimator {has the same first-order oracle covariance as its centralized counterpart when $\varsigma(\tau)$ and $\vartheta(\tau)$ have limit $\varrho(\tau)$, as explained in Remark~\ref{remark:initial}. This is an asymptotic covariance comparison, not equality of finite-sample fits.}
\end{remark}

\noindent

In the next theorem, we establish the oracle property of the adaptive LASSO penalized $L^{k}\mathrm{QR}$ estimator $\widehat{\boldsymbol{\beta}}^{\text{(A)}}$ under ~\mbox{Assumption~\ref{assumption 4}}, as stated in ~\mbox{Theorem~\ref{theorem al}}.

\begin{theorem}[Oracle properties of Adaptive LASSO estimator] \label{theorem al}
Under Assumptions~\ref{assumption 1}, \ref{assumption 2}, and \ref{assumption 4}, the $\sqrt{N}$-consistent local minimizer $\widehat{\boldsymbol{\beta}}^{\text{(A)}} = \begin{pmatrix} 
\widehat{\boldsymbol{\beta}}^{\text{(A)}}_{1} \\[2pt]
\widehat{\boldsymbol{\beta}}^{\text{(A)}}_{2}
\end{pmatrix}$
satisfies:
\begin{enumerate}[(i)]
    \item \textbf{Sparsity {(with probability tending to one)}:} 
    $\widehat{\boldsymbol{\beta}}^{\text{(A)}}_{2}
    = \boldsymbol{0}_{p-m}$;
    
    \item \textbf{Asymptotic normality:} 
    \[
     \sqrt{N}\,
     \boldsymbol{\tilde \Sigma}_{11}^{-1/2}
     \big(
       \widehat{\boldsymbol{\beta}}^{\text{(A)}}_{1}
       - \boldsymbol{\beta}^{\star}_{1}
     \big)
     \xrightarrow{\mathcal{D}}
     \mathcal{N}\!\left(\boldsymbol{0}_{m},\, \mathbb{I}_{m}\right),
    \]
    where $\boldsymbol{\tilde \Sigma}_{11}$ is defined in ~\mbox{Theorem~\ref{theorem scad}}.
\end{enumerate}
\end{theorem}

\begin{remark}
The adaptive LASSO enjoys a conceptual advantage over SCAD in the derivation of
its asymptotic properties.  
Because the adaptive LASSO penalty is convex, the overall penalized $L^{k}\mathrm{QR}$ objective remains convex, which guaranties a unique global minimizer. 
Consequently, once sparsity and asymptotic normality are established in $~\mbox{Theorem~\ref{theorem al}}$, $\sqrt{N}$-consistency follows automatically.

In contrast, the SCAD penalty is nonconvex, and the penalized objective is also nonconvex. The LLA method used to approximate the SCAD estimator replaces the nonconvex penalty with a locally linear approximation, which is convex only at each iteration. As a result, additional conditions and separate arguments are required to guaranty the existence of a $\sqrt{N}$-consistent local minimizer before establishing the oracle property. 
\end{remark}

\section{Simulation studies and real data applications} 
\subsection{Simulation studies} \label{Sec. 5}
\noindent We conduct simulation studies to evaluate the finite-sample performance of the proposed method. The strategies compared are as follows: 
\begin{enumerate}[a)]
    \item

The averaging method (Ave), which uses the standard ADMM  in \mbox{Algorithm~\ref{alg:admm}} to solve the penalized variant of the objective function in \eqref{local loss function} based on datasets from each machine, takes the average over the number of machines;
\item The proposed Algorithm $\ref{alg:distributed}$ (Csl) (For ease of presentation, we suppress $k$ from $\mathrm{CSL}^k$);
\item The distributed optimization method (Sub), {the first-machine version of Csl described in \mbox{Remark~\ref{remark sub method}}, without aggregation of the local fitted parameters};
\item
The global optimization method (Cen) uses the full dataset as a {pooled-data reference} to tackle the penalized version of the true objective function
in \eqref{global loss function} by implementing the standard ADMM algorithm.
\end{enumerate}
{\par \label{loc:definitions}
\begin{center}\begin{tabular}{lp{.72\linewidth}}\toprule
Cen & Pooled-data penalized fit.\\
Ave & Local fits followed by coefficient averaging.\\
Sub & First-machine surrogate fit, without averaging local fitted parameters.\\
Csl & Site-specific surrogate fits followed by aggregation of fitted parameters and auxiliary states.\\\bottomrule
\end{tabular}\end{center}
\noindent The Sub estimator can be viewed as a particular version of Csl in which Step~\ref{STEP 4} of Algorithm~\ref{alg:distributed} is performed only on the first machine and the averaging operation in Step~\ref{STEP 5} is omitted. The remaining machines contribute only through their local gradients. In contrast, Csl exploits the computational capacity of all machines by performing local optimization in parallel and aggregating the resulting estimates. Implementation details are given in Appendix~\ref{rev:additional}.}
Scalar responses are generated according to the heteroscedastic model \eqref{hetero}:
\begin{equation}\label{hetero} Y = X_6 + X_{12} + X_{15} + X_{20} + 0.7X_1\varepsilon, \end{equation} where the predictors $X_1, X_2, \cdots, X_p$ are generated by $X_1 = \Phi (\tilde{X}_1)$ and $X_l = \tilde{X}_l$ for $l = 2, 3, \cdots, p$, where $\Phi(\cdot)$ denotes the cumulative distribution function of the standard normal distribution and $(\tilde{X}_1, \tilde{X}_2, \cdots, \tilde{X}_p)^\top \sim \mathcal{N}(\boldsymbol{0}_{p}, \boldsymbol{\Sigma}),$ $\boldsymbol{\Sigma} = \left( 0.5^{|i-j|}\right)_{(i,j) \in [p]\times [p]}$. It is crucial to note that $X_1$ significantly affects the conditional distribution of $Y$ given the predictors, but does not directly affect the mean of the conditional distribution. To ensure a thorough comparison, the error term $\varepsilon_{i}$ is drawn independently from one of the following distributions:
\begin{itemize}
    \item[(i)] Normal distribution: $\varepsilon \sim \mathcal{N}(0, 1)$;
    \item[(ii)] Student's distribution with $5$ degrees of freedom: $\varepsilon \sim t(5)$.
\end{itemize}

\noindent 
We consider $k=1.20, 1.50,1.80$ and $2.00$, with $\tau=0.25,0.50,$ and $\tau=0.75$. Set  parameter \( p\) to \(200 \), and set the number of machines \( V\) to \(15 \). Assign a sample size of $100$ to each machine. We choose the tuning parameter $\lambda$ by minimizing the test error using the $V$-fold cross-validation approach. Specifically, we select the tuning parameter \(\lambda\) by grid search on a validation set, using the log-spaced grid between \(\lambda_{\min}=\varpi\,\lambda_{\max}\) and \(\lambda_{\max}\) as described in  \mbox{Section~\ref{sec:51-implementation}}. 
\noindent
We partition the data into a training set and a test set. We use the training set to estimate $\boldsymbol{\hat \beta}_{\text{train}}$ and use the test set to determine the value $\lambda$ minimizing $\sum \limits_{ i\in \text{test}} \ell_{\tau,k}(Y_i - \boldsymbol{X}_i^\top \boldsymbol{\hat \beta}_{\text{train}})$.
\noindent
Based on $100$ replications, we evaluate the four methods using the following criteria, with the selection measures following \citet{wen2025feature}. 
\begin{enumerate}[1)]
    \item {Size}:
{The average number of nonzero estimated coefficients over the simulation replications.}
\item { P1}: {The percentage of simulation runs in which all four location predictors $X_6,X_{12},X_{15},X_{20}$ are simultaneously selected, excluding $X_1$ at every considered level $\tau$.}
\item { P2}: {The percentage of simulation runs in which the heteroscedasticity predictor $X_1$ is selected.}
\item { ER}: The estimated error 
$\| \boldsymbol{\hat{\beta}} - \boldsymbol{\beta}^\star\|_2^2$.
\end{enumerate}
The sample standard deviations are shown in parentheses in the columns labeled {Size}, {P1}, {P2}, and {ER}.
{\par \label{loc:metrics}
\noindent The structural mean model contains four active predictors. Under the considered symmetric errors, $X_1$ contributes to the conditional $L^k$-quantile at asymmetric levels but not at the median. Consequently, P1 measures simultaneous recovery of the four location predictors, whereas P2 measures selection of the heteroscedasticity predictor $X_1$. Ideally, P1 is close to 100\% at all levels, and P2 is close to zero at $\tau=0.50$ and 100\% at $\tau=0.25,0.75$. Size describes overall sparsity, while P1 and P2 describe the selection of specific predictors. These measures and ER provide complementary descriptive summaries. Appendix D uses tolerance $10^{-6}$ for its selection diagnostics.}

\noindent
{Tables~\ref{table scad normal} and~\ref{table adaptive lasso normal} report the Gaussian-error results for SCAD and adaptive LASSO, respectively. The corresponding Student's $t(5)$ results are presented in Tables~\ref{table scad student} and~\ref{table adaptive lasso student}.}
\noindent
For ease of comparison, all rows corresponding to the Csl estimator  are typeset in \textbf{bold} in these tables.

\begin{figure}[htbp]
  \centering
  \includegraphics[width=\linewidth]{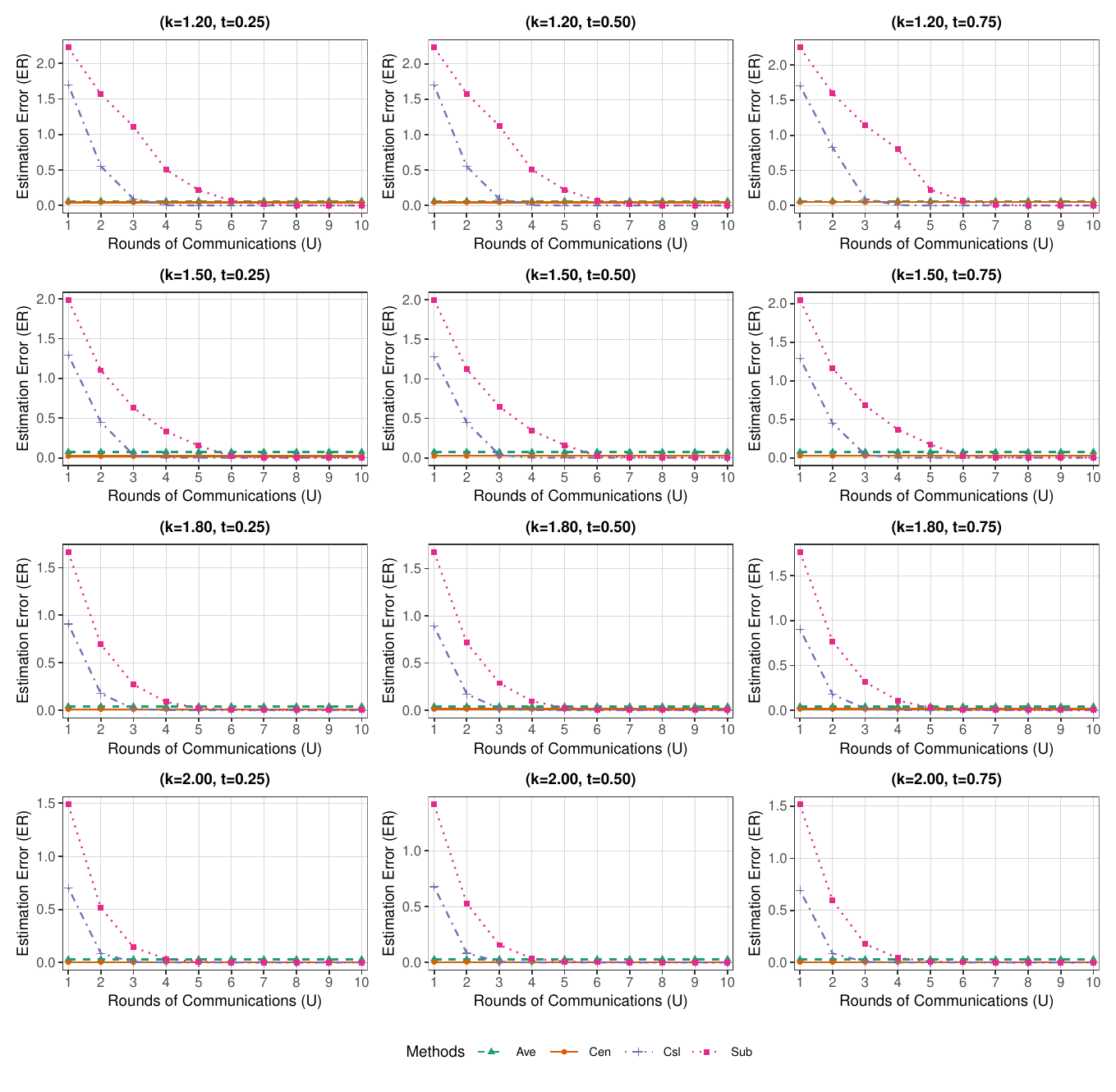} %
  \caption{Plots of the estimation error $\| \boldsymbol{\hat{\beta}} - \boldsymbol{\beta}^\star\|_2^2$ versus the number of communication rounds for the SCAD penalized $L^k\mathrm{QR}$ for different $k=1.20, 1.50, 1.80,2.00$ and  $  \tau=0.25,0.50,0.75$ when the error distribution $\varepsilon \sim \mathcal{N}(0,1)$.}  \label{fig scad normal}
\end{figure}

\begin{figure}[htbp]
    \centering
     \includegraphics[width=\linewidth]{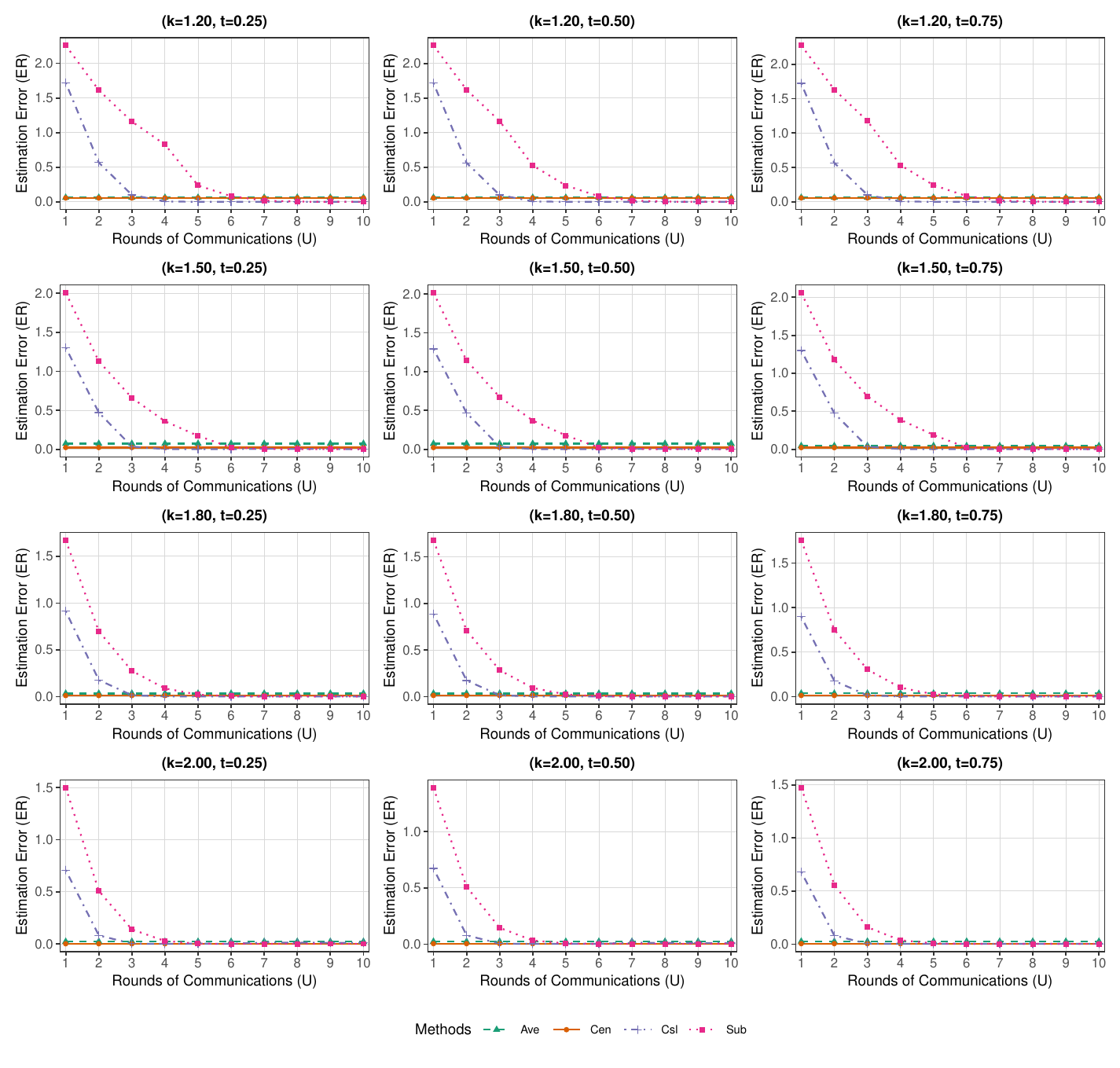} %
    \caption{Plots of the estimation error $\| \boldsymbol{\hat{\beta}} - \boldsymbol{\beta}^\star \|_2^2$ versus the number of communication rounds for the SCAD penalized $L^k\mathrm{QR}$ for different $k=1.20, 1.50, 1.80,2.00$ and  $  \tau=0.25,0.50,0.75$   when the error distribution $\varepsilon \sim t(5)$.}
    \label{fig scad student}
\end{figure}

\begin{figure}[htbp]
  \centering
  \includegraphics[width=\linewidth]{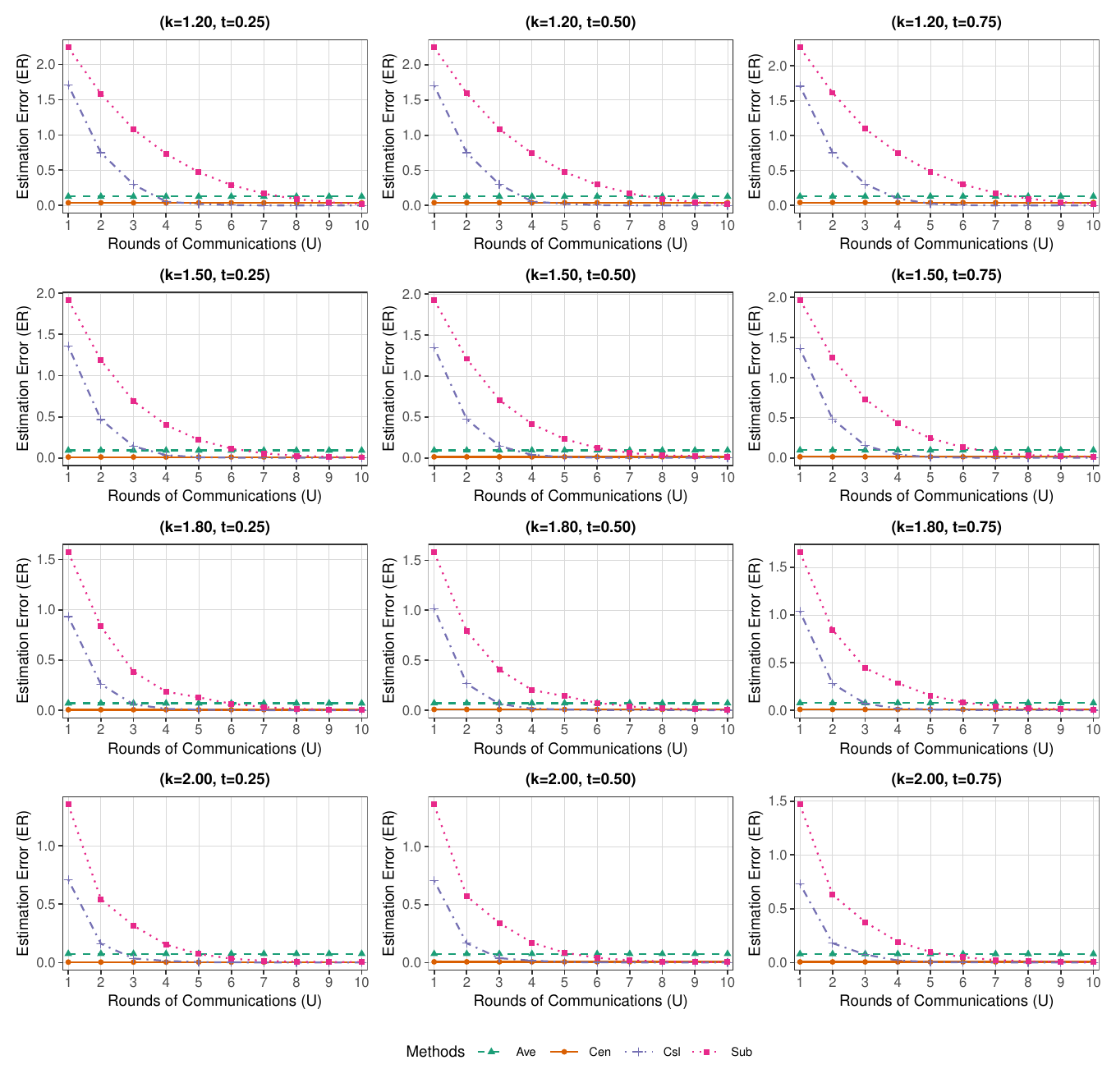} %
   \caption{Plots of the estimation error $\| \boldsymbol{\hat{\beta}} - \boldsymbol{\beta}^\star \|_2^2$ versus the number of communication rounds for the adaptive LASSO penalized $L^k\mathrm{QR}$ for different $k=1.20, 1.50, 1.80,2.00$ and  $  \tau=0.25,0.50,0.75$ when the error distribution $\varepsilon \sim \mathcal{N}(0,1)$.}
    \label{fig alasso normal}
\end{figure}

\begin{figure}[htbp]
  \centering
  \includegraphics[width=\linewidth]{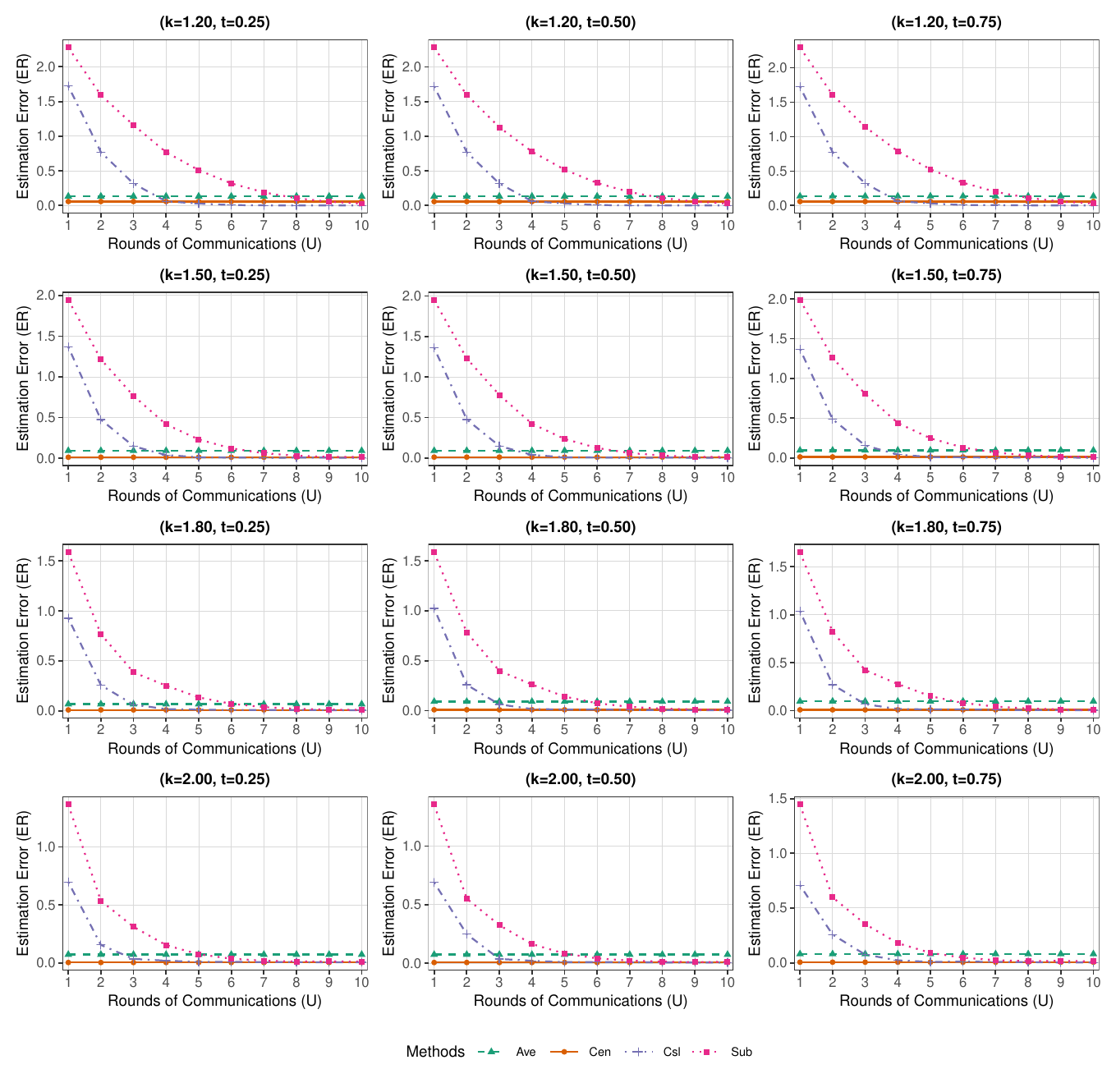} %
 \caption{Plots of the estimation error $\| \boldsymbol{\hat{\beta}} - \boldsymbol{\beta}^\star \|_2^2$ versus the number of communication rounds for the adaptive LASSO penalized $L^k\mathrm{QR}$ for various $k=1.20, 1.50, 1.80,2.00$ and  $  \tau=0.25,0.50,0.75$   when the error distribution $\varepsilon \sim t(5)$.}
    \label{fig alasso student}
\end{figure}

\begin{enumerate}[(i)]
\item Estimation accuracy.
{\label{loc:interpret}The rankings vary across levels and powers. At the Gaussian median with $k=1.50$, Table~\ref{table scad normal} reports ER $0.000280$ for Cen and $0.000290$ for Csl. At $\tau=0.25,k=1.50$, however, Ave has ER $0.037712$ and mean Size $114.50$, compared with ER $0.071032$ and Size $5.00$ for Csl. A larger selected model can have lower coefficient error when reduced shrinkage bias offsets the errors contributed by small additional coefficients. Neither sparsity nor a pooled fit guarantees the smallest realized ER in each finite sample. The comparisons do not imply uniform superiority of Csl over Ave or Sub. In Table~\ref{table adaptive lasso normal}, at $\tau=0.50$ and $k=1.50$, Csl and Cen have nearly identical reported ER ($0.000207$ and $0.000209$, respectively), whereas Ave and Sub have larger errors ($0.001824$ and $0.000850$). The mean sizes of Csl and Ave are $6.90$ and $14.10$, respectively, illustrating that the estimation accuracy of Csl can be accompanied by a substantially smaller fitted model than Ave.}
\item Support recovery across quantiles.
{P2 reaches 100\% in several asymmetric-level configurations. A low P2 is desirable at the median but represents a missed target variable at asymmetric levels. A small model can omit signals, and P1 alone does not penalize extra selections. P1 and P2 assess distinct aspects of signal recovery; neither alone measures exact support recovery.}
\item Communication efficiency.
{Figure~\ref{fig scad normal} shows a rapid decrease in ER for Csl during the first few communication rounds, while Sub generally requires more rounds to approach the low-error region. Cen and Ave serve as round-independent references. Figures~\ref{fig scad student}, \ref{fig alasso normal}, and \ref{fig alasso student} extend this comparison to Student errors and adaptive LASSO; the curves describe the displayed configurations and do not establish a universal round budget. The fixed-budget sensitivity and payload experiments in Appendix~\ref{rev:additional} separately examine runtime, local sample size and transferred bytes.}
\item Choosing the power $k$.
{The tables describe the selected grid and do not imply a distribution-independent optimal power. A common training-only validation criterion permits joint tuning of $k$ and $\lambda$, as discussed in Section~\ref{rev:tuning}. Losses with different powers have different scales, so their raw values should not be compared as if they were the same score.}
\end{enumerate}

\noindent {Tables~\ref{table scad student} and~\ref{table adaptive lasso student} present the results under Student's $t(5)$ errors. In Table~\ref{table scad student}, at $\tau=0.50$ and $k=1.50$, the reported ER of Csl ($0.000389$) is close to that of Cen ($0.000401$), with a mean model size of four for both. At $k=2$, Sub has a markedly larger ER ($0.400857$) than Csl ($0.001287$), despite the same reported mean size. Table~\ref{table adaptive lasso student} shows that, at $\tau=0.50$ and $k=1.50$, Csl has ER $0.000503$ and mean Size $4.00$, compared with $0.006174$ and $38.70$ for Ave. These examples illustrate the distinction between coefficient accuracy and sparsity under heavy-tailed errors.

\noindent Figure~\ref{fig scad student} shows that the early decrease in Csl ER observed under Gaussian errors also appears in the Student-error setting; Sub approaches the low-error region more gradually, especially at $k=1.20$. Figure~\ref{fig alasso normal} exhibits a similar early reduction for Csl under adaptive LASSO, with Sub decreasing over a longer sequence of rounds. In Figure~\ref{fig alasso student}, the same qualitative separation between the early Csl and Sub trajectories is visible under heavy-tailed errors. These comparisons concern the plotted finite-round errors, rather than a proof of algorithmic convergence.}

\scriptsize
\begin{longtable}{ccccccc}
\caption{Simulation results for the SCAD penalized $L^k\mathrm{QR}$ model for $k\in\{1.20,1.50,1.80,2.00\}$ and $\tau\in\{0.25,0.50,0.75\}$ under $\varepsilon \sim \mathcal{N}(0,1)$. }\label{table scad normal}\\
\toprule
$\tau$ & $k$ & Method & Size & P1  & P2  & ER \\
\midrule
\endfirsthead
\caption*{\textit{Table \thetable\ (continued)}} \\
\toprule
$\tau$ & $k$ & Method & Size & P1  & P2  & ER \\
\midrule
\endhead
\midrule
\multicolumn{7}{r}{\textit{Continued on next page}} \\
\endfoot
\bottomrule
\endlastfoot

0.50 & 1.20 & Cen & 4.00(0.00) & 100.00(0.00) & 0.00(0.00) & 0.001588(0.000573) \\
 &  & Ave & 40.00(2.83) & 100.00(0.00) & 0.00(0.00) & 0.005008(0.002847) \\
 &  & Sub & 4.00(0.00) & 100.00(0.00) & 0.00(0.00) & 0.000371(0.000136) \\
 &  & \textbf{Csl} & \textbf{4.00(0.00)} & \textbf{100.00(0.00)} & \textbf{0.00(0.00)} & \textbf{0.000200(0.000056)} \\
\addlinespace

 & 1.50 & Cen & 4.00(0.00) & 100.00(0.00) & 0.00(0.00) & 0.000280(0.000044) \\
 &  & Ave & 10.50(0.71) & 100.00(0.00) & 0.00(0.00) & 0.004181(0.002742) \\
 &  & Sub & 4.00(0.00) & 100.00(0.00) & 0.00(0.00) & 0.000297(0.000205) \\
 &  & \textbf{Csl} & \textbf{4.00(0.00)} & \textbf{100.00(0.00)} & \textbf{0.00(0.00)} & \textbf{0.000290(0.000212)} \\
\addlinespace

 & 1.80 & Cen & 4.00(0.00) & 100.00(0.00) & 0.00(0.00) & 0.000415(0.000320) \\
 &  & Ave & 35.00(12.73) & 100.00(0.00) & 0.00(0.00) & 0.002669(0.001678) \\
 &  & Sub & 4.00(0.00) & 100.00(0.00) & 0.00(0.00) & 0.003058(0.000293) \\
 &  & \textbf{Csl} & \textbf{4.00(0.00)} & \textbf{100.00(0.00)} & \textbf{0.00(0.00)} & \textbf{0.000399(0.000281)} \\
\addlinespace

 & 2.00 & Cen & 4.00(0.00) & 100.00(0.00) & 0.00(0.00) & 0.000500(0.000366) \\
 &  & Ave & 11.00(4.24) & 100.00(0.00) & 0.00(0.00) & 0.170062(0.001174) \\
 &  & Sub & 4.00(0.00) & 100.00(0.00) & 0.00(0.00) & 0.196209(0.276134) \\
 &  & \textbf{Csl} & \textbf{4.00(0.00)} & \textbf{100.00(0.00)} & \textbf{0.00(0.00)} & \textbf{0.000476(0.000308)} \\

\midrule

0.25 & 1.20 & Cen & 5.00(0.00) & 100.00(0.00) & 100.00(0.00) & 0.163879(0.017517) \\
 &  & Ave & 148.00(12.29) & 100.00(0.00) & 100.00(0.00) & 0.126024(0.018387) \\
 &  & Sub & 5.00(0.13) & 100.00(0.00) & 100.00(0.00) & 0.03006(0.093392) \\
 &  & \textbf{Csl} & \textbf{5.00(0.00)} & \textbf{100.00(0.00)} & \textbf{100.00(0.00)} & \textbf{0.01056(0.021204)} \\

\addlinespace
 & 1.50 & Cen & 5.50(0.71) & 100.00(0.00) & 100.00(0.00) & 0.122201(0.009086) \\
 &  & Ave & 114.50(16.26) & 100.00(0.00) & 100.00(0.00) & 0.037712(0.041025) \\
 &  & Sub & 5.00(0.00) & 100.00(0.00) & 100.00(0.00) & 0.106345(0.017307) \\
 &  & \textbf{Csl} & \textbf{5.00(0.00)} & \textbf{100.00(0.00)} & \textbf{100.00(0.00)} & \textbf{0.071032(0.017708)} \\

\addlinespace
  & 1.80 & Cen & 5.00(0.00) & 100.00(0.00) & 100.00(0.00) & 0.091680(0.006681) \\
 &  & Ave & 134.50(86.97) & 100.00(0.00) & 100.00(0.00) & 0.059283(0.055494) \\
 &  & Sub & 5.00(0.00) & 100.00(0.00) & 100.00(0.00) & 0.099727(0.006234) \\
 &  & \textbf{Csl} & \textbf{5.00(0.00)} & \textbf{100.00(0.00)} & \textbf{100.00(0.00)} & \textbf{0.078441(0.004201)} \\

\addlinespace
  & 2.00 & Cen & 5.00(0.00) & 100.00(0.00) & 100.00(0.00) & 0.074276(0.007612) \\
 &  & Ave & 109.50(98.29) & 100.00(0.00) & 50.00(0.00) & 0.031347(0.037989) \\
 &  & Sub & 4.00(0.00) & 100.00(0.00) & 0.00(0.00) & 0.414007(0.000415) \\
 &  & \textbf{Csl} & \textbf{5.00(0.00)} & \textbf{100.00(0.00)} & \textbf{100.00(0.00)} & \textbf{0.000538(0.004550)} \\

\midrule

0.75 & 1.20 & Cen & 11.50(0.71) & 100.00(0.00) & 100.00(0.00) & 0.190769(0.006272) \\
 &  & Ave & 197.00(0.00) & 100.00(0.00) & 100.00(0.00) & 0.188847(0.013625) \\
 &  & Sub & 17.00(1.41) & 100.00(0.00) & 100.00(0.00) & 0.180926(0.005275) \\
 &  & \textbf{Csl} & \textbf{5.00(0.00)} & \textbf{100.00(0.00)} & \textbf{100.00(0.00)} & \textbf{0.165167(0.001198)} \\
\addlinespace

 & 1.50 & Cen & 5.50(0.71) & 100.00(0.00) & 100.00(0.00) & 0.067528(0.003482) \\
 &  & Ave & 136.50(0.71) & 100.00(0.00) & 100.00(0.00) & 0.052343(0.011964) \\
 &  & Sub & 5.00(0.00) & 100.00(0.00) & 100.00(0.00) & 0.097868(0.056328) \\
 &  & \textbf{Csl} & \textbf{5.00(0.00)} & \textbf{100.00(0.00)} & \textbf{100.00(0.00)} & \textbf{0.013273(0.000724)} \\
\addlinespace

 & 1.80 & Cen & 10.00(0.00) & 100.00(0.00) & 100.00(0.00) & 0.109124(0.005617) \\
 &  & Ave & 189.00(2.83) & 100.00(0.00) & 100.00(0.00) & 0.059911(0.013082) \\
 &  & Sub & 4.00(0.00) & 100.00(0.00) & 0.00(0.00) & 0.000411(0.000122) \\
 &  & \textbf{Csl} & \textbf{5.00(0.00)} & \textbf{100.00(0.00)} & \textbf{100.00(0.00)} & \textbf{0.000942(0.000587)} \\
\addlinespace

 & 2.00 & Cen & 5.50(0.71) & 100.00(0.00) & 100.00(0.00) & 0.077658(0.008322) \\
 &  & Ave & 163.00(7.07) & 100.00(0.00) & 100.00(0.00) & 0.027327(0.019664) \\
 &  & Sub & 4.00(0.00) & 100.00(0.00) & 0.00(0.00) & 0.409248(0.015876) \\
 &  & \textbf{Csl} & \textbf{4.00(0.00)} & \textbf{100.00(0.00)} & \textbf{100.00(0.00)} & \textbf{0.000644(0.000034)} \\

\end{longtable}
\normalsize

\scriptsize
\begin{longtable}{ccccccc}
\caption{Simulation results for the aLASSO penalized $L^k\mathrm{QR}$ model for $k\in\{1.20,1.50,1.80,2.00\}$ and $\tau\in\{0.25,0.50,0.75\}$ under $\varepsilon \sim \mathcal{N}(0,1)$.}  \label{table adaptive lasso normal} \\
\toprule
$\tau$ & $k$ & Method & Size & P1  & P2  & ER \\
\midrule
\endfirsthead
\caption*{\textit{Table \thetable\ (continued)}} \\
\toprule
$\tau$ & $k$ & Method & Size & P1  & P2  & ER \\
\midrule
\endhead
\midrule
\multicolumn{7}{r}{\textit{Continued on next page}} \\
\endfoot
\bottomrule
\endlastfoot

0.50 & 1.20 & Cen & 4.00(0.00) & 100.00(0.00) & 0.00(0.00) & 0.000185(0.000174) \\
      &      & Ave & 10.20(10.89) & 100.00(0.00) & 0.00(0.00) & 0.002107(0.000590) \\
      &      & Sub & 4.10(0.32) & 100.00(0.00) & 0.00(0.00)  & 0.000729(0.000863) \\
      &      & \textbf{Csl} & \textbf{4.20(0.63)} & \textbf{100.00(0.00)} & \textbf{0.00(0.00)} & \textbf{0.000220(0.000197)} \\
\addlinespace
      & 1.50 & Cen & 4.10(0.32) & 100.00(0.00) & 0.00(0.00) & 0.000209(0.000182) \\
      &      & Ave & 14.10(3.38) & 100.00(0.00) & 0.00(0.00) & 0.001824(0.000698) \\
      &      & Sub & 4.00(0.00) & 100.00(0.00) & 0.00(0.00) & 0.000850(0.000601) \\
      &      & \textbf{Csl} & \textbf{6.90(1.91)} & \textbf{100.00(0.00)} & \textbf{0.00(0.00)} & \textbf{0.000207(0.000199)} \\
\addlinespace
      & 1.80 & Cen & 4.00(0.00) & 100.00(0.00) & 0.00(0.00) & 0.000274(0.000232) \\
      &      & Ave & 15.20(16.18) & 100.00(0.00) & 0.00(0.00) & 0.002146(0.000610) \\
      &      & Sub & 4.00(0.00) & 100.00(0.00) & 0.00(0.00) & 0.001944(0.000795) \\
      &      & \textbf{Csl} & \textbf{4.00(0.00)} & \textbf{100.00(0.00)} & \textbf{0.00(0.00)} & \textbf{0.000388(0.000325)} \\
\addlinespace
      & 2.00 & Cen & 4.60(1.26) & 100.00(0.00) & 0.00(0.00) & 0.000398(0.000277) \\
      &      & Ave & 20.70(3.95) & 100.00(0.00) & 0.00(0.00) & 0.002183(0.000967) \\
      &      & Sub & 4.00(0.00) & 100.00(0.00) & 0.00(0.00) & 0.003977(0.001909) \\
      &      & \textbf{Csl} & \textbf{4.00(0.00)} & \textbf{100.00(0.00)} & \textbf{0.00(0.00)} & \textbf{0.000636(0.000447)} \\
\midrule

0.25 & 1.20 & Cen & 5.80(1.32) & 100.00(0.00) & 100.00(0.00) & 0.053828(0.019467) \\
      &      & Ave & 55.20(33.70) & 100.00(0.00) & 100.00(0.00) & 0.077782(0.029197) \\
      &      & Sub & 5.50(0.71) & 100.00(0.00) & 100.00(0.00) & 0.070429(0.039880) \\
      &      & \textbf{Csl} & \textbf{5.50(0.12)} & \textbf{100.00(0.00)} & \textbf{100.00(0.00)} & \textbf{0.031834(0.030200)} \\
\addlinespace
      & 1.50 & Cen & 5.00(0.00) & 100.00(0.00) & 100.00(0.00) & 0.118455(0.014861) \\
      &      & Ave & 73.20(6.53) & 100.00(0.00) & 100.00(0.00) & 0.056765(0.031424) \\
      &      & Sub & 4.80(0.42) & 100.00(0.00) & 70.00(0.00)  & 0.019696(0.018784) \\
      &      & \textbf{Csl} & \textbf{4.00(0.09)} & \textbf{100.00(0.00)} & \textbf{100.00(0.00)} & \textbf{0.069488(0.037274)} \\
\addlinespace
      & 1.80 & Cen & 5.30(0.95) & 100.00(0.00) & 100.00(0.00) & 0.089898(0.012785) \\
      &      & Ave & 54.60(36.90) & 100.00(0.00) & 80.00(0.00)  & 0.034716(0.020005) \\
      &      & Sub & 4.40(0.52) & 100.00(0.00) & 40.00(0.00)  & 0.004283(0.002108) \\
      &      & \textbf{Csl} & \textbf{4.80(0.42)} & \textbf{100.00(0.00)} & \textbf{70.00(0.00)} & \textbf{0.032337(0.028454)} \\
\addlinespace
      & 2.00 & Cen & 5.40(0.97) & 100.00(0.00) & 100.00(0.00) & 0.078830(0.013182) \\
      &      & Ave & 65.30(29.50) & 100.00(0.00) & 70.00(0.00)  & 0.032121(0.022917) \\
      &      & Sub & 4.00(0.00) & 100.00(0.00) & 0.00(0.00)     & 0.05803(0.002622) \\
      &      & \textbf{Csl} & \textbf{4.60(0.52)} & \textbf{100.00(0.00)} & \textbf{80.00(0.00)} & \textbf{0.017761(0.017493)} \\
\midrule

0.75 & 1.20 & Cen & 6.00(1.33) & 100.00(0.00) & 100.00(0.00) & 0.152657(0.018607) \\
      &      & Ave & 62.40(32.55) & 100.00(0.00) & 100.00(0.00) & 0.063401(0.019444) \\
      &      & Sub & 5.70(0.82) & 100.00(0.00) & 100.00(0.00) & 0.055694(0.036077) \\
      &      & \textbf{Csl} & \textbf{5.90(0.95)} & \textbf{100.00(0.00)} & \textbf{100.00(0.00)} & \textbf{0.118019(0.024494)} \\
\addlinespace
      & 1.50 & Cen & 5.10(0.32) & 100.00(0.00) & 100.00(0.00) & 0.111900(0.012969) \\
      &      & Ave & 86.00(39.80) & 100.00(0.00) & 100.00(0.00) & 0.047813(0.018218) \\
      &      & Sub & 4.70(0.48) & 100.00(0.00) & 60.00(0.00)  & 0.014016(0.015724) \\
      &      & \textbf{Csl} & \textbf{5.00(0.98)} & \textbf{100.00(0.00)} & \textbf{100.00(0.00)} & \textbf{0.063030(0.032698)} \\
\addlinespace
      & 1.80 & Cen & 5.00(0.00) & 100.00(0.00) & 100.00(0.00) & 0.082578(0.011051) \\
      &      & Ave & 44.60(35.60) & 100.00(0.00) & 70.00(0.00)  & 0.021220(0.016010) \\
      &      & Sub & 4.30(0.48) & 100.00(0.00) & 20.00(0.00)  & 0.003606(0.002834) \\
      &      & \textbf{Csl} & \textbf{4.90(0.74)} & \textbf{100.00(0.00)} & \textbf{60.00(0.00)} & \textbf{0.021321(0.024086)} \\
\addlinespace
      & 2.00 & Cen & 5.70(1.06) & 100.00(0.00) & 100.00(0.00) & 0.074645(0.014344) \\
      &      & Ave & 54.40(29.50) & 100.00(0.00) & 60.00(0.00)  & 0.019652(0.019622) \\
      &      & Sub & 4.20(0.42) & 100.00(0.00) & 10.00(0.00)  & 0.006549(0.003754) \\
      &      & \textbf{Csl} & \textbf{5.50(0.53)} & \textbf{100.00(0.00)} & \textbf{70.00(0.00)} & \textbf{0.010401(0.013406)} \\

\end{longtable}
\normalsize

\scriptsize
\begin{longtable}{ccccccc}
\caption{Simulation results for the SCAD penalized $L^k\mathrm{QR}$ model for $k\in\{1.20,1.50,1.80,2.00\}$ and $\tau\in\{0.25,0.50,0.75\}$ under $\varepsilon \sim t(5)$.}\label{table scad student}  \\
\toprule
$\tau$ & $k$ & Method & Size & P1 & P2 & ER \\
\midrule
\endfirsthead
\caption*{\textit{Table \thetable\ (continued)}} \\
\toprule
$\tau$ & $k$ & Method & Size & P1 & P2 & ER \\
\midrule
\endhead
\midrule
\multicolumn{7}{r}{\textit{Continued on next page}} \\
\endfoot
\bottomrule
\endlastfoot

0.50 & 1.20 & Cen & 4.00(0.00)   & 100.00(0.00) & 0.00(0.00) & 0.000210(0.000132) \\
      &      & Ave & 35.40(8.21)  & 100.00(0.00) & 0.00(0.00) & 0.001315(0.000571) \\
      &      & Sub & 4.00(0.00)   & 100.00(0.00) & 0.00(0.00) & 0.001257(0.001578) \\
      &      & \textbf{Csl} & \textbf{4.00(0.00)} & \textbf{100.00(0.00)} & \textbf{0.00(0.00)} & \textbf{0.000220(0.000086)} \\
\addlinespace
      & 1.50 & Cen & 4.00(0.00)   & 100.00(0.00) & 0.00(0.00) & 0.000401(0.000272) \\
      &      & Ave & 33.30(59.04) & 100.00(0.00) & 0.00(0.00) & 0.001455(0.001140) \\
      &      & Sub & 4.00(0.00)   & 100.00(0.00) & 0.00(0.00) & 0.000817(0.000992) \\
      &      & \textbf{Csl} & \textbf{4.00(0.00)} & \textbf{100.00(0.00)} & \textbf{0.00(0.00)} & \textbf{0.000389(0.000258)} \\
\addlinespace
      & 1.80 & Cen & 4.00(0.00)   & 100.00(0.00) & 0.00(0.00) & 0.000753(0.000581) \\
      &      & Ave & 52.50(12.13) & 100.00(0.00) & 0.00(0.00) & 0.001278(0.000621) \\
      &      & Sub & 4.00(0.00)   & 100.00(0.00) & 0.00(0.00) & 0.037243(0.025783) \\
      &      & \textbf{Csl} & \textbf{4.00(0.00)} & \textbf{100.00(0.00)} & \textbf{0.00(0.00)} & \textbf{0.000788(0.000614)} \\
\addlinespace
      & 2.00 & Cen & 4.00(0.00)   & 100.00(0.00) & 0.00(0.00) & 0.001167(0.000933) \\
      &      & Ave & 19.40(8.82)  & 100.00(0.00) & 0.00(0.00) & 0.001324(0.000832) \\
      &      & Sub & 4.00(0.00)   & 100.00(0.00) & 0.00(0.00) & 0.400857(0.107353) \\
      &      & \textbf{Csl} & \textbf{4.00(0.02)} & \textbf{100.00(0.00)} & \textbf{0.00(0.00)} & \textbf{0.001287(0.001076)} \\
\midrule

0.25 & 1.20 & Cen & 5.00(0.00)   & 100.00(0.00) & 100.00(0.00) & 0.194919(0.021251) \\
      &      & Ave & 196.80(1.48) & 100.00(0.00) & 100.00(0.00) & 0.123644(0.026632) \\
      &      & Sub & 4.60(0.52)   & 100.00(0.00) &  60.00(0.00) & 0.026860(0.003147) \\
      &      & \textbf{Csl} & \textbf{5.00(0.00)} & \textbf{100.00(0.00)} & \textbf{100.00(0.00)} & \textbf{0.023877(0.022533)} \\
\addlinespace
      & 1.50 & Cen & 6.70(3.33)   & 100.00(0.00) & 100.00(0.00) & 0.141454(0.024626) \\
      &      & Ave & 138.20(14.81) & 100.00(0.00) & 100.00(0.00) & 0.095943(0.044819) \\
      &      & Sub & 4.00(0.00)   & 100.00(0.00) &   0.00(0.00)  & 0.000954(0.000909) \\
      &      & \textbf{Csl} & \textbf{5.00(0.08)} & \textbf{100.00(0.00)} & \textbf{100.00(0.00)} & \textbf{0.006703(0.021744)} \\
\addlinespace
      & 1.80 & Cen & 5.30(0.67)   & 100.00(0.00) & 100.00(0.00) & 0.136878(0.019282) \\
      &      & Ave & 193.20(2.74) & 100.00(0.00) & 100.00(0.00) & 0.072155(0.027854) \\
      &      & Sub & 4.00(0.00)   & 100.00(0.00) &   0.00(0.00)  & 0.045559(0.037483) \\
      &      & \textbf{Csl} & \textbf{5.00(0.00)} & \textbf{100.00(0.00)} & \textbf{100.00(0.00)} & \textbf{0.098801(0.069741)} \\
\addlinespace
      & 2.00 & Cen & 5.00(0.00)   & 100.00(0.00) & 100.00(0.00) & 0.122508(0.021653) \\
      &      & Ave & 174.90(9.21) & 100.00(0.00) & 100.00(0.00) & 0.049974(0.030411) \\
      &      & Sub & 4.00(0.00)   & 100.00(0.00) &   0.00(0.00)  & 0.422744(0.135372) \\
      &      & \textbf{Csl} & \textbf{5.90(0.00)} & \textbf{100.00(0.00)} & \textbf{80.00(0.00)} & \textbf{0.035419(0.057071)} \\
\midrule

0.75 & 1.20 & Cen & 5.00(0.00)   & 100.00(0.00) & 100.00(0.00) & 0.119566(0.021155) \\
      &      & Ave & 195.20(2.25) & 100.00(0.00) & 100.00(0.00) & 0.124463(0.019390) \\
      &      & Sub & 4.40(0.52)   & 100.00(0.00) &  40.00(0.00) & 0.003188(0.005375) \\
      &      & \textbf{Csl} & \textbf{5.00(0.04)} & \textbf{100.00(0.00)} & \textbf{100.00(0.00)} & \textbf{0.006069(0.025533)} \\
\addlinespace
      & 1.50 & Cen & 5.80(1.75)   & 100.00(0.00) & 100.00(0.00) & 0.131313(0.023926) \\
      &      & Ave & 139.20(8.95) & 100.00(0.00) & 100.00(0.00) & 0.084200(0.035489) \\
      &      & Sub & 4.00(0.00)   & 100.00(0.00) &   0.00(0.00)  & 0.001413(0.002614) \\
      &      & \textbf{Csl} & \textbf{5.50(0.71)} & \textbf{100.00(0.00)} & \textbf{100.00(0.00)} & \textbf{0.00162(0.023898)} \\
\addlinespace
      & 1.80 & Cen & 5.10(0.32)   & 100.00(0.00) & 100.00(0.00) & 0.132760(0.018992) \\
      &      & Ave & 193.70(2.16) & 100.00(0.00) & 100.00(0.00) & 0.073757(0.024202) \\
      &      & Sub & 4.00(0.00)   & 100.00(0.00) &   0.00(0.00)  & 0.038895(0.027359) \\
      &      & \textbf{Csl} & \textbf{5.00(0.00)} & \textbf{100.00(0.00)} & \textbf{100.00(0.00)} & \textbf{0.091125(0.069981)} \\
\addlinespace
      & 2.00 & Cen & 5.00(0.00)   & 100.00(0.00) & 100.00(0.00) & 0.120477(0.019936) \\
      &      & Ave & 179.20(6.89) & 100.00(0.00) & 100.00(0.00) & 0.049912(0.029081) \\
      &      & Sub & 5.00(0.00)   & 100.00(0.00) &   0.00(0.00)  & 0.395736(0.115689) \\
      &      & \textbf{Csl} & \textbf{5.00(0.00)} & \textbf{100.00(0.00)} & \textbf{80.00(0.00)} & \textbf{0.035882(0.063970)} \\

\end{longtable}
\normalsize

\scriptsize
\begin{longtable}{ccccccc}
\caption{Simulation results for the aLASSO penalized $L^k\mathrm{QR}$ model for $k\in\{1.20,1.50,1.80,2.00\}$ and $\tau\in\{0.25,0.50,0.75\}$ under $\varepsilon \sim t(5)$.}  
\label{table adaptive lasso student}  \\
\toprule
$\tau$ & $k$ & Method & Size & P1 & P2 & ER \\
\midrule
\endfirsthead

\caption*{\textit{Table \thetable\ (continued)}} \\
\toprule
$\tau$ & $k$ & Method & Size & P1 & P2 & ER \\
\midrule
\endhead

\midrule
\multicolumn{7}{r}{\textit{Continued on next page}} \\
\endfoot

\bottomrule
\endlastfoot

0.50 & 1.20 & Cen & 4.00(0.00) & 100.00(0.00) & 0.00(0.00) & 0.000216(0.000055) \\
 &  & Ave & 33.90(33.04) & 100.00(0.00) & 0.00(0.00) & 0.006609(0.002461) \\
 &  & Sub & 4.00(0.00) & 100.00(0.00) & 0.00(0.00) & 0.001229(0.001673) \\
 &  & \textbf{Csl} & \textbf{4.50(0.71)} & \textbf{100.00(0.00)} & \textbf{0.00(0.00)} & \textbf{0.000296(0.000090)} \\
\addlinespace

 & 1.50 & Cen & 4.00(0.00) & 100.00(0.00) & 0.00(0.00) & 0.000386(0.000233) \\
 &  & Ave & 38.70(7.47) & 100.00(0.00) & 0.00(0.00) & 0.006174(0.001899) \\
 &  & Sub & 4.00(0.00) & 100.00(0.00) & 0.00(0.00) & 0.000928(0.000503) \\
 &  & \textbf{Csl} & \textbf{4.00(0.05)} & \textbf{100.00(0.00)} & \textbf{0.00(0.00)} & \textbf{0.000503(0.000264)} \\
\addlinespace

 & 1.80 & Cen & 4.10(0.32) & 100.00(0.00) & 0.00(0.00) & 0.000808(0.000504) \\
 &  & Ave & 31.20(28.81) & 100.00(0.00) & 0.00(0.00) & 0.007147(0.003609) \\
 &  & Sub & 4.00(0.00) & 100.00(0.00) & 0.00(0.00) & 0.002569(0.001225) \\
 &  & \textbf{Csl} & \textbf{4.70(0.82)} & \textbf{100.00(0.00)} & \textbf{0.00(0.00)} & \textbf{0.000989(0.000555)} \\
\addlinespace

 & 2.00 & Cen & 4.10(0.32) & 100.00(0.00) & 0.00(0.00) & 0.001173(0.000821) \\
 &  & Ave & 50.20(21.30) & 100.00(0.00) & 0.00(0.00) & 0.008388(0.003025) \\
 &  & Sub & 4.00(0.00) & 100.00(0.00) & 0.00(0.00) & 0.009848(0.008169) \\
 &  & \textbf{Csl} & \textbf{4.20(0.42)} & \textbf{100.00(0.00)} & \textbf{0.00(0.00)} & \textbf{0.001627(0.000892)} \\
\midrule

0.25 & 1.20 & Cen & 7.30(7.27) & 100.00(0.00) & 100.00(0.00) & 0.222801(0.031557) \\
 &  & Ave & 191.40(2.91) & 100.00(0.00) & 100.00(0.00) & 0.122552(0.032786) \\
 &  & Sub & 4.60(0.52) & 100.00(0.00) & 60.00(0.00) & 0.048924(0.102138) \\
 &  & \textbf{Csl} & \textbf{4.80(0.42)} & \textbf{100.00(0.00)} & \textbf{90.00(0.00)} & \textbf{0.026103(0.080763)} \\
\addlinespace

 & 1.50 & Cen & 6.30(1.57) & 100.00(0.00) & 100.00(0.00) & 0.142842(0.022804) \\
 &  & Ave & 101.30(28.83) & 100.00(0.00) & 90.00(0.00) & 0.040231(0.023966) \\
 &  & Sub & 4.50(0.53) & 100.00(0.00) & 50.00(0.00) & 0.013887(0.018501) \\
 &  & \textbf{Csl} & \textbf{7.40(2.17)} & \textbf{100.00(0.00)} & \textbf{90.00(0.00)} & \textbf{0.061748(0.040445)} \\
\addlinespace

 & 1.80 & Cen & 5.40(0.70) & 100.00(0.00) & 100.00(0.00) & 0.121039(0.019056) \\
 &  & Ave & 74.30(49.76) & 100.00(0.00) & 60.00(0.00) & 0.026195(0.017139) \\
 &  & Sub & 4.10(0.32) & 100.00(0.00) & 10.00(0.00) & 0.04053(0.003718) \\
 &  & \textbf{Csl} & \textbf{5.20(0.79)} & \textbf{100.00(0.00)} & \textbf{70.00(0.00)} & \textbf{0.022452(0.027521)} \\
\addlinespace

 & 2.00 & Cen & 6.00(2.49) & 100.00(0.00) & 100.00(0.00) & 0.109785(0.021123) \\
 &  & Ave & 89.50(30.56) & 100.00(0.00) & 50.00(0.00) & 0.024153(0.015992) \\
 &  & Sub & 4.00(0.00) & 100.00(0.00) & 0.00(0.00) & 0.013430(0.010501) \\
 &  & \textbf{Csl} & \textbf{4.60(0.52)} & \textbf{100.00(0.00)} & \textbf{80.00(0.00)} & \textbf{0.010925(0.017583)} \\
\midrule

0.75 & 1.20 & Cen & 7.20(6.96) & 100.00(0.00) & 100.00(0.00) & 0.212352(0.024344) \\
 &  & Ave & 174.40(51.81) & 100.00(0.00) & 100.00(0.00) & 0.119337(0.021236) \\
 &  & Sub & 4.60(0.52) & 100.00(0.00) & 60.00(0.00) & 0.082047(0.123558) \\
 &  & \textbf{Csl} & \textbf{4.90(0.32)} & \textbf{100.00(0.00)} & \textbf{90.00(0.00)} & \textbf{0.000525(0.000587)} \\
\addlinespace

 & 1.50 & Cen & 5.50(0.97) & 100.00(0.00) & 100.00(0.00) & 0.133649(0.018400) \\
 &  & Ave & 94.30(35.81) & 100.00(0.00) & 70.00(0.00) & 0.032174(0.024843) \\
 &  & Sub & 4.30(0.48) & 100.00(0.00) & 30.00(0.00) & 0.005486(0.009466) \\
 &  & \textbf{Csl} & \textbf{7.20(1.81)} & \textbf{100.00(0.00)} & \textbf{70.00(0.00)} & \textbf{0.042997(0.035345)} \\
\addlinespace

 & 1.80 & Cen & 5.50(0.53) & 100.00(0.00) & 100.00(0.00) & 0.119011(0.018351) \\
 &  & Ave & 65.80(43.92) & 100.00(0.00) & 60.00(0.00) & 0.018335(0.016259) \\
 &  & Sub & 4.00(0.00) & 100.00(0.00) & 0.00(0.00) & 0.020862(0.001707) \\
 &  & \textbf{Csl} & \textbf{5.00(0.67)} & \textbf{100.00(0.00)} & \textbf{60.00(0.00)} & \textbf{0.010936(0.017971)} \\
\addlinespace

 & 2.00 & Cen & 5.30(0.48) & 100.00(0.00) & 100.00(0.00) & 0.108415(0.018911) \\
 &  & Ave & 80.20(48.71) & 100.00(0.00) & 40.00(0.00) & 0.021664(0.018859) \\
 &  & Sub & 4.00(0.00) & 100.00(0.00) & 0.00(0.00) & 0.011617(0.012174) \\
 &  & \textbf{Csl} & \textbf{4.50(0.53)} & \textbf{100.00(0.00)} & \textbf{80.00(0.00)} & \textbf{0.004731(0.008906)} \\

\end{longtable}
\normalsize


\clearpage
\subsection{Real data analysis} \label{Sec. 6}
\begin{figure}[htbp]\centering
\captionsetup{labelfont={color=black},textfont={color=black}}
\includegraphics[width=\linewidth]{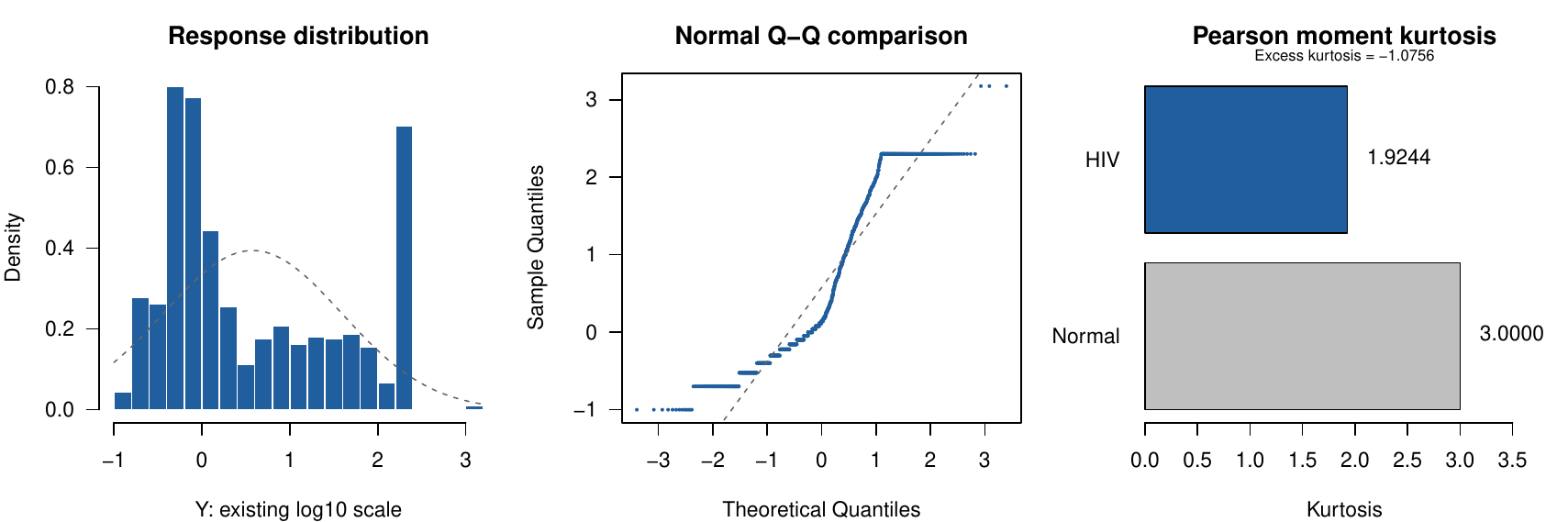}
\caption{{Distribution of the HIV response $Y$ ($n=1{,}472$,  $\log_{10}$ scale). (a) Histogram with a normal reference having the same mean and sample standard deviation. (b) Normal Q--Q plot; the dashed line is fitted through the ordered-response versus theoretical-quantile comparison. (c) Pearson moment kurtosis ($1.9244$) versus the normal reference ($3$); moment excess kurtosis is $-1.0756$ and bias-adjusted excess kurtosis is $-1.0751$. These are sample descriptive statistics, not a formal test of tail behavior.}}
\label{fig:hiv_descriptive}
\end{figure}

\noindent
We apply the communication-efficient penalized $L^k\mathrm{QR}$ model to the human immunodeficiency virus (HIV) drug resistance cohort study, available from the HIV Drug Resistance Database (\url{http://hivdb.stanford.edu}). The data consist of $1,472$ viral isolates (observations), each observation representing a laboratory-tested sample of HIV obtained from an infected individual. For every isolate, laboratory assays measure a fold reduction in drug susceptibility, defined as how many times more drug is required to inhibit viral replication compared to the normal, drug-sensitive (``wild-type'') virus. {The fold-change values range from $0.1$ to $1{,}500$, with higher values indicating greater resistance.} The genetic composition of each isolate is summarized by 197 binary mutation indicators (predictors), taking the value $1$ if a specific amino-acid substitution is present and 0 otherwise. These predictors are often correlated since many resistance mutations tend to occur together as part of characteristic mutational patterns or pathways that the virus develops under selective drug pressure. To analyze these data, we model the logarithm ($log_{10}$) of the fold-reduction as a continuous outcome in the proposed penalized $L^k\mathrm{QR}$ regression framework. {\label{loc:hivstats}The response is on the $\log_{10}$ scale. Its moment skewness is $0.5650$ and its Pearson moment kurtosis is $1.9244$ (excess $-1.0756$). The response is moderately right-skewed, but these marginal summaries do not support a heavier-than-Gaussian tail claim. The asymmetric loss is motivated by asymmetry and interest in different conditional features. Table~\ref{tab:hiv_descriptive} and Figure~\ref{fig:hiv_descriptive} summarize the response distribution.}

Data on HIV drug resistance is often stored in many locations.
 Consequently, communication-efficient $L^k\mathrm{QR}$ is an important approach  for analyzing this data. In practice, the sites of viral mutations are typically infrequent, resulting in a sparse solution through the application of penalties. The analysis of the uppermost quantile of sensitivity is particularly important because it is related to increased drug resistance. Therefore, we analyze the two quantiles: $\tau = 0.50$ and $\tau = 0.75$, as well as $\tau = 0.25$ for illustrative purposes.\\
Our analysis uses SCAD and aLASSO-penalized models and the Csl approach, as well as three competing approaches (Ave, Cen, Sub)  to identify relevant mutations. We perform $10$ random partitions. For each partition, we randomly select 1200 HIV isolates for training data and the remaining $272$ for testing data. We configure $1200$ training data on $V=12$ machines, assigning $n=100$ data to each machine ($n = 100$, $V = 12$). Twelve-fold cross-validation is used on the training data to determine the tuning parameters. We use a standard $5$--fold cross-validation for selection while using the centralized technique.\\

\scriptsize
\begin{longtable}{cc l cc cc}
\caption{Results of the analysis of the HIV drug resistance data.}
\label{tab:hiv_results}   \\
\toprule
\multirow{2}{*}{$\tau$} & \multirow{2}{*}{$k$} & \multirow{2}{*}{Method} & \multicolumn{2}{c}{SCAD} & \multicolumn{2}{c}{aLASSO}\\
\cmidrule(lr){4-5}\cmidrule(lr){6-7}
 &  &  & Size & PE & Size & PE \\
\midrule
\endfirsthead
\caption*{\textit{Table \thetable\ (continued)}}\\
\toprule
\multirow{2}{*}{$\tau$} & \multirow{2}{*}{$k$} & \multirow{2}{*}{Method} & \multicolumn{2}{c}{SCAD} & \multicolumn{2}{c}{aLASSO}\\
\cmidrule(lr){4-5}\cmidrule(lr){6-7}
 &  &  & Size & PE & Size & PE \\
\midrule
\endhead
\midrule
\multicolumn{7}{r}{\textit{Continued on next page}}\\
\endfoot
\bottomrule
\endlastfoot

\addlinespace[2pt]
\multicolumn{7}{l}{\textbf{$\tau=0.25$}}\\
\addlinespace[2pt]
 & 1.20 & Cen & 23.30(2.58) & 0.195844(0.011858) & 19.70(5.46) & 0.199822(0.013200)\\
 &      & Ave & 132.80(59.30) & 0.187753(0.010408) & 184.30(11.45) & 0.187366(0.009967)\\
 &      & Sub & 3.00(0.82) & 0.214315(0.013491) & 9.40(7.17) & 0.206188(0.012382)\\
 &      & \textbf{Csl} & \textbf{41.90(5.28)} & \textbf{0.186376(0.008939)} & \textbf{65.90(8.18)} & \textbf{0.188996(0.009109)}\\
\addlinespace
 & 1.50 & Cen & 20.60(13.54) & 0.166736(0.013076) & 28.50(9.40) & 0.165990(0.012080)\\
 &      & Ave & 183.30(9.55) & 0.155322(0.012700) & 190.20(12.93) & 0.154973(0.012739)\\
 &      & Sub & 4.90(6.69) & 0.196514(0.027729) & 11.90(5.07) & 0.174472(0.016496)\\
 &      & \textbf{Csl} & \textbf{124.30(35.01)} & \textbf{0.154541(0.012023)} & \textbf{126.90(40.23)} & \textbf{0.155431(0.013557)}\\
\addlinespace
 & 1.80 & Cen & 45.80(3.36) & 0.130916(0.012666) & 47.60(4.22) & 0.130949(0.012918)\\
 &      & Ave & 156.20(36.91) & 0.121671(0.011638) & 185.40(14.06) & 0.121925(0.011006)\\
 &      & Sub & 1.00(0.00) & 0.222452(0.024248) & 9.40(1.43) & 0.155118(0.016250)\\
 &      & \textbf{Csl} & \textbf{51.10(54.44)} & \textbf{0.135635(0.015481)} & \textbf{75.80(36.90)} & \textbf{0.122202(0.013312)}\\
\addlinespace
 & 2.00 & Cen & 42.10(2.38) & 0.121208(0.022586) & 56.20(17.17) & 0.118059(0.020679)\\
 &      & Ave & 163.50(12.70) & 0.112905(0.018965) & 186.40(10.63) & 0.111095(0.019266)\\
 &      & Sub & 2.60(3.37) & 0.207989(0.034631) & 11.30(5.42) & 0.147987(0.033585)\\
 &      & \textbf{Csl} & \textbf{28.20(37.20)} & \textbf{0.130658(0.031966)} & \textbf{85.60(28.74)} & \textbf{0.115767(0.027229)}\\
\midrule

\addlinespace[2pt]
\multicolumn{7}{l}{\textbf{$\tau=0.50$}}\\
\addlinespace[2pt]
 & 1.20 & Cen & 53.30(3.43) & 0.189464(0.013931) & 59.70(2.50) & 0.189089(0.012992)\\
 &      & Ave & 128.40(59.84) & 0.190062(0.013499) & 170.80(18.57) & 0.186593(0.015102)\\
 &      & Sub & 8.30(8.86) & 0.267054(0.037013) & 14.80(4.18) & 0.219353(0.021623)\\
 &      & \textbf{Csl} & \textbf{22.90(2.13)} & \textbf{0.191948(0.013156)} & \textbf{63.30(38.96)} & \textbf{0.192884(0.017861)}\\
\addlinespace
 & 1.50 & Cen & 94.60(36.80) & 0.143275(0.018339) & 82.60(19.25) & 0.140675(0.017364)\\
 &      & Ave & 167.90(37.85) & 0.135571(0.019748) & 180.20(16.16) & 0.136968(0.016002)\\
 &      & Sub & 8.50(5.56) & 0.229423(0.070421) & 12.60(1.90) & 0.176265(0.028953)\\
 &      & \textbf{Csl} & \textbf{17.00(1.63)} & \textbf{0.168779(0.030881)} & \textbf{72.70(44.77)} & \textbf{0.143181(0.020675)}\\
\addlinespace
 & 1.80 & Cen & 138.20(42.59) & 0.132081(0.014628) & 87.50(13.88) & 0.129078(0.015176)\\
 &      & Ave & 160.60(4.22) & 0.125426(0.010750) & 175.90(12.78) & 0.125783(0.009848)\\
 &      & Sub & 4.80(4.73) & 0.265843(0.099998) & 12.30(2.00) & 0.174668(0.039216)\\
 &      & \textbf{Csl} & \textbf{14.80(2.25)} & \textbf{0.167269(0.035063)} & \textbf{60.00(38.46)} & \textbf{0.135633(0.020620)}\\
\addlinespace
 & 2.00 & Cen & 113.00(24.30) & 0.126540(0.015431) & 84.40(13.11) & 0.125888(0.016271)\\
 &      & Ave & 150.50(7.00) & 0.121087(0.016751) & 187.20(12.25) & 0.122729(0.019753)\\
 &      & Sub & 2.80(3.33) & 0.303726(0.071703) & 10.90(3.41) & 0.189335(0.048302)\\
 &      & \textbf{Csl} & \textbf{13.60(1.26)} & \textbf{0.165441(0.033621)} & \textbf{28.00(4.67)} & \textbf{0.143514(0.029659)}\\
\midrule

\addlinespace[2pt]
\multicolumn{7}{l}{\textbf{$\tau=0.75$}}\\
\addlinespace[2pt]
 & 1.20 & Cen & 34.10(25.06) & 0.237125(0.010183) & 37.00(12.89) & 0.234982(0.011371)\\
 &      & Ave & 147.10(49.35) & 0.228947(0.009800) & 196.10(0.99) & 0.231566(0.010208)\\
 &      & Sub & 21.00(6.94) & 0.238643(0.028567) & 24.90(14.36) & 0.245396(0.011684)\\
 &      & \textbf{Csl} & \textbf{35.00(4.88)} & \textbf{0.221803(0.008940)} & \textbf{88.20(18.41)} & \textbf{0.234125(0.013788)}\\
\addlinespace
 & 1.50 & Cen & 55.00(4.40) & 0.192875(0.012800) & 42.50(15.59) & 0.195336(0.012293)\\
 &      & Ave & 163.50(33.39) & 0.185321(0.012049) & 195.90(2.51) & 0.185645(0.010938)\\
 &      & Sub & 9.00(7.42) & 0.245072(0.056687) & 17.20(7.74) & 0.215694(0.021925)\\
 &      & \textbf{Csl} & \textbf{77.80(58.71)} & \textbf{0.184890(0.020956)} & \textbf{117.90(42.39)} & \textbf{0.183309(0.016964)}\\
\addlinespace
 & 1.80 & Cen & 66.80(34.96) & 0.163478(0.012802) & 57.90(4.28) & 0.161782(0.012656)\\
 &      & Ave & 163.20(18.50) & 0.156122(0.011711) & 194.70(4.99) & 0.157433(0.012435)\\
 &      & Sub & 3.10(3.51) & 0.283606(0.059313) & 13.70(7.07) & 0.200759(0.023308)\\
 &      & \textbf{Csl} & \textbf{52.30(52.51)} & \textbf{0.164170(0.019003)} & \textbf{85.40(45.01)} & \textbf{0.163582(0.025141)}\\
\addlinespace
 & 2.00 & Cen & 89.10(35.66) & 0.159399(0.015026) & 57.00(4.24) & 0.160411(0.015485)\\
 &      & Ave & 153.50(15.53) & 0.154278(0.015585) & 192.30(11.61) & 0.157000(0.013026)\\
 &      & Sub & 1.70(0.48) & 0.323278(0.031021) & 11.60(3.27) & 0.207641(0.032025)\\
 &      & \textbf{Csl} & \textbf{18.90(2.56)} & \textbf{0.170213(0.017190)} & \textbf{68.50(9.32)} & \textbf{0.165433(0.020520)}\\

\end{longtable}
\normalsize
{
\begin{table}[htbp]\centering\captionsetup{labelfont={color=black},textfont={color=black}}
\caption{{Descriptive statistics of the HIV response $Y$ ($\log_{10}$ scale).}}
\label{tab:hiv_descriptive}
\begin{tabular}{lr}\toprule
Statistic & Value\\\midrule
Number of isolates & 1,472\\
Number of binary predictors & 197\\
Mean & 0.5767\\
Sample standard deviation & 1.0114\\
Minimum & -1.0000\\
First quartile & -0.2218\\
Median & 0.1461\\
Third quartile & 1.4624\\
Maximum & 3.1761\\
Moment skewness & 0.5650\\
Adjusted Fisher--Pearson skewness & 0.5656\\
Pearson moment kurtosis & 1.9244\\
Moment excess kurtosis & -1.0756\\
Bias-adjusted excess kurtosis & -1.0751\\\bottomrule
\end{tabular}
\par\smallskip\begin{minipage}{.85\linewidth}\small
With $m_j=n^{-1}\sum_i(Y_i-\bar Y)^j$, moment skewness is $m_3/m_2^{3/2}$ and Pearson moment kurtosis is $m_4/m_2^2$; excess kurtosis subtracts 3. The adjusted statistics use the standard finite-sample Fisher--Pearson corrections. Normal reference values are zero for skewness and excess kurtosis, and 3 for Pearson kurtosis.\end{minipage}
\end{table}
}

\normalsize

\noindent
~\mbox{Table~\ref{tab:hiv_results}} summarizes the performance of the distributed $L^k$--quantile regression estimators across three quantile levels, $\tau \in \{0.25, 0.50, 0.75\}$, and four power parameters, $k \in \{1.20, 1.50, 1.80, 2.00\}$.  
For each configuration, we report the average number of nonzero regression coefficients (Size) and the average prediction error (PE) defined by
\[
\mathrm{PE} = \frac{1}{272} \sum_{i=1}^{272} \ell_{\tau,k}(Y_i - \widehat{Y}_i),
\]
where $\ell_{\tau,k}(\cdot)$ denotes the asymmetric $L^k$ loss function.  
Numbers in parentheses are standard deviations calculated over ten random data partitions.  
Both the SCAD and adaptive LASSO  penalties are evaluated for the centralized (Cen), naive average (Ave), {first-site surrogate (Sub)}, and consensus (Csl) estimators.

\medskip
\noindent
The main empirical observations are as follows:

{\par \label{loc:hivinterpret}
\noindent The relative ranking in Table~\ref{tab:hiv_results} depends on $(\tau,k)$ and the penalty. For example, under SCAD at $\tau=0.25,k=1.50$, Csl has PE $0.154541$ and Ave has PE $0.155322$; at $k=1.80$, Ave has PE $0.121671$ and Csl has PE $0.135635$. Larger models can improve prediction through smaller shrinkage bias or reduced prediction variance after averaging; model size alone is not evidence of overfitting. Predictive performance and support interpretation are distinct.

\par \label{loc:hivsize}
\noindent The observed differences in model size may reflect the distinct weighting mechanisms of the two penalties. Adaptive weights depend on the pilot, whereas the SCAD derivative decreases to zero for sufficiently large coefficients. Tuning and predictor correlation may also influence selection. For Csl at $\tau=0.25,k=1.50$, the reported mean sizes are $124.30$ under SCAD and $126.90$ under aLASSO, illustrating this pattern in the HIV data.

\par \label{loc:hivpower}
\noindent At $\tau=0.25$, the Csl SCAD PE values decrease from $0.186376$ to $0.154541$, $0.135635$ and $0.130658$ as $k$ increases through the displayed grid. These powers use differently scaled losses and target different functionals, so this sequence does not by itself identify the best $k$. A common MAE criterion, evaluated using training-only cross-validation, provides a way to compare candidate powers as described in Section~\ref{rev:tuning}.

}
\section{Discussion} \label{Sec. 7}

 In this work, we developed a new framework for penalized large-scale $L^k$--quantile regression in distributed environments.  
Our approach accommodates SCAD and adaptive LASSO penalties, and we established oracle properties analogous to those obtained in the classical centralized setting \citep{fan2001}.  
A key contribution of this article is the construction of a communication-efficient surrogate loss, denoted $\mathrm{CSL}^k$, which enables statistically optimal estimation while maintaining low communication cost.  
{The theoretical guarantees here apply to fixed dimension and a fixed number of sites.}

To optimize the penalized $\mathrm{CSL}^k$ objective, we introduced an augmented proximal ADMM algorithm tailored to the structure of our surrogate loss.  
{The proximal updates are tractable; this alone does not establish convergence of the complete nonconvex procedure.}  
Simulation studies demonstrate that our estimator performs competitively in favorable finite-sample settings: after a few rounds of communication, the estimation quality of $\mathrm{CSL}^k$ nearly matches that of the centralized estimator.  
Furthermore, the empirical application confirms that the method is straightforward to implement in real-world large-scale problems. To facilitate practical use, the proposed algorithm has been implemented in the dedicated \textsf{R} package \texttt{dlkqr}, {publicly available at}  \url{https://github.com/abdellah-atanane/dlkqr}. 

 Despite these promising results, several important challenges remain.  
First, the variable selection procedure studied here focuses on the regime where the number of covariates $p$ is fixed while the sample size $N$ is large.  
Extending the theory to the high-dimensional settings $p \to \infty$ or $p \gg N$ is nontrivial.  
In such cases, the stability of nonconvex penalties, the behavior of the ADMM iterates, and the statistical properties of the distributed surrogate loss require new analysis.  
A common remedy in the $p \gg N$ context is a two--stage strategy:  
(i) an initial screening step reduces the dimensionality to a manageable scale, and  
(ii) penalized methods perform the final selection and estimation \citep{singh2023subdata}.  
Investigating how such screening can be integrated with $\mathrm{CSL}^k$ and whether the resulting pipeline retains the oracle properties is an open and challenging question.

 {Second, the fixed-budget experiments show sensitivity to local sample size, and} rigorous convergence guarantees for nonconvex ADMM in distributed penalized quantile-type problems remain largely undeveloped.  
Recent work on ADMM for nonconvex optimization (e.g., \citealp{guo2017convergence} and  \citealp{WangYinZeng2019}) provides conditions ensuring convergence to stationary points, but these conditions are not directly applicable to our setting due to the structure of the linear constraints induced by the surrogate function.  
Developing a dedicated theoretical analysis for the proposed augmented ADMM, in particular under SCAD and adaptive LASSO penalties, is therefore an important avenue for future research.

 We expect that the ideas developed here---surrogate loss construction, proximal ADMM optimization, and distributed regularization---will contribute to further advances in scalable quantile-type estimation and motivate additional theoretical investigations.

\section*{Declarations}
\noindent The authors declare that they have no conflicts of interest or competing interests.

\appendix
\renewcommand{\thefigure}{\arabic{figure}}\renewcommand{\thetable}{\arabic{table}}

\section{Proofs }  \label{proofs}

\noindent This section contains the main lemmas along with their corresponding proofs. Suppose the sample set \(\{\boldsymbol{X}_i, Y_i\}_{i=1}^N\) is generated according to process \eqref{process without lost of generality}; for simplicity, the dataset stored on the \(\nu\)-th machine is denoted by \(\{\boldsymbol{X}_{\nu i}, Y_{\nu i}\} \equiv \{\boldsymbol{X}_i, Y_i\}_{i=1}^{n}\), for ease of notation and when there is no risk of confusion. We also suppress $\boldsymbol{\beta}^0$ and $N$ from $\mathcal{R}_{N}(\boldsymbol{\beta};\boldsymbol{\beta}^0)$ when there is no possible confusion.\\ 
The following lemmas are essential for the proofs of \mbox{Theorems~\ref{theorem scad} and \ref{theorem al}}.
\begin{lm}\label{lemma 3}
Under $\mbox{Assumptions~\ref{assumption 1} and \ref{assumption 2}}$, denote   $ \varepsilon^0_{\nu i} = Y_{\nu i} - \boldsymbol{X}_{ \nu i}^\top \boldsymbol{\beta}^{0}$, then,
\begin{enumerate}[i)]

\item $\varrho(\tau) = \mathbb{E}([\ell'_{\tau,k}(\varepsilon_i) ]^2)$ and $\varkappa(\tau)=\mathbb{E}([\ell_{\tau,k}''(\varepsilon_{i})])$  are well-defined, i.e., $\varrho(\tau)<\infty$,  $\varkappa(\tau)<\infty$.

 \item For any constant $\mathcal{C}$  large enough, and for any $\boldsymbol{\delta}$ such that \( \|\boldsymbol{\delta}\|_2 = \mathcal{C}\), if we set

\begin{equation}\label{a1}
 T_n (\boldsymbol{\delta}) = \sum_{i=1}^n \left[ \ell_{\tau,k} \left( \varepsilon_{ i} - \boldsymbol{X}_{ i}^\top \boldsymbol{\delta}/\sqrt{n} \right) - \ell_{\tau,k} (\varepsilon_{ i}) \right] + n  [\nabla \mathcal{R}_{N,\nu} (\boldsymbol{\beta}^0) ]^\top \boldsymbol{\delta}/\sqrt{n},   
\end{equation}

then, one has 
\[
T_n (\boldsymbol{\delta}) = \varkappa (\tau)\boldsymbol{\delta}^\top \boldsymbol \Sigma_n \boldsymbol{\delta}/2 + \boldsymbol{\Delta}_n^\top \boldsymbol{\delta} + o_p (1),
\]
where 
$\boldsymbol{H}_i = \boldsymbol{b}_i - \boldsymbol{e}_i/V - \boldsymbol{d}_i, \, i=1,2,\ldots,n $, $\boldsymbol{b}_i = \ell'_{\tau,k} ( \varepsilon^0_{ i}) \boldsymbol{X}_{ i},  \; \boldsymbol{e}_i = \sum_{\nu=1}^V \ell'_{\tau,k} (\varepsilon^0_{\nu i}) \boldsymbol{X}_{\nu i}$, $\boldsymbol{d}_i = \ell'_{\tau ,k}(\varepsilon_{ i}) \boldsymbol{X}_{ i}$,  $\boldsymbol{\Delta}_n = \frac{1}{\sqrt{n}} \sum_{i=1}^n \boldsymbol{H}_i,$ and $V$ is the number of machines.

\end{enumerate}
 
\end{lm}

\subsubsection*{Proof of \mbox{Lemma~\ref{lemma 3}} }

\begin{enumerate}[i)]
    \item

 We have \( \mathbb{E}(\ell'_{\tau,k}(\varepsilon_i) \mid \boldsymbol{X}_i) = 0 \), since by  \mbox{Assumption~\ref{assumption 1}}, $Q_{\tau,k}(\varepsilon_i\mid \boldsymbol{X}_i):=\arg\min\limits_{\theta \in \mathbb{R}} \mathbb{E}[\ell(\varepsilon_i-\theta)-\ell(\varepsilon_i) \mid \boldsymbol{X}_i]=0.$
Then, one has, since $\Psi_\tau(\varepsilon_i)\in{\{\tau,1-\tau\}}$, we have the uniform bound
\[
\Psi_\tau^2(\varepsilon_i)\le \max\{\tau^2,(1-\tau)^2\}
= \big(\max\{\tau,1-\tau\}\big)^2.
\]
Therefore,
\[
\begin{aligned}
\varrho(\tau)
& =\mathbb{E}\left[  \ell'^2_{\tau,k}(\varepsilon_{ i})\mid\boldsymbol X_i \right]-\left( \mathbb{E}\left[  \ell'_{\tau,k}(\varepsilon_{ i}) \mid\boldsymbol X_i\right] \right)^2 =\mathbb{E}\!\left(k^2\Psi_\tau^2(\varepsilon_i)\,|\varepsilon_i|^{2(k-1)}\mid X_i\right)\\
&\le k^2\big(\max\{\tau,1-\tau\}\big)^2\,
\mathbb{E}\!\left(|\varepsilon_i|^{2(k-1)}\mid X_i\right)\\
&= k^2\big(\max\{\tau,1-\tau\}\big)^2\,
\mathbb{E}\!\left(
|\varepsilon_i|^{2(k-1)}\mathds{1}\{|\varepsilon_i|<1\}
+
|\varepsilon_i|^{2(k-1)}\mathds{1}\{|\varepsilon_i|\ge 1\}
\mid X_i
\right).
\end{aligned}
\]

On the event $\{|\varepsilon_i|<1\}$, we have
$|\varepsilon_i|^{2(k-1)}\le 1$ since $2(k-1)>0$.
On the event $\{|\varepsilon_i|\ge 1\}$, note that the function
$s\mapsto a^{s}$ is nondecreasing  for any $a>1$; hence,
because $2(k-1)\le 2k$ for $k\in(1,2]$, we obtain
\[
|\varepsilon_i|^{2(k-1)}\le |\varepsilon_i|^{2k}.
\]
Consequently,
\[
\begin{aligned}
\varrho(\tau)
&\le k^2\big(\max\{\tau,1-\tau\}\big)^2
\mathbb{E}\!\left(
\mathds{1}\{|\varepsilon_i|<1\}
+
|\varepsilon_i|^{2k}\mathds{1}\{|\varepsilon_i|\ge 1\}
\mid \boldsymbol{X}_i
\right)\\
&\le c_0
\Big(1+\mathbb{E}\!\left(|\varepsilon_i|^{2k}\mid \boldsymbol{X}_i\right)\Big),
\end{aligned}
\]
where we used the monotonicity of $s\mapsto a^{s}$ on $[1,\infty)$ for $a>1$, with $c_0=k^2\max^2 \left\{ \tau, 1-\tau \right\}.$ Then, \( \varrho(\tau) < \infty \), since \( 1<k\leq2 \), according to  \mbox{Assumption~\ref{assumption 1}}. \\

Now, we want to prove that \( \varkappa(\tau) < \infty \). It is easy to see that this is true if \( k = 2 \); then, we only consider the case \( 1 < k < 2 \). We have

\begin{align}\label{A1}
 \varkappa(\tau) = \int_{0}^{\infty} k(k - 1) \tau t^{k - 2} f_{\varepsilon}(t) \, dt + \int_{-\infty}^{0} k(k - 1)(1 - \tau) |t|^{k - 2} f_{\varepsilon}(t) \, dt. \quad    
\end{align}

The first term on the right-hand side of \eqref{A1} satisfies
\begin{align*}
 \int_{0}^{\infty} k(k - 1) \tau t^{k - 2} f_{\varepsilon}(t) \, dt &\leq k(k - 1) \tau \left[ \int_{0}^{1} c_1\, t^{k - 2} \, dt + \int_{1}^{\infty} f_{\varepsilon}(t) \, dt \right] \\
 &\leq k(k - 1) \tau \left( \frac{c_1}{k - 1} + 1 \right) \\
 &< \infty.   
\end{align*}

With \( c_1 = \max_{t \in [0,1]} f_{\varepsilon}(t) \), $c_1<\infty$ due to the continuous nature  of \(f_{\varepsilon}(\cdot)\) on the closed interval \([0,1]\) 
and $\int_{1}^{\infty} f_{\varepsilon}(t) \, dt=F(\infty)-F_{\varepsilon}(1)=1-F_{\varepsilon}(1)<1,$ due to the continuous nature of the density of the density $f_{\varepsilon}(\cdot)$.\\

Similarly, the second term on the right-hand side of \eqref{A1}  is finite. i.e.;

\begin{align*}
 \int_{-\infty}^{0} k(k - 1) (1-\tau) (-t)^{k - 2} f_{\varepsilon}(t) \, dt &=  \int_{0}^{+\infty} k(k - 1) (1-\tau) (s)^{k - 2} f_{\varepsilon}(-s) \, ds\\
 &\leq k(k - 1) (1-\tau) \left[ \int_{0}^{1} c_2 s^{k - 2} \, ds + \int_{1}^{+\infty} f_{\varepsilon}(-s) \, ds \right] \\
 &\leq k(k - 1) {(1-\tau)} \left( \frac{c_2}{k - 1} + {F_{\varepsilon}(-1)} \right) \\
 &< \infty,  
\end{align*}
with \( c_2= \max_{t \in [0,1]} f_{\varepsilon}(-t)<\infty \). In conclusion, we have \( \varkappa(\tau) < \infty \).

\item Define $h_{\tau,k}(\varepsilon):=\frac{\partial}{\partial t}(\ell_{\tau,k}(\varepsilon-t))\mid_{t=0}$, i.e., the derivative of \( \ell_{\tau,k}(\varepsilon - t) \) with respect to \( t \) at \( t = 0 \).

For simplicity, we introduce the notations

\[
G_{i,n}(\boldsymbol{\delta}) := \ell_{\tau,k} \left(\varepsilon_{i} - \boldsymbol{X}_{i}^\top \boldsymbol{\delta}/\sqrt{n}\right) - \ell_{\tau,k} (\varepsilon_{i}) - h_{\tau,k} (\varepsilon_{i}) \boldsymbol{X}_{i}^\top \boldsymbol{\delta}/\sqrt{n}.
\]
and \[
\hslash_n(\boldsymbol{\delta})=\sum_{i=1}^{n}{\big[}\ell_{\tau,k} \left(\varepsilon_{i} - \boldsymbol{X}_{i}^\top \boldsymbol{\delta}/\sqrt{n}\right) - \ell_{\tau,k} (\varepsilon_{i}){\big]}.
\]
Next, we define the function \[
 J(t) = \mathbb{E}\left( \ell_{\tau,k} (\varepsilon_{i} - t) - \ell_{\tau,k} (\varepsilon_{i}) \right).
\]
Its second-order Taylor expansion has the following form
\[
J(t) = \varkappa(\tau) t^2/2 + o(t^2).
\]

So for a large \( n \), we have

\begin{align*}
\mathbb{E}(\hslash_n(\boldsymbol{\delta})) &= \sum_{i=1}^{n} J\left( \boldsymbol{X}_{i}^\top \boldsymbol{\delta} /\sqrt{n}\right) \notag \\
&= \sum_{i=1}^{n} \left[ \varkappa(\tau)\left(  \boldsymbol{X}_{i}^\top \boldsymbol{\delta}/\sqrt{n} \right)^2 /2+ o\left( \left(  \boldsymbol{X}_{i}^\top \boldsymbol{\delta} \right/\sqrt{n})^2\right) \right] \\
&= \varkappa(\tau) \boldsymbol{\delta}^\top \boldsymbol{\Sigma}_n \boldsymbol{\delta}/2 + o\left( \boldsymbol{\delta}^\top \boldsymbol{\Sigma}_n \boldsymbol{\delta} \right),
\end{align*}
where $\boldsymbol \Sigma_n=\frac{1}{n}\sum_{i=1}^n\boldsymbol{X}_i\boldsymbol{X}_i^\top.$
{Since $\max_i\|\boldsymbol X_i\|_2/\sqrt n\to0$ and $\boldsymbol\delta^\top\boldsymbol\Sigma_n\boldsymbol\delta=O(1)$ for fixed $\boldsymbol\delta$, the sum of the deterministic remainders is $o(1)$.}

Thus, under {\mbox{Assumptions~\ref{assumption 1} and \ref{assumption 2}}}, we have
\begin{align*}
\mathbb{E}(\hslash_n(\boldsymbol{\delta})) = \varkappa(\tau) \boldsymbol{\delta}^\top \boldsymbol{\Sigma}_n \boldsymbol{\delta}/2 + o(1).
\end{align*}
Hence, one has
\begin{align*}
\hslash_n(\boldsymbol{\delta}) &= \mathbb{E}(\hslash_n(\boldsymbol{\delta})) + \sum_{i=1}^{n} h_{\tau,k} (\varepsilon_{i}) \boldsymbol{X}_{i}^\top \boldsymbol{\delta}/ \sqrt{n} + \sum_{i=1}^{n} \left[ G_{i,n}(\boldsymbol{\delta}) - \mathbb{E} (G_{i,n}(\boldsymbol{\delta}))\right].
\end{align*}
To conclude, we need to prove that
\begin{align*}
\sum_{i=1}^{n} \left[ G_{i,n}(\boldsymbol{\delta}) - \mathbb{E}( G_{i,n}(\boldsymbol{\delta})) \right] = o_p(1).
\end{align*}

For a fixed \( \boldsymbol{\delta} \), we have

\begin{align*}
\mathbb{E}{\Big[}\left( \sum_{i=1}^{n} \left[ G_{i,n}(\boldsymbol{\delta}) - \mathbb{E} (G_{i,n}(\boldsymbol{\delta})) \right] \right)^2{\Big]} &= \sum_{i=1}^{n} \mathbb{E} \left[ \left( G_{i,n}(\boldsymbol{\delta}) - \mathbb{E}[ G_{i,n}(\boldsymbol{\delta})] \right)^2 \right] \notag \\
&\leq \sum_{i=1}^{n} \mathbb{E}\left[ \left( G_{i,n}(\boldsymbol{\delta}) \right)^2 \right],
\end{align*}
because 
\begingroup\small
\begin{align*}
\mathbb{E}\left[\left(\sum_{i=1}^n (G_{i,n}(\boldsymbol{\delta}) - \mathbb{E}[G_{i,n}(\boldsymbol{\delta})])\right)^2\right] 
&= \mathbb{E}\left[ \sum_{i=1}^n (G_{i,n}(\boldsymbol{\delta}) - \mathbb{E}[G_{i,n}(\boldsymbol{\delta})])^2 \right. \\
&\quad + \left. 2 \sum_{1 \leq i < j \leq n} (G_{i,n}(\boldsymbol{\delta}) - \mathbb{E}[G_{i,n}(\boldsymbol{\delta})]) (G_{j,n}(\boldsymbol{\delta}) - \mathbb{E}[G_{j,n}(\boldsymbol{\delta})]) \right],
\end{align*}
\endgroup
then 
\begingroup\footnotesize
\begin{eqnarray*}
\mathbb{E}\left[\left(\sum_{i=1}^n (G_{i,n}(\boldsymbol{\delta}) - \mathbb{E}[G_{i,n}(\boldsymbol{\delta})])\right)^2\right] 
& = & \sum_{i=1}^n \mathbb{E}\left[  (G_{i,n}(\boldsymbol{\delta}) - \mathbb{E}[G_{i,n}(\boldsymbol{\delta})])^2 \right] \\
& & + 2 \sum_{1\leq i<j\leq n} \left( \mathbb{E}[G_{i,n}(\boldsymbol{\delta})] - \mathbb{E}[G_{i,n}(\boldsymbol{\delta})] \right)  \left( \mathbb{E}[G_{{j},n}(\boldsymbol{\delta})] - \mathbb{E}[G_{j,n}(\boldsymbol{\delta})] \right),
\end{eqnarray*}
\endgroup
hence 
\begin{align}
\mathbb{E}\left[\left(\sum_{i=1}^n (G_{i,n}(\boldsymbol{\delta}) - \mathbb{E}[G_{i,n}(\boldsymbol{\delta})])\right)^2\right] 
&= \sum_{i=1}^n  \mathbb{V}\mathrm{ar}[  (G_{i,n}(\boldsymbol{\delta}) )] \notag \\
&= \sum_{i=1}^n {\big\{}\mathbb{E}[  (G_{i,n}(\boldsymbol{\delta}) )^2]-  (\mathbb{E}[  G_{i,n}(\boldsymbol{\delta} )])^2{\big\}} \notag \\
&\leq \sum_{i=1}^n  \mathbb{E}[  (G_{i,n}(\boldsymbol{\delta}) )^2]. \label{last equality1} 
\end{align}

The last part of  inequality \eqref{last equality1} can be written as

\begin{align}
\sum_{i=1}^{n} \mathbb{E}\left[ \left( G_{i,n}(\boldsymbol{\delta}) \right)^2\right]&= \sum_{i=1}^{n} \mathbb{E}\left[ \left(G_{i,n}(\boldsymbol{\delta})\right)^2 \mathds{1}\left\{|\varepsilon_{i}| \leq \left|\boldsymbol{X}_{i}^\top \boldsymbol{\delta}/ \sqrt{n}\right|\right\} \right] \notag \\
&\quad + \sum_{i=1}^{n} \mathbb{E} \left[  \left(G_{i,n}(\boldsymbol{\delta})\right)^2  \mathds{1}\left\{|\varepsilon_{i}| > \left|\boldsymbol{X}_{i}^\top \boldsymbol{\delta}/ \sqrt{n}\right|\right\} \right]  \notag \\
&:= I_{1n}(\boldsymbol{\delta}) + I_{2n}(\boldsymbol{\delta}),\label{I_1 I_2}
\end{align}
the components in equation \eqref{I_1 I_2} are such that
\begin{align}
    I_{1n}(\boldsymbol{\delta})=\sum_{i=1}^{n} \mathbb{E}\left[ \left(G_{i,n}(\boldsymbol{\delta})\right)^2 \mathds{1}\left\{|\varepsilon_{i}| \leq \left|\boldsymbol{X}_{i}^\top \boldsymbol{\delta}/ \sqrt{n}\right|\right\} \right],\label{I_1}\\
    I_{2n}(\boldsymbol{\delta})=\sum_{i=1}^{n} \mathbb{E} \left[  \left(G_{i,n}(\boldsymbol{\delta})\right)^2  \mathds{1}\left\{|\varepsilon_{i}| > \left|\boldsymbol{X}_{i}^\top \boldsymbol{\delta}/ \sqrt{n}\right|\right\} \right]\label{I2}.
\end{align}
We begin by analyzing the term \( I_{1n} \) in equation \eqref{I_1}. Using a second-order Taylor expansion of \( \ell_{\tau,k}(\varepsilon_i - t) \) at \( t = 0 \), {for a fixed nonzero residual and a sufficiently small perturbation,} we obtain (see \mbox{Remark~\ref{remark residual}})

\[
\ell_{\tau,k}(\varepsilon_i - \boldsymbol{X}_i^\top \boldsymbol{\delta} / \sqrt{n}) = \ell_{\tau,k}(\varepsilon_i) + h_{\tau,k}(\varepsilon_i) \boldsymbol{X}_i^\top \boldsymbol{\delta} / \sqrt{n} + \frac{1}{2} \ell_{\tau,k}''(\varepsilon_i) (\boldsymbol{X}_i^\top \boldsymbol{\delta} / \sqrt{n})^2 + O(|\boldsymbol{X}_i^\top \boldsymbol{\delta} / \sqrt{n}|^3).
\]

\noindent {The constant in the remainder may depend on the fixed residual.} Subtracting \( \ell_{\tau,k}(\varepsilon_i) \) and \( h_{\tau,k}(\varepsilon_i) \boldsymbol{X}_i^\top \boldsymbol{\delta} / \sqrt{n} \), we get

\[
G_{i,n}(\boldsymbol{\delta}) = \frac{1}{2} \ell_{\tau,k}''(\varepsilon_i) (\boldsymbol{X}_i^\top \boldsymbol{\delta} / \sqrt{n})^2 + O(|\boldsymbol{X}_i^\top \boldsymbol{\delta} / \sqrt{n}|^3).
\]

Since the dominant term is quadratic, it follows {for each fixed $i$,} that

\[
G_{i,n}(\boldsymbol{\delta}) = O_p((\boldsymbol{X}_i^\top \boldsymbol{\delta} / \sqrt{n})^2),
\]

{The pointwise $O_p$ statement alone does not give a deterministic inequality.} {For the bound below, the H\"older continuity of the loss derivative and integration give $|G_{i,n}(\boldsymbol{\delta})|\le C_{\tau,k}|\boldsymbol{X}_i^\top\boldsymbol{\delta}/\sqrt n|^k\le C_{\tau,k}|\boldsymbol{X}_i^\top\boldsymbol{\delta}/\sqrt n|$ when $|\boldsymbol{X}_i^\top\boldsymbol{\delta}/\sqrt n|\le1$, with $C_{\tau,k}=2\max(\tau,1-\tau)$. We have}
\[
|G_{i,n}(\boldsymbol{\delta})| \mathds{1}\left\{|\varepsilon_{i}| \leq \left|\boldsymbol{X}_{i}^\top \boldsymbol{\delta}/ \sqrt{n}\right|\right\} \leq {C_{\tau,k}}\left|\boldsymbol{X}_{i}^\top \boldsymbol{\delta}/ \sqrt{n}\right| \mathds{1}\left\{|\varepsilon_{i}| \leq \left|\boldsymbol{X}_{i}^\top \boldsymbol{\delta}/ \sqrt{n}\right|\right\},
\]
therefore 
\begin{align}
I_{1n}(\boldsymbol{\delta}) &\leq {C_{\tau,k}^{2}}\sum_{i=1}^{n} \left[\left|\boldsymbol{X}_{i}^\top \boldsymbol{\delta}/ \sqrt{n}\right|^2 \mathbb{P}\left(|\varepsilon_{i}| \leq  \|\boldsymbol{\delta}\|_2 \max_{1 \leq i \leq n} \|\boldsymbol{X}_{i}\|_2/\sqrt{n} 
 \right) \right] \notag \\
&\leq {C_{\tau,k}^{2}}\sum_{i=1}^{n} \left|\boldsymbol{X}_{i}^\top \boldsymbol{\delta}/ \sqrt{n}\right|^2 \mathbb{P}\left(|\varepsilon_{i}| \leq \|\boldsymbol{\delta}\|_2 \max_{1 \leq i \leq n} \|\boldsymbol{X}_{i}\|_2/\sqrt{n} \right) \notag\\
&\to 0, \label{deal I1}
\end{align}
equation \eqref{deal I1} follows under the common error law and Assumption~\ref{assumption 2}; the stronger bound also gives $I_{1n}\le C_{\tau,k}^2\sum_i |\boldsymbol X_i^\top\boldsymbol\delta/\sqrt n|^{2k}\to0$.
Next, we analyze the term \( I_{2n} \) in equation \eqref{I2}. Put $d_{i,n}=\boldsymbol{X}_i^\top\boldsymbol{\delta}/\sqrt n$. On $|\varepsilon_i|>|d_{i,n}|$, the preceding H\"older bound and $|\ell''_{\tau,k}(\varepsilon_i)d_{i,n}^2|\le k(k-1)|d_{i,n}|^k$ give
\[
G_{i,n}^2(\boldsymbol{\delta})=\frac14[\ell''_{\tau,k}(\varepsilon_i)]^2d_{i,n}^4+O(|d_{i,n}|^{2k}).
\]
The remainder bound is uniform on this event, and $\sum_i|d_{i,n}|^{2k}=o(1)$ by Assumption~\ref{assumption 2}. 
So, if \( 1 < k \leq 2 \), we have

\begin{align}
I_{2n}(\boldsymbol{\delta}) &\leq \frac{1}{4}k^2(k-1)^2 \sum_{i=1}^{n} \left| \boldsymbol{X}_{i}^\top \boldsymbol{\delta}/\sqrt{n}\right|^4 \int_{|t| > \left| \boldsymbol{X}_{i}^\top \boldsymbol{\delta}/\sqrt{n}\right|} |t|^{2(k-2)} f_{\varepsilon}(t) dt +o(1) \notag \\
&\leq \frac{1}{4}k^2(k-1)^2 \sum_{i=1}^{n} \left| \boldsymbol{X}_{i}^\top \boldsymbol{\delta}/\sqrt{n}\right|^{2k} +o(1) \notag \\
&=  \frac{1}{4}k^2(k-1)^2 \sum_{i=1}^{n} \left| \boldsymbol{X}_{i}^\top \boldsymbol{\delta}/\sqrt{n}\right|^{2}\left| \boldsymbol{X}_{i}^\top \boldsymbol{\delta}/\sqrt{n}\right|^{2(k-1)} +o(1) \notag \\
&\leq  \frac{1}{4}k^2(k-1)^2 \sum_{i=1}^{n} \left| \boldsymbol{X}_{i}^\top \boldsymbol{\delta}/\sqrt{n}\right|^{2}\max_{i\in[n]}\left| \boldsymbol{X}_{i}^\top \boldsymbol{\delta}/\sqrt{n}\right|^{2(k-1)} +o(1) \notag \\
& \leq \frac{1}{4}k^2(k-1)^2 [\boldsymbol{\delta}^\top   \boldsymbol{\Sigma}_n \boldsymbol{\delta}] \left|  (\max_{i\in [n]} \| \boldsymbol{X}_i \|_2/{\sqrt n}) \|\boldsymbol{\delta} \|_2 \right|^{2(k-1)} +o(1) \notag \\
&\leq  \frac{1}{4}k^2(k-1)^2 [\boldsymbol{\delta}^\top   \boldsymbol{\Sigma}_n \boldsymbol{\delta}] \left(  \left[ \max_{i\in [n]}  \| \boldsymbol{X}_i \|_2/\sqrt{n}\right]^{2(k-1)}  \|\boldsymbol{\delta} \|^{2(k-1)}_2 \right) +o(1) \notag \\
&\to 0,\label{deal I_2}
\end{align}
equation \eqref{deal I_2} follows since \( \boldsymbol{\delta}^\top   \boldsymbol{\Sigma}_n\, \boldsymbol{\delta}  \to \boldsymbol{\delta}^\top\boldsymbol{\Sigma}\boldsymbol{\delta} \) and \(\left[ \frac{\max_{i\in [n]}  \| \boldsymbol{X}_i \|_2}{\sqrt{n}}\right]^{2(k-1)} \|\boldsymbol{\delta} \|^{2(k-1)}_2 \to 0\) for \(\|\boldsymbol{\delta} \|_2=\mathcal{C}.\)
As a result of \eqref{deal I1} and \eqref{deal I_2}, we conclude that

\[
\sum_{i=1}^{n}\left[G_{i,n}(\boldsymbol{\delta}) - \mathbb{E} (G_{i,n}(\boldsymbol{\delta}))\right] = o_p\left(1\right) \text{ for } 1 < k {\le} 2.
\]

Consequently, we derive

\begin{align}\label{a6}
\sum_{i=1}^n \left[\ell_{\tau,k} \left(\varepsilon_{i} - \boldsymbol{X}_{i}^\top \boldsymbol{\delta}/\sqrt{n}\right) - \ell_{\tau,k} (\varepsilon_{i}) \right] &= \varkappa(\tau) \boldsymbol{\delta}^\top \boldsymbol{\Sigma}_n \boldsymbol{\delta}/2 -\sum_{i=1}^{n} \ell'_{\tau,k} (\varepsilon_{i}) \boldsymbol{X}_{i}^\top \boldsymbol{\delta}/ \sqrt{n}  + o_p (1).
\end{align}

By a straightforward calculation, we obtain

\begin{equation}\label{a7}
\nabla \mathcal{R}_{N,\nu} (\boldsymbol{ \beta}^0) = \frac{1}{n} \sum_{i=1}^n \left[ \ell'_{\tau,k} ( \varepsilon^0_{ i}) \boldsymbol{X}_{ i} - \frac{1}{V} \sum_{\nu=1}^V \ell'_{\tau,k} ( \varepsilon^0_{\nu i}) \boldsymbol{X}_{\nu i} \right].
\end{equation}

Combining the  equations \eqref{a1}, \eqref{a6}, and \eqref{a7}, we obtain
\begin{equation*}
T_n (\boldsymbol{\delta}) = \varkappa (\tau)\boldsymbol{\delta}^\top \boldsymbol{\Sigma}_n \boldsymbol{\delta}/2 + \boldsymbol{\Delta}_n^\top \boldsymbol{\delta} + o_p (1).
\end{equation*}
This concludes the proof.

\begin{lm}\label{lemma 4}
  For \(\boldsymbol{\Delta}_n\) defined in  \mbox{Lemma~\ref{lemma 3}} and under  $\mbox{Assumptions~\ref{assumption 1} and \ref{assumption 2}}$, one has
\[
\boldsymbol{\Delta}_n = \frac{1}{\sqrt{n}} \sum_{i=1}^n \boldsymbol{H}_i \xrightarrow{\mathcal D} \mathcal N \left( \boldsymbol{0}, \upsilon (\tau) \boldsymbol{\Sigma}/V \right).
\]

\end{lm}

\subsubsection*{Proof of $\mbox{Lemma~\ref{lemma 4}}$}
 \mbox{Assumption~\ref{assumption 1}} implies that the \(\tau\)-th \( L^k \)-quantile of the error term is zero, which leads to  
\begin{align}\label{Exp lprime}
\mathbb{E} \left( \ell'_{\tau,k} (\varepsilon_{i}) \mid \boldsymbol{X}_{i} \right) = \mathbb{E} \left( \ell'_{\tau,k} (\varepsilon^0_{i}) \mid \boldsymbol{X}_{i} \right) = 0.
\end{align}

Since
\begin{align*}
\boldsymbol{\Delta}_n &= \frac{1}{\sqrt{n}} \sum_{i=1}^n \boldsymbol{H}_i, \quad \text{where} \quad \boldsymbol{H}_i = \begin{pmatrix} \boldsymbol{b}_i -  \boldsymbol{e}_i/V - \boldsymbol{d}_i \end{pmatrix} \\
{\boldsymbol H_i}&= \begin{pmatrix} \mathbb{I}_p, - \mathbb{I}_p/V,  -\mathbb{I}_p \end{pmatrix} \begin{pmatrix} \boldsymbol{b}_i^\top ,\boldsymbol{e}_i^\top, \boldsymbol{d}_i^\top\end{pmatrix}^\top,
\end{align*}
a straightforward calculation yields
\begin{equation}\label{a9}
\mathbb{C}\mathrm{ov} \begin{pmatrix} \boldsymbol{b}_i \\ \boldsymbol{e}_i \\ \boldsymbol{d}_i \end{pmatrix} =
\begin{pmatrix}
\boldsymbol{B}_{11} & \boldsymbol{B}_{12} & \boldsymbol{B}_{13} \\
\boldsymbol{B}_{12} & \boldsymbol{B}_{22} & \boldsymbol{B}_{13} \\
\boldsymbol{B}_{13} & \boldsymbol{B}_{13} & \boldsymbol{B}_{33}
\end{pmatrix},
\end{equation}
where
\[
\boldsymbol{B}_{11} = \varsigma(\tau) \boldsymbol{X}_{i} \boldsymbol{X}_{i}^\top, \quad
\boldsymbol{B}_{22} = \varsigma(\tau) {\sum_{\nu=1}^{V}} \boldsymbol{X}_{\nu i} \boldsymbol{X}_{\nu i}^\top, \quad
\boldsymbol{B}_{33} = \varrho(\tau) \boldsymbol{X}_{i} \boldsymbol{X}_{i}^\top,
\]
\[
\boldsymbol{B}_{12} = \varsigma(\tau) \boldsymbol{X}_{i} \boldsymbol{X}_{i}^\top, \quad
\boldsymbol{B}_{13} = \vartheta(\tau) \boldsymbol{X}_{i} \boldsymbol{X}_{i}^\top,
\]
based on \eqref{Exp lprime},
with
$
\varsigma(\tau) = \mathbb{V}\mathrm{ar} \left[\ell'_{\tau,k}(\varepsilon^0_{i})\right],$ and $\vartheta(\tau) = \mathbb{C}\mathrm{ov} \left[\ell'_{\tau,k}(\varepsilon^0_{i}), \ell'_{\tau,k}(\varepsilon_{i})\right]$, where \( \ell'_{\tau,k}(\cdot) \) is defined in \eqref{ell prime}.

We now verify the Lindeberg--Feller condition for the central limit theorem \citep{vanderVaart1998}. By \eqref{a9}, \mbox{Assumption~\ref{assumption 1}}, and the fact that \( \boldsymbol{\Delta}_n \) is a sum of independent  zero-mean random vectors, we obtain
\begingroup\small
\begin{align}
\mathbb{C}\mathrm{ov}(\boldsymbol{\Delta}_n) &= \frac{1}{n} \sum_{i=1}^n \begin{pmatrix} \mathbb{I}_p, {-\mathbb{I}_p/V, -\mathbb{I}_p} \end{pmatrix} \mathbb{V}\mathrm{ar}  \begin{pmatrix} \boldsymbol{b}_i \\ \boldsymbol{e}_i \\ \boldsymbol{d}_i \end{pmatrix} \begin{pmatrix} \mathbb{I}_p, {-\mathbb{I}_p/V, -\mathbb{I}_p} \end{pmatrix}^\top \notag \\
&= \frac{1}{n} \sum_{i=1}^n \left\{  \left( \varrho(\tau) + (V-2) \varsigma(\tau)/V + 2(1-V)\vartheta(\tau)/V \right) \boldsymbol{X}_{ i} \boldsymbol{X}_{ i}^\top + \varsigma(\tau)/V^2 \sum_{\nu=1}^V \boldsymbol{X}_{\nu i} \boldsymbol{X}_{\nu i}^\top \right\} \notag \\
&= \frac{1}{V} \left\{ \left\{ \varrho(\tau)V + (V-2) \varsigma(\tau) + 2(1-V) \vartheta(\tau) \right \} \frac{1}{n} \sum_{i=1}^n   \boldsymbol{X}_{ i} \boldsymbol{X}_{ i}^\top + \frac{\varsigma(\tau)}{nV} \sum_{i=1}^{n} \sum_{\nu=1}^V \boldsymbol{X}_{\nu i} \boldsymbol{X}_{ \nu i}^\top  \right\}  \notag \\
&\overset{}{\rightarrow} \upsilon(\tau) \boldsymbol{\Sigma}/V, \textit{as } n\to \infty,\label{sum matrices}
\end{align}
\endgroup
equation \eqref{sum matrices} follows from \mbox{Assumption~\ref{assumption 2}}-\ref{assumption 2-i}), i.e.,
 $\frac{1}{n}\sum_{i=1}^n   \boldsymbol{X}_{ i} \boldsymbol{X}_{ i}^\top\to\boldsymbol{\Sigma}$ and\linebreak $  \frac{1}{nV}\sum_{i=1}^{n}\sum_{\nu=1}^{V} \boldsymbol{X}_{\nu i}\boldsymbol{X}^\top_{\nu i}\to \boldsymbol \Sigma,  $
with $\upsilon(\tau) =\left\{ \varrho(\tau)V + (V-1) \varsigma(\tau) + 2(1-V) \vartheta(\tau) \right \}$.

\mbox{Assumption~\ref{assumption 1}} implies 
\(\mathbb{E}\!\big(\ell'_{\tau,k}(\varepsilon_i)\mid \boldsymbol X_i\big)
=\mathbb{E}\!\big(\ell'_{\tau,k}(\varepsilon^0_i)\mid \boldsymbol X_i\big)=0\),
hence \(\mathbb{E}(\boldsymbol H_i)=\boldsymbol 0\).
To apply the Lindeberg--Feller central limit theorem for vector triangular arrays,
it suffices to verify the vector Lindeberg condition: for every \(\xi>0\),

\[
\frac{1}{n}\sum_{i=1}^n
\mathbb{E}\!\left[\,
\|\boldsymbol H_i\|^{2}\,
\mathds{1}\!\left\{\|\boldsymbol H_i\|>\xi\sqrt{n}\right\}
\right]
\ \longrightarrow\ 0.
\]
Using \( z^{2}\mathds{1}\{z>a\}\le z^{4}/a^{2} \) with \( z=\|\boldsymbol H_i\| \) and \( a=\xi\sqrt{n} \),
\begin{align}\label{LIND}
\frac{1}{n}\sum_{i=1}^n
\mathbb{E}\!\left[
\|\boldsymbol H_i\|^{2}\,\mathds{1}\{\|\boldsymbol H_i\|>\xi\sqrt{n}\}
\right]
&\;\le\;
\frac{1}{\xi^{2} n^{2}}\sum_{i=1}^n \mathbb{E}(\,\|\boldsymbol H_i\|^{4}).
\end{align}

We now bound the fourth moments. Since \(1<k\le 2\),
\[
\ell'_{\tau,k}(s)=k\,\Psi_\tau(s)\,|s|^{k-1}\mathrm{sgn}(s),
\quad 0\le \Psi_\tau\le 1,
\quad\Rightarrow\quad
\ell'_{\tau,k}(s)^{4}\le k^{4}\,|s|^{4(k-1)}.
\]
Because \(0\le 4(k-1)\le 2k\) and \(\mathbb{E}(|\varepsilon_i|^{2k})<\infty\), there is \(C_\xi<\infty\) with
\[
 \mathbb{E}\,[\ell'_{\tau,k}(\varepsilon_i)]^{4}\le C_\xi,
\qquad
 \mathbb{E}\,[\ell'_{\tau,k}(\varepsilon^0_i)]^{4}\le C_\xi.
\]
Hence, using determinism of the design,
\[
\mathbb{E}(\|\boldsymbol b_i\|^{4})
=\|\boldsymbol X_i\|^{4}\,\mathbb{E}[\,\ell'_{\tau,k}(\varepsilon^0_i)^{4}]
\le C_\xi \|\boldsymbol X_i\|^{4},
\quad
\mathbb{E}(\|\boldsymbol d_i\|^{4})
\le C_\xi \|\boldsymbol X_i\|^{4},
\]
and, by \((\sum_{l=1}^{m} a_l)^r\le m^{r-1}\sum_{l=1}^{m} a_l^r;\,m>0, {r\ge1}, a_l>0\). One has
\[
\mathbb{E}(\|\boldsymbol e_i\|^{4})
\ \le\ V^{3}C_\xi \sum_{\nu=1}^{V}\|\boldsymbol X_{\nu i}\|^{4}.
\]
Finally, with \((a+b+c)^{4}\le 3^{3}(a^{4}+b^{4}+c^{4})\),
\[
\mathbb{E}(\|\boldsymbol H_i\|^{4})
\ \le\ C\,C_\xi\left(
2\|\boldsymbol X_i\|^{4}
+\frac{V^3}{V^4}\sum_{\nu=1}^{V}\|\boldsymbol X_{\nu i}\|^{4}
\right),
\quad C:=3^3>0.
\]
Based on \mbox{Assumption~\ref{assumption 2}}-\ref{assumption 2-ii}), we have 
\[
\frac{1}{n}\sum_{i=1}^{n}\|\boldsymbol X_i\|^{4}=O(1)
\quad\text{and}\quad
\frac{1}{nV}\sum_{i=1}^{n}\sum_{\nu=1}^{V}\|\boldsymbol X_{\nu i}\|^{4}=O(1),
\]
then
\[
\frac{1}{n^{2}}\sum_{i=1}^{n}\mathbb{E}(\|\boldsymbol H_i\|^{4})
\ =\ O\!\left(\frac{1}{n}\right).
\]
Together with  \eqref{LIND} implies the Lindeberg condition.
Therefore, the Lindeberg--Feller CLT applies to the vector sum \( \boldsymbol\Delta_n \).

Hence,
\[
\boldsymbol\Delta_n\ \xrightarrow{\ \mathcal{D}\ }\ \mathcal{N}\!\big(\boldsymbol 0,\ \upsilon(\tau)\,\boldsymbol\Sigma/V\big),
\qquad n\to\infty.
\]

This completes the proof.

\begin{lm} \label{lemma 5}
Under  $\mbox{Assumptions~\ref{assumption 1} and \ref{assumption 2}}$ {and \mbox{Assumption~\ref{assumption 3}}}, with probability tending to $1$, for any given \(\boldsymbol{\beta}_1\) that satisfies \(\|\boldsymbol{\beta}_1 - \boldsymbol{\beta}^\star_{1}\|_2 = O_p (n^{-1/2})\) and any positive constant \(\mathcal{C}\), we obtain
\[
{\boldsymbol{0}_{p-m}} = \arg \min_{\|\boldsymbol{\beta}_2\|_2 \leq \mathcal{C} n^{-1/2}} \mathcal{R}^{\text{(S)}} \left( (\boldsymbol{\beta}_1^\top, \boldsymbol{\beta}_2^\top)^\top \right).
\]   
\end{lm}

\subsubsection*{Proof of  $\mbox{Lemma~\ref{lemma 5}}$}
For any \( \boldsymbol{\beta}_1 \) and \( \boldsymbol{\beta}_2 \) that satisfy 
\[
\|\boldsymbol{\beta}_1 - \boldsymbol{\beta}^\star_{1}\|_2 = O_p (n^{-1/2}), \quad \|\boldsymbol{\beta}_2\|_2 \leq \mathcal{C} n^{-1/2},
\]
and by \mbox{Lemma~\ref{lemma 3}}, one has

\begingroup\footnotesize
\begin{align}\label{a14}
&n \left\{ \mathcal{R}^{\text{(S)}} \left( (\boldsymbol{\beta}_1^\top, \boldsymbol{0}^\top)^\top \right) - \mathcal{R}^{\text{(S)}} \left( (\boldsymbol{\beta}_1^\top, \boldsymbol{\beta}_2^\top)^\top \right) \right\}\notag \\
&= n \left\{ {\mathcal{R}^{\text{(S)}}} \left( (\boldsymbol{\beta}_1^\top, \boldsymbol{0}^\top)^\top \right) - \mathcal{R}_\nu (\boldsymbol{\beta}^\star_{1}) \right\} - n \left\{ \mathcal{R}^{\text{(S)}} \left( (\boldsymbol{\beta}_1^\top, \boldsymbol{\beta}_2^\top)^\top \right) - \mathcal{R}_\nu (\boldsymbol{\beta}^\star_{1}) \right\} \notag\\
&= T_n \left( \sqrt{n} \left[ ((\boldsymbol{\beta}_1 - \boldsymbol{\beta}^\star_{1})^\top, \boldsymbol{0}^\top)^\top\right] \right) - T_n \left( \sqrt{n} \left[ ((\boldsymbol{\beta}_1 - \boldsymbol{\beta}^\star_{1})^\top, \boldsymbol{\beta}_2^\top)^\top \right] \right) - n \sum_{l=m+1}^p \gamma_{\lambda} (|\beta_l|) \notag\\
&= \frac{\varkappa (\tau)}{2} \sqrt{n} ((\boldsymbol{\beta}_1 - \boldsymbol{\beta}^\star_{1})^\top, \boldsymbol{0}^\top)^\top \left( \frac{1}{n} \sum_{i=1}^n \boldsymbol{X}_{ i} \boldsymbol{X}_{ i}^\top \right) \sqrt{n} ((\boldsymbol{\beta}_1 - \boldsymbol{\beta}^\star_{1})^\top, \boldsymbol{0}^\top)^\top + \sqrt{n} ((\boldsymbol{\beta}_1 - \boldsymbol{\beta}^\star_{1})^\top, \boldsymbol{0}^\top)^\top \boldsymbol{\Delta}_n \notag\\
&\quad - \frac{\varkappa (\tau)}{2} \sqrt{n} ((\boldsymbol{\beta}_1 - \boldsymbol{\beta}^\star_{1})^\top, \boldsymbol{\beta}_2^\top)^\top \left( \frac{1}{n} \sum_{i=1}^n \boldsymbol{X}_{ i} \boldsymbol{X}_{ i}^\top \right) \sqrt{n} ((\boldsymbol{\beta}_1 - \boldsymbol{\beta}^\star_{1})^\top, \boldsymbol{\beta}_2^\top)^\top - \sqrt{n} ((\boldsymbol{\beta}_1 - \boldsymbol{\beta}^\star_{1})^\top, \boldsymbol{\beta}_2^\top)^\top\boldsymbol{\Delta}_n \notag\\
&\quad - n \sum_{l=m+1}^p \gamma_{\lambda} (|\beta_l|) + o_p (1).
\end{align}
\endgroup

The conditions \(\|\boldsymbol{\beta}_1 - \boldsymbol{\beta}^\star_{1}\|_2 = O_p (n^{-1/2})\) and \(\|\boldsymbol{\beta}_2\|_2 \leq \mathcal{C} n^{-1/2}\) imply that
\[
\begin{aligned}
&\frac{\varkappa (\tau)}{2} \sqrt{n} ((\boldsymbol{\beta}_1 - \boldsymbol{\beta}^\star_{1})^\top, \boldsymbol{0}^\top)^\top \left( \frac{1}{n} \sum_{i=1}^n \boldsymbol{X}_{ i} \boldsymbol{X}_{ i}^\top \right) \sqrt{n} ((\boldsymbol{\beta}_1 - \boldsymbol{\beta}^\star_{1})^\top, \boldsymbol{0}^\top)^\top = O_p (1), \\
&\frac{\varkappa (\tau)}{2} \sqrt{n} ((\boldsymbol{\beta}_1 - \boldsymbol{\beta}^\star_{1})^\top, \boldsymbol{\beta}_2^\top)^\top \left( \frac{1}{n} \sum_{i=1}^n \boldsymbol{X}_{ i} \boldsymbol{X}_{ i}^\top \right) \sqrt{n} ((\boldsymbol{\beta}_1 - \boldsymbol{\beta}^\star_{1})^\top, \boldsymbol{\beta}_2^\top)^\top = O_p (1),
\end{aligned}
\]
and
\begin{align*}
\sqrt{n} ((\boldsymbol{\beta}_1 - \boldsymbol{\beta}^\star_{1})^\top, \boldsymbol{0}^\top)^\top \boldsymbol{\Delta}_n - \sqrt{n} ((\boldsymbol{\beta}_1 - \boldsymbol{\beta}^\star_{1})^\top, \boldsymbol{\beta}_2^\top)^\top \boldsymbol{\Delta}_n &= -\sqrt{n} (\boldsymbol{0}^\top, \boldsymbol{\beta}_2^\top)^\top \boldsymbol{\Delta}_n \\
&={O_p\left(\sqrt n\,\|\boldsymbol\beta_2\|_2\right)} \\
&= O_p (1),  
\end{align*}
{The stochastic bound follows from $\|\boldsymbol{\Delta}_n\|_2=O_p(1)$ and Cauchy--Schwarz. The corresponding quadratic-form bound is deterministic uniformly over $\|\beta_2\|_2\le\mathcal C/\sqrt n$.}
since 
\begin{align}\label{ineq}
   \left| n  \varsigma(\tau) \boldsymbol{\beta}_2^{\top}  \boldsymbol{\Sigma}_{22}  \boldsymbol{\beta}_2/V \right| &\leq (p-m)  \varsigma(\tau) \left\| \boldsymbol{\Sigma}_{22}  \right\|_\infty n\|\boldsymbol{\beta}_2 \|^2_2/V \notag\\
   &={O(1)},  
\end{align}
where $\boldsymbol{\Sigma}_{22}$ is the right lower $(p-m)$ by $(p-m)$ submatrix
of $\boldsymbol{\Sigma}$ and  $\|\boldsymbol{\Sigma}_{22}\|_{\infty}$ denotes the matrix $\ell_\infty$-norm, i.e., the maximum absolute row sum, $\|\boldsymbol A\|_\infty:=\max_{1\le i\le d}\sum_{l=1}^d |a_{il}|$ for a $(d\!\times\! d)$-matrix.\\
To justify the inequality \eqref{ineq}, write, by the triangle inequality,
\[
|\boldsymbol{\beta}_2^\top \boldsymbol{\Sigma}_{22}\boldsymbol{\beta}_2|
= \Big|\sum_{i=1}^{p-m}\sum_{l=1}^{p-m} \beta_{2,i}(\boldsymbol{\Sigma}_{22})_{il}\beta_{2,l}\Big|
\;\le\; \sum_{i=1}^{p-m} |\beta_{2,i}|\sum_{l=1}^{p-m}|(\boldsymbol{\Sigma}_{22})_{il}|\,|\beta_{2,l}|.
\]
For each $i$, $\sum_{l}|(\boldsymbol{\Sigma}_{22})_{il}|\le \|\boldsymbol{\Sigma}_{22}\|_\infty$, hence
\[
|\boldsymbol{\beta}_2^\top \boldsymbol{\Sigma}_{22}\boldsymbol{\beta}_2|
\;\le\; \|\boldsymbol{\Sigma}_{22}\|_\infty\, \sum_{i=1}^{p-m} |\beta_{2,i}| \,\max_{1\le j\le p-m} |\beta_{2,{j}}|
\;=\; \|\boldsymbol{\Sigma}_{22}\|_\infty\, \|\boldsymbol{\beta}_2\|_1\,\|\boldsymbol{\beta}_2\|_\infty.
\]
Next, use the standard norm relations (for any $d$-vector $\boldsymbol{v}$):
$\|\boldsymbol{v}\|_1 \le d\,\|\boldsymbol{v}\|_\infty$ and $\|\boldsymbol{v}\|_\infty \le \|\boldsymbol{v}\|_2$.
With $d=p-m$ and $\boldsymbol{v}=\boldsymbol{\beta}_2$ this yields
\[
\|\boldsymbol{\beta}_2\|_1\,\|\boldsymbol{\beta}_2\|_\infty
\;\le\; (p-m)\,\|\boldsymbol{\beta}_2\|_\infty^2
\;\le\; (p-m)\,\|\boldsymbol{\beta}_2\|_2^2.
\]
Combining the displays gives
\[
|\boldsymbol{\beta}_2^\top \boldsymbol{\Sigma}_{22}\boldsymbol{\beta}_2|
\;\le\; (p-m)\,\|\boldsymbol{\Sigma}_{22}\|_\infty\,\|\boldsymbol{\beta}_2\|_2^2.
\]
\noindent
Given that  $ \liminf\limits_{\lambda \to 0} \liminf\limits_{\xi \to 0^+} [\gamma_{\lambda}'(\xi)/\lambda ]= 1$ \citep{fan2001}, {for all sufficiently large $n$, $|\beta_l|\le\mathcal C/\sqrt n<\lambda_n$ on the inactive-coordinate ball, so the SCAD contribution is exactly}
\begin{align*}
  n \sum_{l=m+1}^p \gamma_{\lambda} (|{\beta}_l|)
  &={n \sum_{l=m+1}^{p}\lambda_n |\beta_l|}\\
  &={n\lambda_n\sum_{l=m+1}^{p}|\beta_l|}\\
  &={\sqrt n\lambda_n\|\boldsymbol{v}\|_1,\qquad \boldsymbol{v}=\sqrt n\boldsymbol{\beta}_2}.
\end{align*}
{To compare the orders uniformly, put $\boldsymbol{u}=\sqrt n(\boldsymbol{\beta}_1-\boldsymbol{\beta}_1^\star)$ and $\boldsymbol{v}=\sqrt n\boldsymbol{\beta}_2$. Lemmas~\ref*{lemma 3} and~\ref*{lemma 1}, applied to the convex function $T_n(\boldsymbol{\delta})-\boldsymbol{\Delta}_n^\top\boldsymbol{\delta}$, give the quadratic expansion uniformly on compact sets. The oscillation of $T_n$ on a larger compact ball bounds its subgradients on a smaller ball; since $\boldsymbol{\Delta}_n=O_p(1)$, this gives a random constant $L_n=O_p(1)$ such that
\[
|T_n((\boldsymbol{u}^\top,\boldsymbol{v}^\top)^\top)-T_n((\boldsymbol{u}^\top,\boldsymbol{0}^\top)^\top)|\le L_n\|\boldsymbol{v}\|_1.
\]
The tightness of $\boldsymbol{u}$ permits restriction to such a ball with arbitrarily high probability. Consequently, the objective difference in (\ref*{a14}) is at most $(L_n-\sqrt n\lambda_n)\|\boldsymbol{v}\|_1$, which is strictly negative for every $\boldsymbol{v}\ne\boldsymbol{0}$ with probability tending to one.}
Specifically, we have
\[
\mathcal{R}^{\text{(S)}} \left( (\boldsymbol{\beta}_1^\top, \boldsymbol{0}^\top)^\top \right) - \mathcal{R}^{\text{(S)}} \left( (\boldsymbol{\beta}_1^\top, \boldsymbol{\beta}_2^\top)^\top \right) < 0
\]
{for every nonzero $\boldsymbol{\beta}_2$ in the stated ball, with probability tending to one. At $\boldsymbol{\beta}_2=\boldsymbol{0}$ the difference is exactly zero. Consequently,}
\[
\inf_{\|\boldsymbol{\beta}_2\|_2 \leq \mathcal{C} n^{-1/2}} \mathcal{R}^{\text{(S)}} \left( (\boldsymbol{\beta}_1^\top, \boldsymbol{\beta}_2^\top)^\top \right) {=} \mathcal{R}^{\text{(S)}} \left( (\boldsymbol{\beta}_1^\top, \boldsymbol{0}^\top)^\top \right),
\]
with the probability tending to one. This concludes the proof.

\section*{Proof of  \mbox{Theorem~\ref{theorem scad}}  }

\begin{enumerate}[i)]
    \item Following the approach of \cite{wu2009variable}, in order to establish ~\mbox{Theorem~\ref{theorem scad}}--\ref{consistency scad}), it is sufficient to verify that for any fixed \( \alpha > 0 \), there exists a sufficiently large constant \( \mathcal{C} > 0 \) such that
\[
\mathbb{P} \left[ \inf_{\|\boldsymbol{\delta}\|_2 = \mathcal{C}} 
\mathcal{R}^{\text{(S)}} \!\left( \boldsymbol{\beta}^\star + \frac{\boldsymbol{\delta}}{\sqrt{n}} \right) 
> \mathcal{R}^{\text{(S)}} (\boldsymbol{\beta}^\star) \right] \geq 1 - \alpha.
\]
This condition means that, with probability at least \( 1 - \alpha \), the empirical risk function 
\( \mathcal{R}^{\text{(S)}}(\cdot) \) is strictly larger than its value at the true parameter 
\( \boldsymbol{\beta}^\star \) on the boundary of the ball
\[
\partial \mathbb{B} = \Bigl\{ \boldsymbol{\beta}^\star + \tfrac{\boldsymbol{\delta}}{\sqrt{n}} : \|\boldsymbol{\delta}\|_2 = \mathcal{C} \Bigr\},
\]
where the ball itself is
\[
\mathbb{B} = \Bigl\{ \boldsymbol{\beta}^\star + \tfrac{\boldsymbol{\delta}}{\sqrt{n}} : \|\boldsymbol{\delta}\|_2 \leq \mathcal{C} \Bigr\}.
\]
By the continuity of \( \mathcal{R}^{\text{(S)}} \) and the extreme value theorem, the function attains a minimum over the compact set \( \mathbb{B} \). Since all boundary points have strictly larger values than the center point, the minimum cannot occur on \(\partial \mathbb{B}\); hence, a local minimizer must lie strictly inside \( \mathbb{B} \) with probability at least \(1 - \alpha\).
This argument shows that \( \boldsymbol{\beta}^\star \) is contained in a ``basin of attraction'' where the risk function increases outward, which guarantees the existence of a consistent local solution \citep{basin1989strogatz}.
This consequently suggests the existence of a local minimizer for which \( \|\boldsymbol{\hat \beta}^{\text{(S)}} - \boldsymbol{\beta}^\star\|_2 = O_p(n^{-1/2}) \). By carrying out a straightforward calculation and applying \mbox{Lemma~\ref{lemma 3}}, one obtains 
\begingroup\footnotesize
\begin{align}
n \left\{ \mathcal{R}^{\text{(S)}} (\boldsymbol{\beta}^\star + \boldsymbol{\delta} / \sqrt{n}) - \mathcal{R}^{\text{(S)}} (\boldsymbol{\beta}^\star) \right\} &= n \left\{ \mathcal{R}_\nu (\boldsymbol{\beta}^\star + \boldsymbol{\delta} / \sqrt{n}) - \mathcal{R}_\nu (\boldsymbol{\beta}^\star) \right\} + n   [{\nabla}\mathcal{R}_{N,\nu} (\boldsymbol{ \beta}^0)]^\top \boldsymbol{\delta} / \sqrt{n} \notag\\
&+ n \left\{ \sum_{l=1}^p \gamma_{\lambda} (|{\beta}^\star_{l} + \delta_l / \sqrt{n}|) - \sum_{l=1}^p \gamma_{\lambda} (|{\beta}^\star_{l}|) \right\} \notag\\
&= \sum_{i=1}^n \left[ \ell_{\tau,k} \left( \varepsilon_{ i} - \boldsymbol{X}_{ i}^\top \boldsymbol{\delta}/\sqrt{n} \right) - \ell_{\tau,k} (\varepsilon_{ i}) \right] + n [\nabla \mathcal{R}_{N,\nu} ( \boldsymbol{ \beta}^0)]^\top \boldsymbol{\delta} / \sqrt{n} \notag \\
&+ n \left[ \sum_{l=1}^p \gamma_{\lambda} (| \beta^\star_{l} + \delta_l / \sqrt{n}|) - \sum_{l=1}^p \gamma_{\lambda} (|{\beta}^\star_{l}|) \right] \notag\\
&\geq T_n (\boldsymbol{\delta}) + S_n (\boldsymbol{\delta}),\label{eq theorem}
\end{align}
\endgroup
where   \eqref{eq theorem} is such that
\[
S_n (\boldsymbol{\delta}) = n \sum_{l=1}^m \left[ \gamma_{\lambda} (| \beta^\star_{l} + \delta_l / \sqrt{n}|) - \gamma_{\lambda} (|{\beta}^\star_{l}|) \right].
\]

Since
\[
\gamma'_{\lambda}(s)
= \lambda\,\mathds{1}\{s\le \lambda\}
+ \frac{(a\lambda-s)_+}{{a-1}}\,\mathds{1}\{s>\lambda\} \;\ge\; 0,
\qquad a>2,\ s\ge 0,\qquad \gamma_{\lambda}(0)=0,
\]
it follows that, for the inactive coordinates \(l=m+1,\ldots,p\),
\[
n\!\left(\gamma_{\lambda}\!\bigl(|\beta^\star_{l}+{\delta_{l}}/{\sqrt{n}}|\bigr)
      -\gamma_{\lambda}\!\bigl(|\beta^\star_{l}|\bigr)\right)
= n\!\left(\gamma_{\lambda}\!\bigl(|{\delta_{l}}/{\sqrt{n}}|\bigr)-\gamma_{\lambda}(0)\right)
\;\ge\; 0 .
\]

By \mbox{Lemma~\ref{lemma 4}} , we have \(\boldsymbol{\Delta}_n \xrightarrow{\mathcal{D}} \mathcal N \left(\boldsymbol 0,  \upsilon (\tau) \boldsymbol{\Sigma}/V\right)\). It suffices to  prove that \(\boldsymbol{\Delta}_n^\top \boldsymbol{\delta}\) is bounded in probability, i.e.,
\begin{align}
\boldsymbol{\Delta}_n^\top \boldsymbol{\delta} = O_p \left(  [\upsilon (\tau) \boldsymbol{\delta}^\top \boldsymbol{\Sigma} \boldsymbol{\delta}/V]^{1/2} \right).\label{delta delta}
\end{align}
Since \( \boldsymbol{\Delta}_n^\top \boldsymbol{\delta}  \xrightarrow{\mathcal{D}} \mathcal{N} \left( \boldsymbol{0}, \upsilon (\tau) \boldsymbol{\delta}^\top \boldsymbol{\Sigma} \boldsymbol{\delta}/V \right) \), and for a Gaussian random variable \( Z \sim \mathcal{N}(0, \sigma^2) \), the tail probability satisfies
\[
\mathbb{P}(|Z| > t) \leq 2 e^{-\frac{t^2}{2\sigma^2}}.
\]
Then, we have the following {asymptotic} bound on the probability
\[
{\limsup_{n\to\infty}}\mathbb{P}\left( |\boldsymbol{\Delta}_n^\top \boldsymbol{\delta}| > M \cdot \left[ \mathbb{V}\mathrm{ar}(\boldsymbol{\Delta}_n^\top \boldsymbol{\delta}) \right]^{1/2} \right) {\le} 2 e^{-\frac{M^2}{2}}.
\]
For any \( \epsilon > 0 \), choose \( M \) sufficiently large such that \( 2 e^{-\frac{M^2}{2}} < \epsilon \). This proves the result in \eqref{delta delta}. {Uniformly over $\|\boldsymbol{\delta}\|_2=\mathcal C$, $|\boldsymbol{\Delta}_n^\top\boldsymbol{\delta}|\le\mathcal C\|\boldsymbol{\Delta}_n\|_2=O_p(\mathcal C)$, whereas the positive-definite quadratic term grows as $\mathcal C^2$.}
By using  \mbox{Lemma~\ref{lemma 1}} (in  Appendix~\ref{appendix B}) on \( \Upsilon_n (\boldsymbol{\delta}) := T_n (\boldsymbol{\delta}) - \boldsymbol{\Delta}_n^\top \boldsymbol{\delta} \), we can improve the point-wise convergence to achieve uniform convergence on compact subsets of \( \mathbb{R}^p \).

\noindent Following~\cite{liao2019penalized}, one has, for large \(n\),
\[
n \sum_{l=1}^m \Bigl[
\gamma_{\lambda}\!\Bigl(\Bigl|\beta^\star_l + \frac{\delta_l}{\sqrt{n}}\Bigr|\Bigr)
- \gamma_{\lambda}\!\bigl(|\beta^\star_l|\bigr)
\Bigr] \;=\; 0
\quad\text{uniformly for }\boldsymbol{\delta}\ \text{in any compact subset of }\mathbb{R}^p.
\]
Assume \(a>2\), \(\lambda\to 0\), and that the active coordinates
\(\{\beta^\star_l\}_{l=1}^m\) are all nonzero with \(\min_{l\le m}|\beta^\star_l|>0\).
Then, for all sufficiently large \(n\), we have \(|\beta^\star_l|\ge a\lambda\) for every
\(l\le m\). Moreover, for \(\boldsymbol{\delta}\) in a compact set $\mathbb{ U}$,
\(\max_l|\delta_l|<\infty\), hence \(\bigl|\beta^\star_l+\delta_l/\sqrt{n}\bigr|\ge a\lambda\) also
holds for all large \(n\). Since the SCAD penalty is flat beyond \(a\lambda\),
i.e. \(\gamma'_{\lambda}(s)=0\) for \(s\ge a\lambda\), it follows that
\[
\gamma_{\lambda}\!\Bigl(\Bigl|\beta^\star_l + \frac{\delta_l}{\sqrt{n}}\Bigr|\Bigr)
=\gamma_{\lambda}\!\bigl(|\beta^\star_l|\bigr)
\quad\text{for all large }n,\ \text{uniformly in }\boldsymbol{\delta}\in \mathbb{ U},
\]
which establishes the desired equality.\\
For the second term in \eqref{eq theorem}, a Taylor expansion at \(|\beta^\star_l|\), \(l=1,\ldots,m\), leads to
\begingroup\small
\begin{align*}
n\!\left(\gamma_{\lambda}\!\bigl(|\beta^\star_{l}+{\delta_{l}}/{\sqrt{n}}|\bigr)
         -\gamma_{\lambda}\!\bigl(|\beta^\star_{l}|\bigr)\right)
&= n\!\left(
\gamma'_{\lambda}\!\bigl(|\beta^\star_{l}|\bigr)\operatorname{sgn}(\beta^\star_{l})\,\frac{\delta_{l}}{\sqrt{n}}
+ \frac{\gamma''_{\lambda}\!\bigl(|\beta^\star_{l}|\bigr)}{2}\left(\frac{\delta_{l}}{\sqrt{n}}\right)^{\!2}
+ {o(n^{-1})}
\right)\\[2mm]
&= {0}.
\end{align*}
\endgroup
{The remainder is in fact zero for all sufficiently large $n$, uniformly on each fixed compact set.}
Since, for SCAD,
\begingroup\small
\begin{align*}
\gamma'_{\lambda}\!\bigl(|\beta^\star_{l}|\bigr)
&= \lambda\!\left(
\mathds{1}\!\bigl\{|\beta^\star_{l}|\le \lambda\bigr\}
+ \frac{(a\lambda-|\beta^\star_{l}|)_+}{(a-1)\lambda}\,
  \mathds{1}\!\bigl\{|\beta^\star_{l}|>\lambda\bigr\}
\right)
= \frac{(a\lambda-|\beta^\star_{l}|)_+}{a-1}
\;\longrightarrow\; 0, \textit{as}\, \lambda\to 0
\\[4pt]
\gamma''_{\lambda}\!\bigl(|\beta^\star_{l}|\bigr)
&= -\frac{1}{a-1}\,
\mathds{1}\!\bigl\{\lambda<|\beta^\star_{l}|<a\lambda\bigr\}
\;\longrightarrow\; 0, \textit{as}\, \lambda\to 0,
\end{align*}
\endgroup
the Taylor contribution vanishes; together with the flatness argument above, this
shows that the active--set contribution is exactly \(0\) uniformly on compacts.

By  \mbox{Assumption~\ref{assumption 1}} and the equations derived earlier, the expression \[ n \left[ \mathcal{R}^{\text{(S)}} (\boldsymbol{\beta}^\star + \boldsymbol{\delta} / \sqrt{n}) - \mathcal{R}^{\text{(S)}} (\boldsymbol{\beta}^\star) \right] \] is primarily determined by the term \( \varkappa (\tau) \boldsymbol{\delta}^\top \boldsymbol{\Sigma} \boldsymbol{\delta}/2 \) when \( \|\boldsymbol{\delta}\|_2 \) is sufficiently large, specifically for some positive constant \( \mathcal{C} \). As a result, we obtain the following probability bound:
\[
\mathbb{P} \left[ \inf_{\|\boldsymbol{\delta}\|_2 = \mathcal{C}} \mathcal{R}^{\text{(S)}} (\boldsymbol{\beta}^\star + \boldsymbol{\delta} / \sqrt{n}) > \mathcal{R}^{\text{(S)}} (\boldsymbol{\beta}^\star) \right] \geq 1 - \alpha.
\]
This implies that \( \|\boldsymbol{\hat{\beta}}^{\text{(S)}} - \boldsymbol{\beta}^\star\|_2 = O_p(n^{-1/2}) \) as \( n \to \infty \). Therefore, we conclude that
\[
\|\boldsymbol{\hat{\beta}}^{\text{(S)}} - \boldsymbol{\beta}^\star\|_2 = O_p(N^{-1/2}),
\]
which completes the proof.

\item 
    As demonstrated in \cite{fan2001} and based on \mbox{Lemma~\ref{lemma 5}}, with \( N = nV \), it is straightforward to show that under \mbox{Assumption~\ref{assumption 3}}, \( \boldsymbol{\hat{\beta}}^{\text{(S)}}_{2} = \boldsymbol{0} \).

   \item  
   Next, we establish the asymptotic normality of \( \boldsymbol{\hat{\beta}}^{\text{(S)}}_{1} \). From \mbox{Theorem~\ref{theorem scad}}-\ref{consistency scad}), we know that there exists a \( \sqrt{N} \)-consistent minimizer \( \boldsymbol{\beta}^{\text{(S)}}_{1} \) for \( \mathcal{R}^{\text{(S)}}((\boldsymbol{\beta}_1^\top, \boldsymbol{0}^\top)^\top) \), which is a function of \( \boldsymbol{\beta}_1 \). The proof of \mbox{Theorem~\ref{theorem scad}}-\ref{consistency scad}) further implies that \( \sqrt{n}(\boldsymbol{\hat{\beta}}^{\text{(S)}}_{1} - \boldsymbol{\beta}_1^\star) \) minimizes
\begin{equation}\label{a17}
T_n \left( \begin{pmatrix}
\boldsymbol{\zeta}^\top, \\
\boldsymbol{0}^\top
\end{pmatrix}^\top \right) + n \sum_{l=1}^m \gamma_{\lambda}\left( \left| \beta^\star_{l} + \zeta_l/\sqrt{n} \right| \right)
\end{equation}
with respect to \( \boldsymbol{\zeta} \), where \( \boldsymbol{\zeta} = (\zeta_1, \zeta_2, \dots, \zeta_m)^\top \in \mathbb{R}^m \). Using the convexity of \( T_n(\cdot) \) and applying   \mbox{Lemma~\ref{lemma 3}} and  \mbox{Lemma~\ref{lemma 1}}, we derive

    \begin{equation}\label{a18}
{T_n((\boldsymbol\zeta^\top,\boldsymbol0^\top)^\top)
=\frac{\varkappa(\tau)}2\boldsymbol\zeta^\top
\left[\frac1n\sum_{i=1}^n\boldsymbol X_i^{(1)}\boldsymbol X_i^{(1)\top}\right]\boldsymbol\zeta
+(\boldsymbol\Delta_n^1)^\top\boldsymbol\zeta+o_p(1)}
    \end{equation}

   Uniformly in any compact subset of \( \mathbb{R}^m \), let \( \boldsymbol{\Delta}^1_n = \frac{1}{\sqrt{n}} \sum_{i=1}^n \boldsymbol{H}^1_i \), where \( \boldsymbol{H}^1_i = \boldsymbol{b}^1_i - \frac{\boldsymbol{e}^1_i}{V} - \boldsymbol{d}^1_i \), and
\[
\boldsymbol{b}^1_i = \ell'_{\tau,k}(\varepsilon^0_{i}) \boldsymbol{X}^{(1)}_{i}, \quad \boldsymbol{e}^1_i = \sum_{\nu=1}^V \ell'_{\tau,k}(\varepsilon^0_{\nu i}) \boldsymbol{X}^{(1)}_{\nu i}, \quad \boldsymbol{d}^1_i = \ell'_{\tau,k}(\varepsilon_{i}) \boldsymbol{X}^{(1)}_{i}.
\]
By a similar argument to the proof of \mbox{Lemma~\ref{lemma 4}}, we obtain the following convergence in distribution
\begin{equation}\label{a19}
\boldsymbol{\Delta}^1_n \xrightarrow{\mathcal{D}} \mathcal{N}\left(\boldsymbol{0}, \upsilon(\tau) \boldsymbol{\Sigma}_{11}/V\right).
\end{equation}

    For large \( n \), and under the condition that \( \lambda = \lambda(n) \to 0 \), it follows that
\begin{equation}
n \sum_{l=1}^m \gamma_{\lambda}\left( |\beta^\star_{l} + \zeta_l/\sqrt{n}| \right) = n \sum_{l=1}^m \gamma_{\lambda}(|\beta^\star_{l}|),
\label{a20}
\end{equation}
uniformly over any compact subset of \( \mathbb{R}^m \). Moreover, this expression is independent of the parameter \( \boldsymbol{\zeta} \).
Based on equations \eqref{a17}, \eqref{a18}, \eqref{a20}, and \mbox{Assumption~\ref{assumption 2}}, we derive
\begingroup\small
    \begin{equation}
   T_n (\begin{pmatrix}
    \boldsymbol{\zeta}^\top,
    \boldsymbol{0}^\top
    \end{pmatrix}^\top) + n\sum_{l=1}^m \gamma_{\lambda}\left(|\beta^\star_{l} + \zeta_l/\sqrt{n}|\right) = \frac{\varkappa(\tau)}{2}\boldsymbol{\zeta}^\top \boldsymbol{\Sigma}_{11}\boldsymbol{\zeta} + {\boldsymbol{\Delta}_{n}^{1\top}}\boldsymbol{\zeta} + n\sum_{l=1}^m \gamma_{\lambda}(|\beta^\star_{l}|) {+o_p(1)}.
    \label{eq:A20}
    \end{equation}
\endgroup
Combining \mbox{Lemma~\ref{lemma 2}} (in Appendix~\ref{appendix B}) with \eqref{a19} and \eqref{eq:A20} gives   
\begin{equation}
    \sqrt{n}(\boldsymbol{ \hat{\beta}}^{\text{(S)}}_1 - \boldsymbol{\beta}^\star_{1}) \xrightarrow{\mathcal{D}} \mathcal{N}\left(\boldsymbol{0}, \frac{\upsilon(\tau)}{V\varkappa^2(\tau)} \boldsymbol{\Sigma}_{11}^{-1}\right).
    \label{normal s}
    \end{equation}
    Since \(N = nV\), and using equation \eqref{normal s} along with Slutsky's theorem, it follows that
    \begin{equation*}
    \sqrt{N}(\boldsymbol{\hat{\beta}}^{\text{(S)}}_1- \boldsymbol{\beta}^\star_{1}) \xrightarrow{\mathcal{D}} \mathcal{N}\left(\boldsymbol{0}, \frac{\upsilon(\tau)}{\varkappa^2(\tau)} \boldsymbol{\Sigma}_{11}^{-1}\right).
    \end{equation*}
Thus, the proof is concluded.

\end{enumerate}

\section*{Proof of \mbox{Theorem~\ref{theorem al}}}

\begin{enumerate}[i)]
    \item 
 For any \( \boldsymbol{\beta}_1 \) and \( \boldsymbol{\beta}_2 \) such that \( \|\boldsymbol{\beta}_1 - \boldsymbol{\beta}^\star_{1}\|_2 = O_p(n^{-1/2}) \) and \( \|\boldsymbol{\beta}_2\| \leq \mathcal{C} n^{-1/2} \), following a similar derivation as in equation \eqref{a14}, one has
  \begin{align}
n \left[ \mathcal{R}^{(A)} \left( \begin{pmatrix} \boldsymbol{\beta}_1, \boldsymbol 0 \end{pmatrix}^\top \right) - \mathcal{R}^{(A)} \left( \begin{pmatrix} \boldsymbol{\beta}_1, \boldsymbol{\beta}_2 \end{pmatrix}^\top \right) \right] 
&= T_n \left( \sqrt{n} \begin{pmatrix} \boldsymbol{\beta}_1 - \boldsymbol{\beta}^\star_{1}, \boldsymbol0 \end{pmatrix}^\top \right) \notag\\
&\quad - T_n \left( \sqrt{n} \begin{pmatrix} \boldsymbol{\beta}_1 - \boldsymbol{\beta}_{1}^\star, \boldsymbol{\beta}_2 \end{pmatrix}^\top \right) \notag\\
&\quad - n\lambda \sum_{l=m+1}^{p}  w_l |{\beta}_l|\notag \\
&=I_{1n}+I_{2n},\label{a24}
\end{align}
equation \eqref{a24} is such that $I_{1n}=\allowbreak T_n \left( \sqrt{n} \begin{pmatrix} \boldsymbol{\beta}_1 - \boldsymbol{\beta}^\star_{1}, \boldsymbol0 \end{pmatrix}^\top \right)
 - T_n \left( \sqrt{n} \begin{pmatrix} \boldsymbol{\beta}_1 - \boldsymbol{\beta}_{1}^\star, \boldsymbol{\beta}_2 \end{pmatrix}^\top \right)$ and $I_{2n}=- n\lambda \sum_{l=m+1}^{p}  w_l |{\beta}_l|$.

Observe that the first term $I_{1n}$ in \eqref{a24} coincides with its counterpart in \eqref{a14} and can therefore be controlled in the same way. In contrast, the second term $I_{2n}$ 
\begin{align}
I_{2n} 
&= - \left( n^{(1+\iota)/2} \lambda \right) \sqrt{n} \times \left( \sum_{l=m+1}^{p} \left( \sqrt{n} {(|\hat{{\beta}}_l|+1/n)} \right)^{-\iota} \left|{\beta}_l\right| \right) {\le -\frac{a_n}{M_n^{\iota}}\|\boldsymbol{v}\|_1} \label{eq:A24}
\end{align}
{where $a_n=n^{(1+\iota)/2}\lambda_n\to\infty$, $\boldsymbol{v}=\sqrt n\boldsymbol{\beta}_2$, and $M_n=1+\max_{l>m}\sqrt n|\widehat\beta_l|=O_p(1)$. Thus $a_n/M_n^\iota\to\infty$ in probability. Using the local bound from the proof of Lemma~\ref*{lemma 5},
\[
I_{1n}+I_{2n}\le\left(L_n-\frac{a_n}{M_n^\iota}\right)\|\boldsymbol{v}\|_1<0
\]
for all nonzero $\boldsymbol{v}$ in the stated ball with probability tending to one. The divergence concerns the penalty coefficient relative to $\|\boldsymbol{v}\|_1$; $I_{2n}$ itself need not diverge for perturbations tending to zero arbitrarily fast.}

Putting these facts together yields
\begin{align}
n \left[ \mathcal{R}^{(A)} \left( \begin{pmatrix} \boldsymbol{\beta}_1, \boldsymbol0 \end{pmatrix}^\top \right) - \mathcal{R}^{(A)} \left( \begin{pmatrix} \boldsymbol{\beta}_1, \boldsymbol{\beta}_2 \end{pmatrix}^\top \right) \right] < 0 \notag%
\end{align}
as \( n \to \infty \). We now obtain
\begin{align}
\mathbb{P} \left[ \inf_{  \|\boldsymbol{\beta}_2\|_2=\mathcal{C} n^{-1/2}    }  \mathcal{{R}}^{\text{(A)}} \left( \begin{pmatrix} \boldsymbol{\beta}_1, \boldsymbol{\beta}_2 \end{pmatrix}^\top \right) > \mathcal{{R}}^{\text{(A)}} \left( \begin{pmatrix} \boldsymbol{\beta}_1, \boldsymbol0 \end{pmatrix}^\top \right) \right] \geq 1-\alpha. \notag
\end{align}

Given \( N = nV \) and \mbox{Assumption~\ref{assumption 4}}, then \( \boldsymbol{\hat{\beta}}^{{\text{(A)}}}_{2} = \boldsymbol{0} \) {with probability tending to one}, this completes the proof of the result.

    \item Proceeding with calculations parallel to those in \eqref{eq theorem}, one derives
\begin{align}
n \left[ \mathcal{R}^{(A)} \left( \boldsymbol{\beta}^\star + \boldsymbol{\delta}/\sqrt{n} \right) - {\mathcal{R}^{(A)}} (\boldsymbol{\beta}^\star) \right] 
&= {\sum_{i=1}^{n}} \left( \ell_{\tau,k} \left( \varepsilon_{i} - \boldsymbol{\delta}^\top \boldsymbol{X}_{i}/\sqrt{n} \right) - \ell_{\tau,k} (\varepsilon_{i}) \right) \notag\\
&\quad + n [\nabla \mathcal{R}_{N,\nu} (\boldsymbol{\beta}^0)]^\top\boldsymbol{\delta}/\sqrt{n}  \notag\\
&\quad + n\lambda \sum_{l=1}^{p} \left[{w}_l \left| {\beta}^\star_l + \delta_l/\sqrt{n} \right| - {w}_l \left| {\beta}^\star_l \right| \right]. \label{eq:A27}
\end{align}

First, consider the third term in \eqref{eq:A27}. We now show that, for any active index $l\le m$ with $\beta_l^\star\neq 0$, one has
\[
 w_l \xrightarrow{\mathcal P} |\beta_l^\star|^{-\iota}.
\]
and 

\begin{align}
n\lambda \left[ {w}_l \left| {\beta}^\star_l + \delta_l/\sqrt{n} \right| - {w}_l |{\beta}^\star_l| \right] 
&\overset{\mathcal{P}}\to 0 \label{eq:A28}
\end{align}
as \( \sqrt{n}\lambda \to 0 \) and \( \sqrt{n} \left( \left| {\beta}^\star_l + \delta_l/\sqrt{n} \right| - | {\beta}^\star_l| \right) \to \delta_l \, \mathrm{sgn}({\beta}^\star_l) \).

Recall the adaptive weights \( w_l = {(|\hat\beta_l|+1/n)^{-\iota}}\), with \(\hat\beta_l \xrightarrow{\mathcal P} \beta_l^\star\) by root-\(n\) consistency of the preliminary estimator. Hence, by the Continuous Mapping Theorem,

(Indeed, since $\hat\beta_l \xrightarrow{\mathcal P} \beta_l^\star$ and the map $s\mapsto |s|^{-\iota}$ is continuous at any $s\neq 0$, the continuous mapping theorem yields
\[
|\hat\beta_l|^{-\iota}\xrightarrow{\mathcal P} |\beta_l^\star|^{-\iota}.
\]
Hence the claim (in particular, if $ w_l=|\hat\beta_l|^{-\iota}$, or more generally $ w_l=(|\hat\beta_l|+\varpi_n)^{-\iota}$ with $\varpi_n\to 0$).)

\[
 w_l \;=\; {(|\hat\beta_l|+1/n)^{-\iota}} \xrightarrow{\mathcal P} |\beta_l^\star|^{-\iota} \qquad (l=1,\dots,m),
\]
because \(|\beta_l^\star|>0\) on the active set.

Next, since \(|\cdot|\) is differentiable at any nonzero point with derivative \(\mathrm{sgn}(\cdot)\), we have the first-order expansion (Taylor with remainder)
\[
\Big| \beta_l^\star + \tfrac{\delta_l}{\sqrt{n}} \Big| - |\beta_l^\star|
= \tfrac{\delta_l}{\sqrt{n}}\,\mathrm{sgn}(\beta_l^\star) \;+\; r_{n,l},
\]
where the remainder satisfies \(r_{n,l}=o\big(\tfrac{1}{\sqrt{n}}\big)\) as \(n\to\infty\). More explicitly, for any fixed \(\delta_l\),
\[
\sqrt{n}\left(\Big| \beta_l^\star + \tfrac{\delta_l}{\sqrt{n}} \Big| - |\beta_l^\star|\right)
\;\longrightarrow\; \delta_l\,\mathrm{sgn}(\beta_l^\star),
\]
and the convergence is uniform for \(\boldsymbol{\delta}\) in any compact set (by standard differentiability arguments away from \(0\)).

Now decompose the penalization term as
\[
n\lambda \left[ w_l \left| \beta_l^\star + \tfrac{\delta_l}{\sqrt{n}}\right| -  w_l |\beta_l^\star| \right]
= \underbrace{(\sqrt{n}\lambda)}_{\to 0}\;
\underbrace{ w_l}_{\xrightarrow{\mathcal P} |\beta_l^\star|^{-\iota}}\;
\underbrace{\sqrt{n}\left(\Big| \beta_l^\star + \tfrac{\delta_l}{\sqrt{n}} \Big| - |\beta_l^\star|\right)}_{\to \delta_l\,\mathrm{sgn}(\beta_l^\star)}.
\]
By Slutsky's theorem, the product converges in probability to \(0\) because \(\sqrt{n}\lambda \to 0\), while the other two factors converge in probability to finite constants. Consequently,
\[
n\lambda \left[ {w}_l \left| {\beta}^\star_l + \delta_l/\sqrt{n} \right| - {w}_l |{\beta}^\star_l| \right] 
\xrightarrow{\mathcal P} 0,
\]
which is exactly \eqref{eq:A28}.

For \( l = m+1, m+2, \dots, p \), the true coefficient \( {\beta}^\star_l = 0 \), and thus

\begin{equation}\label{a27}
n\lambda \!\left[{w}_l \left| \beta^\star_l + \frac{\delta_l}{\sqrt{n}} \right| - {w}_l \left| \beta^\star_l \right| \right]
=
\begin{cases}
0, & \text{if } \delta_l = 0, \\[6pt]
\sqrt{n}\,\lambda\,{w}_l\,|\delta_l|
\;{\xrightarrow[n\to\infty]{\mathcal P}}\; \infty, & \text{if } \delta_l \neq 0,
\end{cases}
\end{equation}

due to \( \sqrt{n}\lambda {w}_l = n^{(1+\iota)/2}\lambda \left( \sqrt{n} {(|\widehat\beta_l|+1/n)} \right)^{-\iota} \) with \( \sqrt{n}{\widehat\beta_l} = O_p(1) \) and \( n^{(1+\iota)/2}\lambda \to \infty \). {The lower bound $\sqrt n\lambda_n w_l\ge a_n/M_n^\iota\to\infty$ in probability proves the claim for each fixed $\delta_l\ne0$.}

By equations \eqref{eq:A27}, \eqref{eq:A28}, \eqref{a27} and using  \mbox{Lemma~\ref{lemma 3}} and  \mbox{Assumption~\ref{assumption 2}}, one has
\begin{align}\label{a31}
\mathcal{M}_n(\boldsymbol \delta):=n \left[ \mathcal{R}^{(A)} \left( \boldsymbol{\beta}^\star + \boldsymbol{\delta}/\sqrt{n} \right) - \mathcal{R}^{(A)} (\boldsymbol{\beta}^\star) \right] 
&\overset{\mathcal D}{\to}  {\mathcal{M}_\infty}(\boldsymbol{\delta})
\end{align}
where {$\boldsymbol{Z}_1\sim\mathcal N(\boldsymbol{0},\upsilon(\tau)\boldsymbol{\Sigma}_{11}/V)$ is the limiting score in (\ref*{a19}),}
\[
\mathcal{M}_\infty(\boldsymbol{\delta}) =
\begin{cases}
\frac{ \varkappa(\tau)}{2} \boldsymbol{\delta}^{(1)\top} \boldsymbol \Sigma_{11} \boldsymbol{\delta}^{(1)} + ({\boldsymbol{Z}_1})^\top \boldsymbol{\delta}^{(1)}, & \text{if } \delta_l = 0 \text{ for } l = m+1, \dots, p, \\
 \infty,& \text{if } \delta_l \neq 0\ {\text{for some }l=m+1,\dots,p}.
\end{cases}
\]
and \( \boldsymbol{\delta}^{(1)} = (\delta_1, \delta_2, \dots, \delta_m) \).

We prove now that \( \mathcal{M}_\infty(\cdot) \) possesses a unique minimizer. On the feasible set 
\(
\mathcal F=\{\boldsymbol{\delta}\in\mathbb{R}^p:\ \delta_l=0\ \text{for }l=m+1,\dots,p\},
\)
we have
\[
\mathcal{M}_\infty(\boldsymbol{\delta})
= \frac{\varkappa(\tau)}{2}\,\boldsymbol{\delta}^{(1)\top}\boldsymbol{\Sigma}_{11}\boldsymbol{\delta}^{(1)}
+ ({\boldsymbol{Z}_1})^\top\boldsymbol{\delta}^{(1)} ,
\quad \boldsymbol{\delta}^{(1)}=(\delta_1,\dots,\delta_m).
\]
Since \(\varkappa(\tau)>0\) and \(\boldsymbol{\Sigma}_{11}\) is positive definite.
Hence the Hessian on \(\mathbb{R}^m\) is
\[
\nabla^2_{\boldsymbol{\delta}^{(1)}} \mathcal{M}_\infty
= \varkappa(\tau)\,\boldsymbol{\Sigma}_{11} \ \succ\ 0,
\]
so \(\mathcal{M}_\infty\) is strictly convex on \(\mathcal F\). Strict convexity implies at most one minimizer; coercivity
(\(\mathcal{M}_\infty(\boldsymbol{\delta})\to\infty\) as \(\|\boldsymbol{\delta}^{(1)}\|\to\infty\))
ensures existence. Therefore, \(\mathcal{M}_\infty\) admits a unique minimizer on \(\mathcal F\).
In fact, the (unique) optimizer is
\[
\boldsymbol{\delta}^{(1)*}
= -\frac{1}{\varkappa(\tau)}\,\boldsymbol{\Sigma}_{11}^{-1}{\boldsymbol{Z}_1},
\qquad
\delta_l^*=0\ \text{for }l=m+1,\dots,p.
\]

Since the  function
$\mathcal{M}_n$
is convex in \( \boldsymbol{\delta} \), one has \citep{geyer}
 \begin{align}\label{a32}
\arg\min\limits_{\boldsymbol{\delta} \in \mathbb{R}^{{p}}}  \mathcal{M}_n(\boldsymbol{\delta})  \overset{\mathcal{D}}\to \arg \min_{\boldsymbol{\delta} \in \mathbb{R}^{{p}}} \mathcal{M}_\infty(\boldsymbol{\delta}).
\end{align}

Combining equations~\eqref{a19}, \eqref{a31}, \eqref{a32} with ~\mbox{Lemma~\ref{lemma 2}} (in Appendix~\ref{appendix B}), we obtain
\begin{align}\label{eq:a00}
\sqrt{n} \left( \boldsymbol{\hat \beta}_1^{(A)} - \boldsymbol{\beta}^\star_1 \right) \overset{\mathcal{D}}\to \mathcal{N} \left( \boldsymbol 0, \frac{\upsilon(\tau)}{V\varkappa^2(\tau)} \boldsymbol{\Sigma}_{11}^{-1} \right),
\end{align}
since by definition of $\boldsymbol{\hat \beta}^{(A)}$, we have $\arg\min\limits_{\boldsymbol{\delta} \in \mathbb{R}^{{p}}}  \mathcal{M}_n(\boldsymbol{\delta}) = \sqrt{n} \left( \boldsymbol{\hat \beta}^{(A)} - \boldsymbol{\beta}^\star \right)$.

Under the asymptotic regime where \( \sqrt{n}\lambda \to 0 \) and \( n^{(\iota+1)/2} \lambda \to \infty \) as \( n \to \infty \), equation~\eqref{eq:a00} yields
\begin{align}
\sqrt{N} \left( \boldsymbol{\hat \beta}_1^{(A)} - \boldsymbol{\beta}^\star_1 \right) \overset{\mathcal{D}}\to \mathcal{N} \left( 0, \frac{\upsilon(\tau)}{\varkappa^2(\tau)} \boldsymbol{\Sigma}_{11}^{-1} \right),\notag
\end{align}
since $N=nV.$\\
With this, the proof of ~\mbox{Theorem~\ref{theorem al}} is concluded.

\end{enumerate}

\section{ Two auxiliary lemmas}\label{appendix B}

The first and second lemma are widely known. Readers who wish to explore the proofs can refer to \cite{pollard1991} and \cite{hjort2011asymptotics}.

\begin{lm} \label{lemma 1}
Let $\left\{\Upsilon_n (\boldsymbol{\delta}) : \boldsymbol{\delta} \in \mathbb{O}\right\}$ be a sequence of random convex functions defined on convex, open subset $\mathbb{O}$ of $\mathbb{R}^p$. Suppose $\Upsilon (\boldsymbol{\delta})$ is a real-valued function on $\mathbb{O}$ for which $\Upsilon_n (\boldsymbol{\delta}) \rightarrow \Upsilon (\boldsymbol{\delta})$ in probability, for each $\boldsymbol{\delta} \in \mathbb{O}$. Then for each compact subset $\mathbb{U}$ of $\mathbb{O}$,
\[
\sup_{\boldsymbol{\delta} \in \mathbb{U}} |\Upsilon_n (\boldsymbol{\delta}) - \Upsilon (\boldsymbol{\delta})| \xrightarrow{\mathcal{P}} 0,
\]
and the function $\Upsilon (\cdot)$ is necessarily convex on $\mathbb{O}$.
\end{lm}

\subsection*{Proof of {\mbox{Lemma~\ref{lemma 1}}}}
The proof of \mbox{Lemma~\ref{lemma 1}} can be found in \cite{hjort2011asymptotics}.

\begin{lm}\label{lemma 2}
Let \(\boldsymbol{A}\) be a symmetric and positive definite matrix, \({\boldsymbol B_n\xrightarrow{\mathcal D}\boldsymbol B}\) be {random vectors} and \(\mathcal{L}_n (\boldsymbol{\delta})\) be a convex objective function with argmin \(\boldsymbol{u}_n\), if
\[
    \mathcal{L}_n (\boldsymbol{\delta}) = \frac{1}{2} \boldsymbol{\delta}^\top \boldsymbol{A} \boldsymbol{\delta} + {\boldsymbol{B}_n^\top} \boldsymbol{\delta} + o_p (1)
\]
then \(\boldsymbol{u}_n \xrightarrow{\mathcal D} -\boldsymbol{A}^{-1} \boldsymbol{B}\).
 
\end{lm}

\subsection*{Proof of Lemma $\ref{lemma 2}$ }

The proof of \mbox{Lemma~\ref{lemma 2}} can be found in  \cite{hjort2011asymptotics}.

\end{enumerate}

\section{ Closed-form expressions for the proximal operator of the \texorpdfstring{$L^{k}$}{Lk}-quantile loss}\label{appendix C}

In this appendix, we derive the closed-form expressions of the proximal operator
defined in~\eqref{eq:prox} for the special cases \(k = 3/2\) and \(k = 2\).
Recall that
\[
\ell_{\tau,k}(s)
=
\Psi_\tau(s)\,|s|^{k}, 
\qquad 
\Psi_\tau(s)= \tau\mathds{1}\{s\ge 0\} + (1-\tau)\mathds{1}\{s<0\},
\]
so that \(\ell_{\tau,k}\) is continuously differentiable for \(1<k\le2\) and strictly
convex on each half-line. The proximal operator is defined as
\[
\mathrm{Prox}_{\ell_{\tau,k}}(s,\xi)
:= 
\arg\min_{\theta\in\mathbb{R}}
\Big\{
\ell_{\tau,k}(\theta)
+
\tfrac{\xi}{2}(\theta - s)^2
\Big\},
\qquad \xi>0.
\]

Because the minimizer has the same sign as $s$, we solve the first-order
conditions separately on $\{\theta\ge 0\}$ and $\{\theta<0\}$.

\subsection*{$C.1.$ Case $k=3/2$}

\medskip
\noindent\textbf{Case $s\ge 0$.}
On $[0,\infty)$,
\[
\ell_{\tau,3/2}(\theta)= \tau\,\theta^{3/2},
\qquad
Q(\theta)=\tau\theta^{3/2} + \frac{\xi}{2}(\theta-s)^2.
\]
Setting $z=\sqrt{\theta}$ and differentiating yields the quadratic equation
\[
\xi z^{2} + \tfrac{3}{2}\tau\,z - \xi s = 0,
\]
whose positive root gives
\[
\theta^\star
=
\frac{\big(k\tau - \sqrt{(k\tau)^2 + 4\xi^{2}s}\big)^{2}}{4\xi^{2}},
\qquad s\ge0,\; k=\tfrac32.
\]

\medskip
\noindent\textbf{Case $s<0$.}
On $(-\infty,0)$,
\[
\ell_{\tau,3/2}(\theta)= (1-\tau)\,(-\theta)^{3/2}.
\]
Putting $\theta=-z^2$ with $z>0$ leads to
\[
{\xi z^{2} + \tfrac{3}{2}(1-\tau)\,z + \xi s = 0,}
\]
whose positive solution gives
\[
\theta^\star
=
-
\frac{\big(k(1-\tau) - \sqrt{(k(1-\tau))^{2}-4\xi^{2}s}\big)^{2}}{4\xi^{2}},
\qquad s<0,\; k=\tfrac32.
\]

\medskip
\noindent
Combining both regions yields
\begin{equation}\label{eq:prox-k-3-2-appendixC}
\mathrm{Prox}_{\ell_{\tau,k}}(s,\xi)
=
\begin{cases}
\displaystyle
\frac{\big( k\tau - \sqrt{(k\tau)^2 + 4\xi^{2} s} \big)^2}{4\xi^{2}},
& s \ge 0,\\[10pt]
\displaystyle
-\dfrac{\big( k(1-\tau) - \sqrt{(k(1-\tau))^{2} - 4\xi^{2} s} \big)^2}{4\xi^{2}},
& s < 0,
\end{cases}
\qquad (k=3/2).
\end{equation}

\subsection*{$C.2$. Case $k=2$}

When $k=2$, the loss is quadratic on each side:
\[
\ell_{\tau,2}(s)
=
\Psi_\tau(s)\, s^{2}.
\]

\medskip
\noindent\textbf{Case $s\ge0$.}
\[
(2\tau + \xi)\theta = \xi s
\quad\Longrightarrow\quad
\theta^\star = \frac{\xi s}{\xi + 2\tau}.
\]

\medskip
\noindent\textbf{Case $s<0$.}
\[
\big(2(1-\tau)+\xi\big)\theta = \xi s
\quad\Longrightarrow\quad
\theta^\star = \frac{\xi s}{\xi + 2(1-\tau)}.
\]

\medskip
\noindent
Thus,
\begin{equation}\label{eq:prox-k-2-appendixC}
\mathrm{Prox}_{\ell_{\tau,2}}(s,\xi)
=
\frac{\xi s}{\xi + 2\big(\tau\mathds{1}\{s\ge0\} + (1-\tau)\mathds{1}\{s<0\}\big)}.
\end{equation}

\medskip
Equations~\eqref{eq:prox-k-3-2-appendixC} and
\eqref{eq:prox-k-2-appendixC} provide the closed-form proximal operators of the $L^k$-quantile loss when \(k=3/2\) and \(k=2\), respectively.  
These expressions allow all scalar proximal updates in \mbox{Algorithm~\ref{alg:admm}} to be computed analytically, leading to substantial computational gains in the proposed distributed \(L^{k}\mathrm{QR}\) method.

\clearpage
\section{Supplementary simulation and computational experiments}\label{rev:additional}
\begin{figure}[htbp]\centering
\includegraphics[width=.92\linewidth]{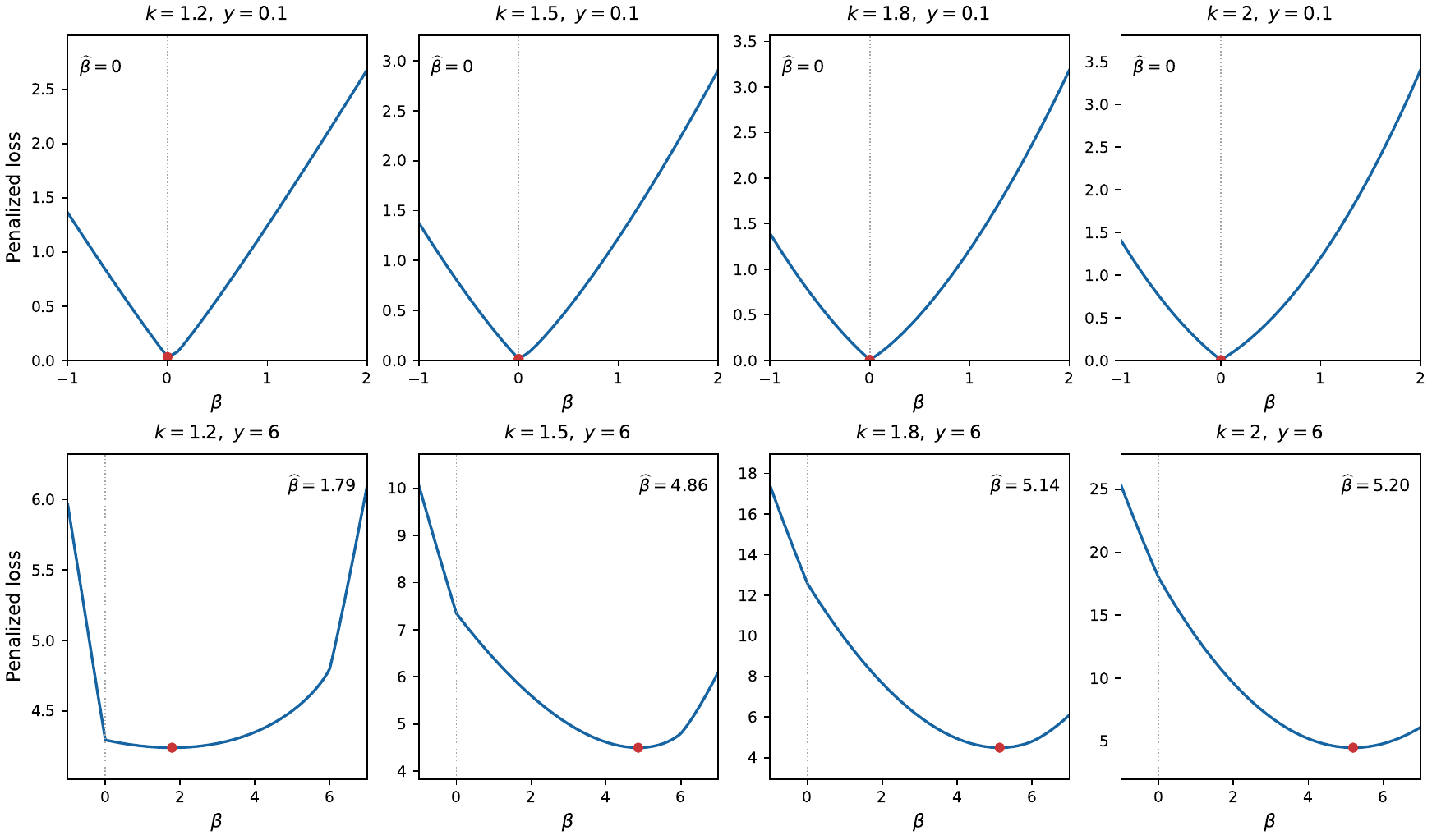}
\caption{{Scalar penalized loss $Q(\beta)=|y-\beta|^k/2+0.8|\beta|$ ($\tau=0.5$) for $k=1.2,1.5,1.8,2$. Red points mark the minimizers: zero when $y=0.1$ (top) and nonzero when $y=6$ (bottom). These fixed-weight examples illustrate sparsity, not the complete SCAD objective.}}
\label{fig:penalty_geometry}\end{figure}

\begin{figure}[htbp]\centering\includegraphics[width=\linewidth]{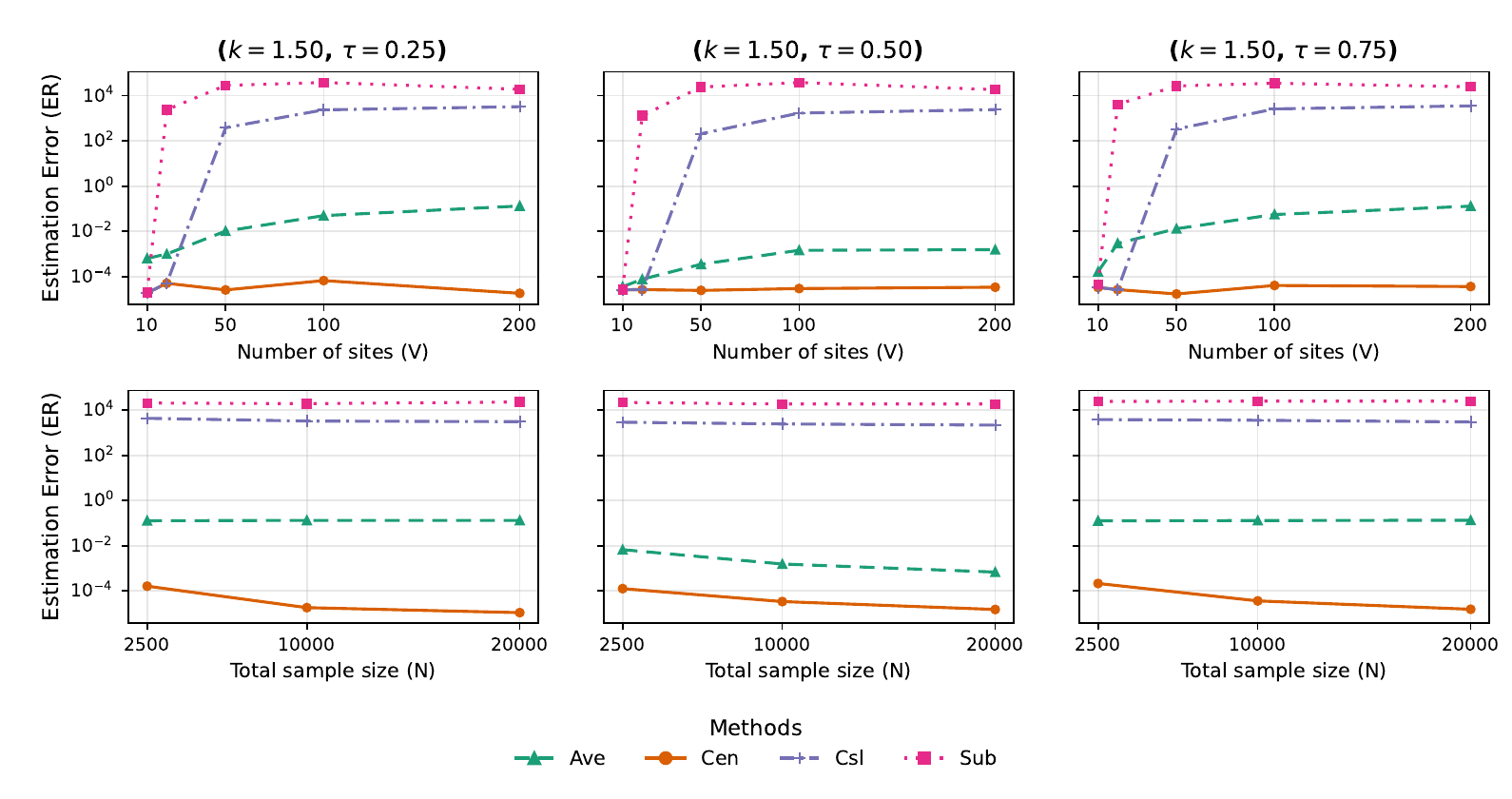}
\caption{{Scaling experiments for SCAD-penalized $L^k$QR with fixed $k=1.5$: mean squared coefficient error relative to the structural vector for $\tau=0.25$, $0.50$ and $0.75$. In each column, the upper panel varies the number of sites $V$ with $N=10{,}000$ and $p=200$, while the lower panel compares total sample sizes under the fixed-budget configurations. The vertical axes use logarithmic scales.}}\label{fig:executed_scaling_er}\end{figure}

\begin{figure}[htbp]\centering\includegraphics[width=\linewidth]{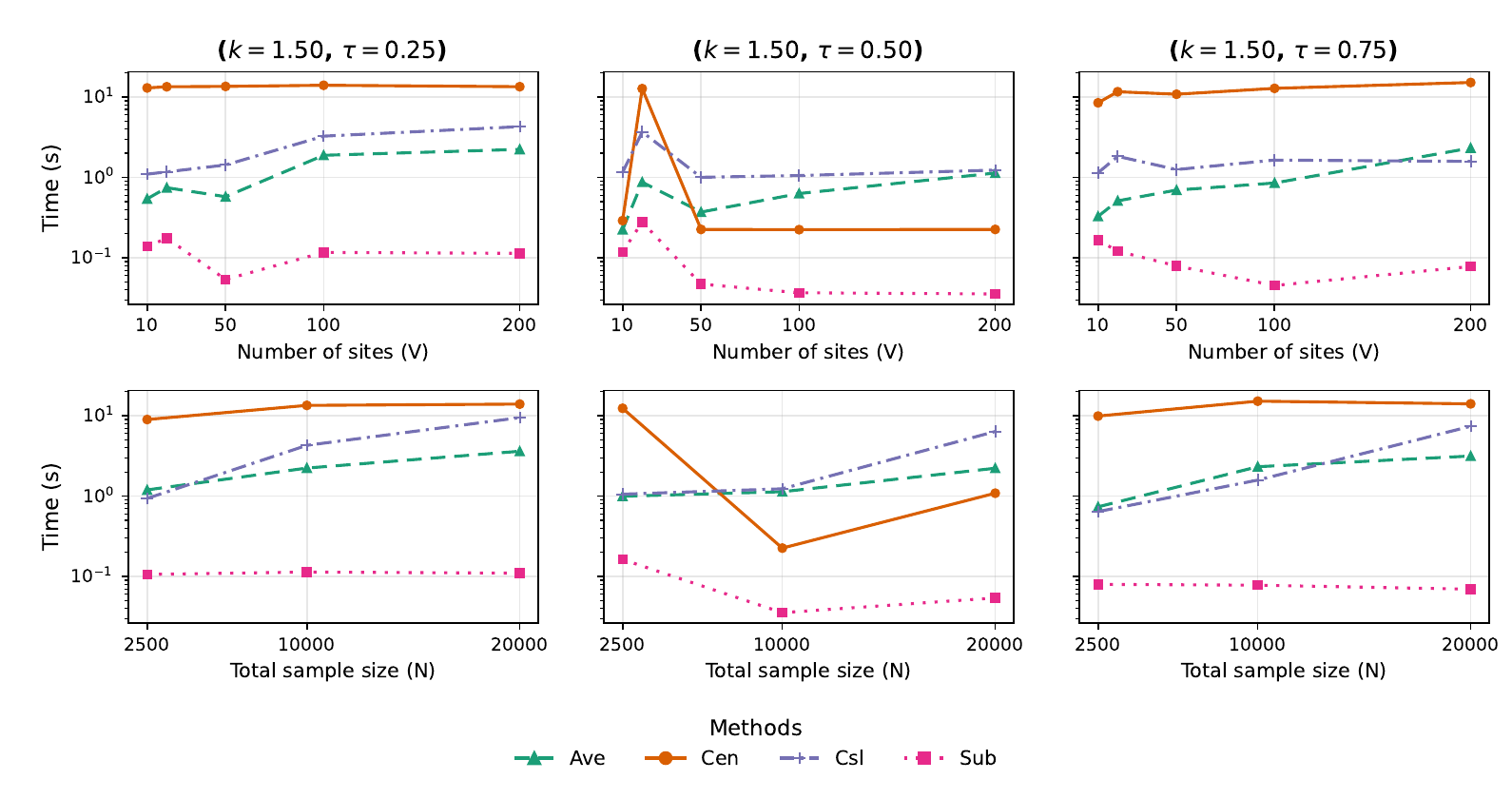}
\caption{{Scaling experiments for SCAD-penalized $L^k$QR with fixed $k=1.5$: mean fitting times for $\tau=0.25$, $0.50$ and $0.75$. In each column, the upper panel varies the number of sites $V$ with $N=10{,}000$ and $p=200$, while the lower panel compares total sample sizes under the fixed-budget configurations. The vertical axes use logarithmic scales. Times are elapsed fitting times recorded in the accuracy runs on one host, excluding tuning and transport; they are not the controlled sequential timing benchmark and may reflect concurrent workloads.}}\label{fig:executed_scaling_time}\end{figure}

\begin{figure}[htbp]\centering\includegraphics[width=.95\linewidth]{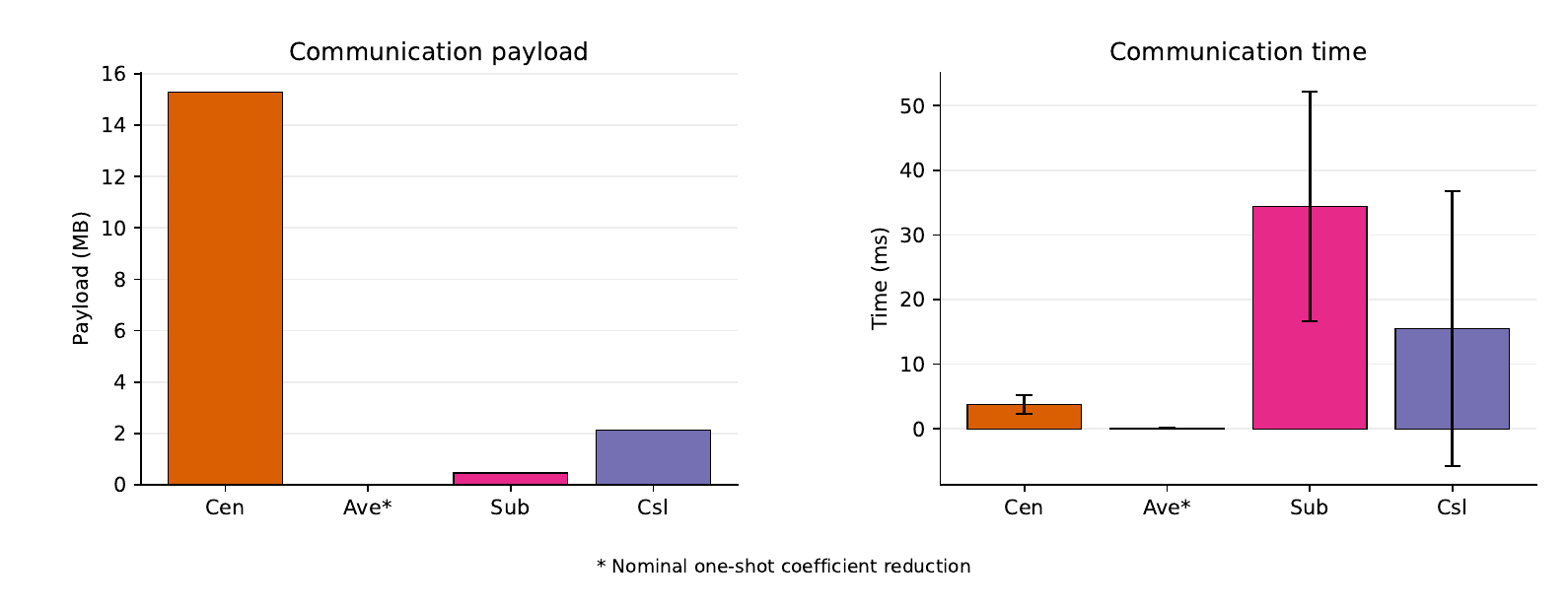}
\caption{{TCP loopback replay at $N=10000,p=200,V=20,U=5$: application payload and mean transfer time with sample standard-deviation bars. Ave denotes the nominal one-shot coefficient reduction; Csl includes residual and dual states. This is a single-host protocol replay, not end-to-end distributed estimator timing.}}\label{fig:executed_errors}\end{figure}

{\par \label{loc:penalty}
\noindent Figure~\ref{fig:penalty_geometry} illustrates a scalar penalized loss with a fixed weight. The weighted $L^1$ penalty creates a kink at zero and the coordinate update is exactly zero when $|\eta_l|\le\lambda w_l/(\mu\zeta)$, including at $k=2$. A nonzero SCAD coefficient with zero LLA weight is not thresholded at that step.}
{\par \label{loc:resources}
\noindent For dense data, matrix-vector work per inner iteration is $O(np)$ per site and $O(Np)$ centrally. Dense data storage is $8n(p+1)$ bytes per site or $8N(p+1)$ bytes centrally for double precision. Spectral setup forms the smaller Gram matrix: with $r=\min(n,p)$, its storage and eigendecomposition costs are $O(r^2)$ and $O(r^3)$, in addition to matrix construction. Thus the iteration cost is not an end-to-end complexity bound.

\par \label{loc:scaling}
\noindent The four methods are compared across a range of settings. The settings use SCAD, fixed $k=1.5$, fixed $\lambda=0.03$, 200 inner iterations and five outer rounds for Sub and Csl. Table~\ref{tab:additional_scaling} reports the fixed-budget scaling comparison at $\tau=0.5$. Figures~\ref{fig:executed_scaling_er} and~\ref{fig:executed_scaling_time} keep $k=1.5$ and extend the visual comparison to $\tau=0.25$, $0.50$ and $0.75$. The accuracy study uses common datasets across methods. At asymmetric levels, Figure~\ref{fig:executed_scaling_er} measures squared error relative to the structural coefficient vector after ordinary empirical quantile centering, not relative to the conditional $L^k$ target. The reproducibility files provide the execution settings and run-level results. In a separate timing benchmark with the same configurations and fixed budgets, local fits were evaluated sequentially on one host. Times include spectral setup and fitting but exclude data generation, tuning and transport.

\noindent The numerical implementation updates SCAD weights within inner iterations. Its Ave routine carries state between sites, and Csl averages residual and dual states indexed by different local observations. These implementation conventions matter because they need not share convergence properties with independently initialized local fits or exact LLA subproblem optimization. All accuracy outputs were finite and retained, but finite output is not a convergence certificate. The separate timing benchmark recorded a process high-water mark of about 329.4 MiB, including the runtime and earlier allocations; this is not incremental memory per fit.
}\begin{table}[htbp]\centering\small
\caption{{Fixed-budget scaling experiments: mean ER for all four methods.}}\label{tab:additional_scaling}
\begin{tabular}{rrrrrrrr}\toprule
$N$ & $p$ & $V$ & $n$ & Cen & Ave & Sub & Csl\\\midrule
1500 & 200 & 15 & 100 & 0.000221 & 0.0112 & 3.89e+04 & 2.38e+03\\
2500 & 200 & 50 & 50 & 0.000127 & 0.0067 & 2.2e+04 & 2.92e+03\\
10000 & 200 & 10 & 1000 & 2.62e-05 & 3.56e-05 & 2.63e-05 & 2.62e-05\\
10000 & 200 & 20 & 500 & 2.65e-05 & 7.64e-05 & 1.33e+03 & 2.65e-05\\
10000 & 200 & 50 & 200 & 2.45e-05 & 0.000352 & 2.43e+04 & 209\\
10000 & 200 & 100 & 100 & 2.96e-05 & 0.00146 & 3.84e+04 & 1.71e+03\\
10000 & 200 & 200 & 50 & 3.39e-05 & 0.00156 & 1.86e+04 & 2.44e+03\\
10000 & 400 & 20 & 500 & 2.62e-05 & 0.000122 & 1.49e+04 & 63.8\\
20000 & 200 & 20 & 1000 & 1.32e-05 & 1.67e-05 & 1.34e-05 & 1.32e-05\\
20000 & 200 & 400 & 50 & 1.51e-05 & 0.000684 & 1.86e+04 & 2.18e+03\\
\bottomrule\end{tabular}\end{table}
\noindent {Table~\ref{tab:additional_scaling} shows that Cen and Csl have the same reported mean ER to the displayed precision at $N=20000$, $p=200$ and $V=20$ ($1.32\times10^{-5}$). At $N=10000,p=200$, increasing $V$ from 20 to 200 reduces $n$ from 500 to 50 and increases mean Csl ER from approximately $2.65\times10^{-5}$ to $2.44\times10^3$. The tested small-local-sample configurations are unstable. At fixed $n=50$, increasing $V$ does not remove this instability. The method is therefore not immune to deterioration from inadequate local information. These experiments provide exploratory sensitivity evidence rather than precise error probabilities. $N=20000$ and $p=400$ were tested in separate configurations; no maximum capacity or parallel multi-host speedup is established. Figure~\ref{fig:executed_scaling_er} displays this sensitivity across the three levels: at fixed $N$, the Csl and Sub curves increase sharply as the number of sites grows, while Cen remains in a low-error region. At fixed $n=50$, increasing $N$ does not eliminate the large structural coefficient errors of Csl and Sub.

\noindent Figure~\ref{fig:executed_scaling_time} shows that Sub has the lowest recorded fitting times in the displayed configurations, whereas the Csl and Ave times tend to increase with $N$ at fixed $n=50$. The low fitting times of Sub must be interpreted alongside its large errors in Figure~\ref{fig:executed_scaling_er}. Because the recorded times may reflect concurrent workloads, the curves describe computational cost in these runs rather than controlled relative speedups.

\par \label{loc:transport}
\noindent For a star with one of the $V$ sites as coordinator and 8 bytes per double, raw-data pooling requires $8(V-1)n(p+1)$ bytes. A nominal one-shot coefficient reduction requires $8(V-1)p$ bytes. Sub requires $24U(V-1)p$ bytes; Csl coefficient/gradient exchanges require $32U(V-1)p$ bytes, plus $32U(V-1)n$ for residual/dual reductions and broadcasts in this implementation. Sequential state-carrying Ave instead requires $16(V-1)(p+n)$ bytes under the stated state-transfer and coefficient-reduction convention.

\noindent To assess communication overhead, we transmitted synthetic messages of the specified sizes using the Transmission Control Protocol (TCP) through a local loopback interface. Every acknowledged byte count matched. The timing excludes connection setup and includes the final byte-count acknowledgement. At $N=10000,p=200,V=20,U=5$, the payloads were 15,276,000 (Cen), 30,400 (nominal one-shot Ave), 456,000 (Sub) and 2,128,000 (Csl) bytes. Mean transfer times (run-level data are provided in the accompanying reproducibility files) were respectively 3.801, 0.157, 34.430 and 15.532 milliseconds; these single-host timings are reported separately from fitting time. Figure~\ref{fig:executed_errors} illustrates the payload reduction of all three distributed protocols relative to raw-data pooling. In particular, at $\tau=0.50$, $k=1.5$, and $N=10{,}000$, $p=200$, $V=20$, Csl and Cen both attain a reported mean ER of $2.65\times10^{-5}$ (Table~\ref{tab:additional_scaling}), to the displayed precision. Under the specified protocol, Csl reduces application payload by approximately 86.1\% relative to Cen. Together, these separate accuracy and communication measurements illustrate the benefit of Csl in this configuration. Csl exchanges more bytes than nominal one-shot Ave and Sub because it uses repeated exchanges, including residual and dual states. The state-carrying Ave count would be 212,800 bytes. Counts exclude headers and retransmissions; the replay is not an end-to-end multi-host fit and its time cannot simply be added to computation time to claim measured networked speedup.}

\clearpage
\bibliographystyle{elsarticle-harv}

\bibliography{mybib}

\end{document}